\documentclass[aps,prd,reprint,superscriptaddress,longbibliography,showpacs]{revtex4-2}

\usepackage{graphicx}
\usepackage{dcolumn}
\usepackage{bm}

\usepackage[figurename=Figure]{caption}
\usepackage{ragged2e}
\DeclareCaptionJustification{plain}{\justifying}
\usepackage{float}    
\usepackage{verbatim} 
\usepackage{amsmath}  
\usepackage{bbold}
\usepackage{upgreek} 
\usepackage{amssymb}  
\usepackage{svg}
\usepackage{enumitem} 
\usepackage{latexsym,epsfig}
\usepackage{subcaption}
\usepackage{stackrel}
\usepackage{mathtools}

\usepackage{placeins} 

\usepackage[colorlinks=true,breaklinks=true,allcolors=blue]{hyperref}
\usepackage{lipsum}
\usepackage{dsfont}

\usepackage[scr=boondox]{mathalfa}
\def\Oo{\ensuremath{{\cal O}}} 
\def\Jj{\ensuremath{{\cal J}}} 
\def\Ll{\ensuremath{{\cal L}}}
\def\Lltil{\ensuremath{{\tilde{\cal{L}}}}}

\def\Hh{\ensuremath{{\cal H}}} 
\def\Nn{\ensuremath{{\cal N}}}

\def\Kk{\ensuremath{{\cal K}}}

\usepackage[scr=boondox]{mathalfa}

\def\p{\ensuremath{{\partial}}}
\def\i{\ensuremath{{\imath}}}

\def\Tr{\ensuremath{{\operatorname{Tr}}}}

\def\p{\ensuremath{{\partial}}}

\newcommand\sbullet[1][.5]{\mathbin{\vcenter{\hbox{\scalebox{#1}{$\bullet$}}}}}

\newcommand{\be}{\begin{equation}}
	\newcommand{\ee}{\end{equation}}
\newcommand{\bea}{\begin{equation}\begin{aligned}}
		\newcommand{\eea}{\end{aligned}\end{equation}}
\newcommand{\ben}{\begin{enumerate}}
	\newcommand{\een}{\end{enumerate}}

\DeclareDocumentCommand{\nint}{ O{} O{} m }{\ensuremath{ \int_{\mbox{\scriptsize $#1$}}^{\mbox{\scriptsize$#2$}}\!\!\! \mbox{\small $\,\mathrm{d}#3$\! }}}

\usepackage{tikz,bm} 
\usepackage{circuitikz}
\usetikzlibrary{decorations.markings}
\usetikzlibrary{decorations.pathreplacing,calligraphy}

\definecolor{mycolor}{rgb}{1,0.2,0.3}
\definecolor{brightgreen}{rgb}{0.4, 1.0, 0.0}
\definecolor{britishracinggreen}{rgb}{0.0, 0.26, 0.15}
\definecolor{cadmiumgreen}{rgb}{0.0, 0.42, 0.24}
\definecolor{ceruleanblue}{rgb}{0.16, 0.32, 0.75}
\definecolor{darkelectricblue}{rgb}{0.33, 0.41, 0.47}
\definecolor{darkpowderblue}{rgb}{0.0, 0.2, 0.6}
\definecolor{darktangerine}{rgb}{1.0, 0.66, 0.07}
\definecolor{emerald}{rgb}{0.31, 0.78, 0.47}
\definecolor{palatinatepurple}{rgb}{0.41, 0.16, 0.38}
\definecolor{pastelviolet}{rgb}{0.8, 0.6, 0.79}

\begin{document}
	
	\preprint{APS/123-QED}
	
	\title{Krylov Delocalization/Localization across Ergodicity Breaking}
	
	\author{Heiko Georg Menzler}
	\email{heiko.menzler@uni-goettingen.de}
	\affiliation{%
		Institute for Theoretical Physics, Georg-August-Universit\"{a}t G\"{o}ttingen, Friedrich-Hund-Platz 1, 37077 G\"{o}ttingen, Germany
	}%
	\author{Rishabh Jha}
	\email{rishabh.jha@uni-goettingen.de}
	\affiliation{%
		Institute for Theoretical Physics, Georg-August-Universit\"{a}t G\"{o}ttingen, Friedrich-Hund-Platz 1, 37077 G\"{o}ttingen, Germany
	}%


	\begin{abstract}
		Krylov complexity has recently gained attention where the growth of operator complexity in time is measured in terms of the off-diagonal operator Lanczos coefficients. 
		The operator Lanczos algorithm reduces the problem of complexity growth to a single-particle semi-infinite tight-binding chain (known as the Krylov chain).
		Employing the phenomenon of Anderson localization, we propose the phenomenology of inverse localization length on the Krylov chain that undergoes delocalization/localization transition on the Krylov chain while the physical system undergoes ergodicity breaking. 
		On the Krylov chain we find delocalization in an ergodic regime, as we show for the SYK model, and localization in case of a weakly ergodicity-broken regime.
		Considering the dynamics beyond scrambling, we find a collapse across different operators in the ergodic regime. 
		We test for two settings: (1) the coupled SYK model, and (2) the quantum East model. 
		Our findings open avenues for mapping ergodicity/weak ergodicity-breaking transitions to delocalization/localization phenomenology on the Krylov chain.
	\end{abstract}
	
	\maketitle
	

	
	\phantomsection\label{sec:introduction}
	\addcontentsline{toc}{section}{Introduction}
	
	\section{Introduction}
	As Landau and Lifshitz remarked years ago \cite{LandauE.M.Lifshitz2014Jan}, ``According to the fundamental principles of statistical physics, the result of statistical averaging does not depend on whether it is with respect to the exact wave function of a stationary state of a closed system or by means of the Gibbs distribution.'' Thermalization in closed quantum systems has long been a topic of deep interest and has led to a wide variety of concepts such as ergodicity, quantum chaos and integrability. 
	The modern cornerstone, namely the Eigenstate Thermalization Hypothesis (ETH), was established in a series of papers \cite{Victor1977Sep, Berry1977Dec, Deutsch1991Feb,Srednicki1994Aug, Srednicki1996Feb, Srednicki1999Feb} and is generally taken as the defining feature of quantum ergodicity \cite{Rigol2008Apr, D'Alessio2016May, Deutsch2018Jul, Abanin2019May}.
	A prerequisite for quantum thermalization is quantum ergodicity that fundamentally relates to the concept of quantum chaos. However, the two are not the same as has been recently highlighted in \cite{Lim2024Jan} where a metric has been proposed for quantum states to differentiate between integrability, chaos and ergodicity.
	Recently, the ``universal operator growth hypothesis'' (UOGH) \cite{Parker2019Oct} has provided a way of detecting quantum chaos and operator complexity via the so-called \emph{Krylov complexity} (K-complexity).
	K-complexity has been extensively studied in a variety of systems ranging from many-body localized systems and non-local spin chains to open quantum systems as well as relativistic systems \cite{journey-to-edge, geometry, mbl, integrability-to-chaos, suppression-of-complexity, saddle-dominated-scrambling, double-scaled-syk, non-local-spin-chains, bose-hubbard, Tang2023Dec, Jeong1, Jeong2, Hashimoto2023Nov, dissipative-open-quantum-systems, open-quantum-systems, bi-lanczos, Liu2023Aug, Dymarsky2021Oct, Heveling2022Jul, Dymarsky2022, Adhikari2023Aug}. 
	However, as was shown in \cite{saddle-dominated-scrambling}, the UOGH can still be satisfied by some (non-chaotic) integrable systems, highlighting the difference between chaos and scrambling. 
	The exponential growth of out-of-time correlator (OTOC) is considered to be a hallmark probe for quantum chaos but it also grows exponentially for integrable semi-classical systems due to the presence of unstable saddle points \cite{Xu2020Apr, Dowling2023Nov}. 
	Therefore there exists a hierarchy in the conceptual foundation of thermalization of closed quantum systems, namely among ergodicity, quantum chaos, scrambling and integrability. 
	
	As seen above, quantum ergodicity (referred to as ergodicity from this point on) is a property of a quantum system that implies that the unitary evolution of the quantum system will eventually reach an equilibrium state in agreement with statistical mechanics. ETH is a sufficient condition for ergodicity where the \textit{conventional} ETH ansatz for an observable $\hat{A}$ is given by \cite{D'Alessio2016May} (we are using natural units throughout, therefore $\hbar = 1$)
	\begin{equation}
		\langle n | \hat{A} | m \rangle = A(\bar{E}) \delta_{m,n} + \frac{1}{\rho(\bar{E})^{1/2}} f(\bar{E}, \omega) R_{nm}
		\label{convention ETH}
	\end{equation}
	where $\bar{E} \equiv (E_n + E_m)/2$, $\omega \equiv E_n - E_m$ and $|n\rangle$ and $|m\rangle$ are eigenstates of the Hamiltonian $\Hh$ satisfying $\Hh |n\rangle = E_n |n\rangle$. $A(\bar{E})$ are the diagonal elements of the matrix while $f$ is the smooth envelope function and $R_{nm}$ is a random number derived from Gaussian ensemble with zero mean and unit variance. The many-body density of states is denoted by $\rho(\bar{E})$ which increases exponentially with increasing system size. Thus any isolated (closed) quantum system satisfying this ansatz thermalizes and thus, is quantum ergodic. The \emph{strong} ETH posits that \emph{all} eigenstates within an energy window satisfy the ETH ansatz that matches with the microcanonical ensemble predictions of statistical mechanics for that given energy window. Accordingly we define the \emph{weak ergodicity-breaking} as any deviation from the aforementioned conventional ETH where there either exists some non-thermal eigenstates that admits slow relaxation dynamics from some initial conditions, thereby showing strong dependency on initial conditions. This is in line with what has been considered in the literature, for example see Refs. \cite{Turner2018Jul, Schecter2019Oct}. Recently, a systematic analysis of how conventional ETH gets violated as one approaches the ergodicity-breaking transition point has been done in Ref. \cite{fading-ergodicity}.
	Note that the ETH ansatz is a sufficient condition for quantum ergodicity for finite systems and accordingly, the entity (called ``Krylov variance'' below) we study in this work undergoes a delocalization/localization across ergodicity breaking for finite systems where the conventional ETH ansatz fails to satisfy.
	
	Therefore, the weak ergodicity-breaking transition as discussed in this work is based on the ETH analysis of matrix elements of operators which in turn is connected to other known approaches for detecting weak ergodicity-breaking transition. For example, in the context of adiabatic gauge potential (AGP) \cite{Pandey2020Oct}, when the ETH scaling fails, the system either admits delayed thermalization \cite{Nandy2022Dec} or does not thermalize at all while being chaotic \cite{Lim2024Jan}. 
	Other examples include quantum scars and Hilbert space fragmentation that can also lead to weak ergodicity-breaking \cite{Sala2020Feb, Khemani2020May, Turner2018Jul, Serbyn2021Jun, Halimeh1, Halimeh2, Halimeh3}. Ref. \cite{fading-ergodicity} also systematically studies the relation between spectral statistics and conventional ETH violation that the authors have dubbed as ``fading ergodicity''. 
	Thus, we emphasize that we are not defining a new transition but study an entity (dubbed as ``Krylov variance'' below) across the standard ergodicity-breaking transition in finite systems (in the same spirit as other probes such as AGP \cite{Pandey2020Oct, Lim2024Jan}, in contrast to the thermodynamic limit) as has been studied in the literature in the aforementioned cited references using the standard measures.
	
	In this work, we investigate this weak breaking of quantum ergodicity in terms of complexity of arbitrary operators and study the delocalization/localization transition on the Krylov chain across the ergodicity breaking. This opens avenues to map the problem of ergodic/weak ergodicity-breaking transitions to delocalization/localization phenomenology on the Krylov chain.
	We showcase this for the Sachdev-Ye-Kitaev model (SYK) \cite{Sachdev1993May, Kitaev2015, Maldacena-syk, Chowdhury2022Sep} that has gained attraction for its analytical tractability despite being a non-integrable and chaotic system.
	The other system we consider is the quantum East model \cite{VanHorssen2015Sept,Crowley2017} which---in contrast to the coupled SYK model---has a local structure and shows at least weakly ergodicity-broken dynamics without the use of disorder \cite{Pancotti2020Jun,Bertini2024Feb}.
	The quantum East model was first introduced in the context of many-body localization (MBL) as an exemplary model that showed indicators for localization without disorder \cite{VanHorssen2015Sept,Nandkishore2015Mar} because the mechanism for weak ergodicity-breaking in the quantum East model is based on kinetic constraints \cite{Pancotti2020Jun,Bertini2024Feb}.
	Hence, there exists a natural connection to other systems that show slow thermalization \cite{Royen2024Feb} and systems with quantum scars \cite{Turner2018Jul,Chandran2023Mar}.
	Recent studies have also shown how quantum East models with particle number conservation are connected to systems observing Hilbert space fragmentation \cite{Khemani2020May, Sala2020Feb, Brighi2023Sep}.
	
	We compare both models at infinite temperature with previous studies \cite{Nandy2022Dec, Pancotti2020Jun} and we find that delocalization/localization on the Krylov chain captures the onset of weak ergodicity-breaking.
	
	\phantomsection\label{sec:setup}
	\addcontentsline{toc}{section}{Setup}
	
	\section{Review and Methodology}
	
	\subsection{Summary of Formal Setup Used}
	
	We consider a vector space of bounded operators in the Hilbert space $\mathbf{H}$ (of dimension~$\Nn$) denoted by $\mathcal{B}(\mathbf{H})$ where $\text{dim}(\mathcal{B}(\mathbf{H})) = \Nn^2$. 
	Krylov complexity is defined in terms of Heisenberg evolution of any initial operator $\Oo_0$ belonging to $\mathcal{B}(\mathbf{H})$ corresponding to a given Hamiltonian $\Hh$. 
	Hence it is easier to use the operator-to-state mapping $\Oo \to |\Oo)$ where the action of the Liouvillian operator $\Ll\equiv \i [\Hh, \sbullet[0.75]]$ leads to time evolution given by $|\Oo(t)) = \sum_{n=0}^{\infty} \frac{ t^n}{n !} |\mathcal{L}^n  \mathcal{O}_0)$. Note that in the literature on Krylov complexity, another physically equivalent definition of Liouvillian has also been used, namely $\Lltil \equiv [\Hh, \sbullet[0.75]]$. We have proved the equivalence of both the definitions and the consequent analyses in Appendix \ref{app. equivalence}. 
	The recursive application of $\Ll$ on $|\Oo_0)$ generates a Krylov space of operators where we apply the Lanczos algorithm to find the Krylov basis \cite{Viswanath1994}. 
	This requires defining an inner-product in $\mathcal{B}(\mathbf{H})$ where we chose the Wightman inner-product given by $(A | B ) \equiv \frac{1}{\Tr[e^{-\beta \Hh}]} \operatorname{Tr}\left[e^{-\beta \Hh / 2} A^{\dagger} e^{-\beta \Hh / 2} B \right]$, where $\beta$ is the inverse temperature.
	Since we investigate the infinite temperature case in this paper, this boils down to $(A |B) = \frac{1	}{\Nn} \Tr[A^\dagger B]$. 
	We generate a set of ortho-normalized basis vectors $\{ |\Oo_n)\}_{n=0}^{\Kk-1}$ where $\Kk$ is the dimension of the Krylov space bounded from above by $\Kk \leq \Nn^2 - \Nn + 1$ \cite{suppression-of-complexity}. 
	The Lanczos algorithm allows for a matrix representation for the Liouvillian operator $\Ll$ in a tri-diagonal form where the off-diagonal elements are given by the Lanczos coefficients $\{b_n\}_{n=1}^{\Kk-1}$ (diagonal elements are vanishing). 
	Details of the algorithm and the associated numerical instabilities are provided later in this section. 
	
	We can now expand any operator in the Krylov space as $|\mathcal{O}(t) )=\sum_{n=0}^{\Kk-1}  \phi_n(t) | \mathcal{O}_n)$ where we have ${| \Oo(t=0) ) = | \Oo_0)}$. The time-dependent coefficients $\phi_n(t)$ capture the spread of the operator over different Krylov basis vectors. Using the Heisenberg equation of motion, it can be shown (derived in Appendix \ref{app. krylov complexity}) that $\phi_n(t)$ satisfies the ``real-wave-equation-type'' differential equation \footnote{As derived in Appendix \ref{app. krylov complexity}, if we would use a different expansion coefficient for the operator such as $|\mathcal{O}(t) )=\sum_{n=0}^{\Kk-1}  \i^n \psi_n(t) | \mathcal{O}_n)$, then we obtain a ``Schro{\"o}dinger-type'' structure, namely $\i \p_t \psi_n(t) = b_n \psi_{n-1}(t) - b_{n+1}\psi_{n+1}(t) $. The coefficients $\phi_n(t)$ and $\psi_n(t)$ are related through $ \phi_n(t) = \i^n \psi_n(t)$.}: $ \p_t \phi_n(t)=b_n \phi_{n-1}(t)+b_{n+1} \phi_{n+1} (t)$ with the initial conditions $\phi_n(t=0) = \delta_{n,0}$, $b_{n=0} =0$ and $ \phi_{-1}(t)= 0$. Therefore solving for $\phi_n(t)$ is equivalent to evolving the operator. Thus, the Lanczos coefficients $\{ b_n\}_{n=1}^{\Kk-1}$ are physically interpreted as ``nearest-neighbor hopping amplitudes'' on this one-dimensional \emph{Krylov chain} with $\phi_n(t)$ being the ``wavefunction'' of the moving ``particle'' along the chain.
	The definition of Krylov complexity then is given by $K(t)\equiv \sum_{n=0}^{\Kk-1} n\left|\phi_n(t)\right|^2.$
	
	The UOGH \cite{Parker2019Oct} implies $b_n \sim n$ for chaotic systems while $b_n \sim \Oo(1)$ for free systems. For integrable systems, the situation is unclear and is still an open problem. If one considers Ref. \cite{Parker2019Oct}, then based on the few numerical tests done on integrable systems, $b_n \sim \sqrt{n}$. However, the situation is far from clear for characterizing integrable systems.
	Then, as shown in \cite{Tang2023Dec}, there is a saturation of $b_n \to 1$, emblematic of random matrix theory \cite{Tang2023Dec}. 
	Recall that the eigenvalues of chaotic systems are extensive in its degrees of freedom while the eigenvalues for a random matrix Hamiltonian is usually scaled to be of $\Oo(1)$. 
	That's why in order to compare the growth of operators with the random matrix behavior, we need to properly scale our results which is explained in Appendix \ref{app. krylov complexity}.

	We now provide a detailed review of various concepts explained here.
	
	\subsection{Krylov space of operators}
	\label{krylov space subsection}
	We consider a vector space of bounded operators in the Hilbert space $\mathbf{H}$ denoted by $\mathcal{B}(\mathbf{H})$. If we consider the dimension of the Hilbert space $\mathbf{H}$ as $\Nn$, then the dimension of the vector space of bounded operators $\mathcal{B}(\mathbf{H})$ is $\Nn^2$. Now if we consider any initial operator $\Oo_0 \in \mathcal{B}(\mathbf{H})$ where $\Oo_0 = \Oo(t=0)$, then the time evolution of the operator in Heisenberg picture for a given time-independent Hamiltonian $\Hh$ is given by 
	\be
	\Oo(t) = e^{\i \Hh t} \Oo_0 e^{-\i \Hh t}
	\ee
	Using the Baker-Campbell-Hausdorff formula, we get
	\begin{equation}
		\begin{aligned}
			\mathcal{O}(t)=&\mathcal{O}_0+\i t[ \Hh, \mathcal{O}_0]+\frac{(\i t)^2}{2 ! }[\Hh,[\Hh, \mathcal{O}_0]] \\
			&+\frac{(\i t)^3}{3 ! }[\Hh,[\Hh,[\Hh, \mathcal{O}_0]]]+\cdots  \\
			=& \sum_{n=0}^{\infty} \frac{t^n}{n !} \mathcal{L}^n \mathcal{O}_0
		\end{aligned}
	\end{equation}
	where Liouvillian operator $\Ll \equiv \i [H, \sbullet[0.75]]$ acts as, for instance, $\Ll\Oo_0 = \i[\Hh, \Oo_0]$ and $\Ll^2 \Oo_0 = \i^2[\Hh, [\Hh, \Oo_0]]$. Since Krylov complexity is defined in terms of evolution of operator in $\mathcal{B}(\mathbf{H})$ for a given Hamiltonian $\Hh$, it's easier to study the vector space by using the operator-to-state mapping $\Oo \to |\Oo)$. In this notation, we have
	\begin{equation}
		| \Oo(t) )=e^{ \mathcal{L} t} | \Oo_0) =\sum_{n=0}^{\infty} \frac{t^n}{n !} | \mathcal{L}^n \Oo_0)
		\label{evolution of operator by Ll}
	\end{equation}
	Each $|\Ll^n \Oo_0 )$ for a particular value of $n$ is independent from others and this forms the basis for Krylov space that is spanned by
	\begin{equation}
		\text{Krylov space} = \mathrm{span}\left[ |\Oo_0 ), |\Ll^1 \Oo_0 ), |\Ll^2 \Oo_0 ), \ldots \right]
	\end{equation}
	We employ the Lanczos algorithm where we first need to define the inner-product on $\mathcal{B}(\mathbf{H})$. We define it as follows:
	\begin{equation}
		(A | B ) =\frac{1}{\Tr[e^{-\beta \Hh}]} \operatorname{Tr}\left[e^{-\beta \Hh / 2} A^{\dagger} e^{-\beta \Hh / 2} B \right]
		\label{inner product for finite temp}
	\end{equation}
	where we have chosen the (\emph{regularized}) Wightman inner-product such that the Boltzmann factor is split equally for both the operators inserted. See Ref. \cite{wightman-vs-standard-inner-product} for a nice comparison of the UOGH with the standard inner-product at finite temperatures, namely $(A|B)^S = \frac{1}{\Tr[e^{-\beta \Hh}]} \operatorname{Tr}\left[e^{-\beta \Hh } A^{\dagger}  B\right]$. Since we are interested in the infinite temperature limit (inverse temperature $\beta = 0$), both the standard and the Wightman inner-producst match and is given by
	\begin{equation}
		(A |B) = \frac{1	}{\Nn} \Tr[A^\dagger B]
		\label{inner product for infinite temp}
	\end{equation}
	where $\Nn$ is the dimension of the Hilbert space $\mathbf{H}$. 
	
	\subsection{Lanczos algorithm and the Krylov basis}
	\label{lanczos algo subsection}
	
	We present the operator Lanczos algorithm for infinite temperature case ($\beta =0$) where inner-product becomes Eq.~\eqref{inner product for infinite temp} but generalizing to finite temperature case is straightforward using Eq.~\eqref{inner product for finite temp}. 
	
	We start with an initial operator that is properly normalized, namely $| \Oo_0) \to \frac{1}{\sqrt{(\Oo_0| \Oo_0)}} | \Oo_0 )$. Then the procedure is
	\begin{enumerate}
		\item Set $b_0 \equiv 0$ and $\Oo_{-1} = 0$.
		\item For $n \geq 1: |A_n)= | \mathcal{L}  \mathcal{O}_{n-1} ) - b_{n-1} |\Oo_{n-2}) $.
		\item Set $b_n = \sqrt{(A_n | A_n)}$.
		\item Stop if $b_n = 0$, else set $| \Oo_n) = \frac{|A_n)}{b_n}$ and jump to step 2.
	\end{enumerate}
	This procedure will construct a set of basis vectors that are normalized and orthogonal to each other
	\begin{equation}
		\text{Krylov basis } = \{ |\Oo_n)\}_{n=0}^{\Kk-1}
		\label{krylov basis}
	\end{equation}
	as well as a set of Lanczos coefficients $\{b_n\}_{n=1}^{\Kk-1}$. Here $\Kk$ is the Krylov space dimension which is bounded from above as follows \cite{suppression-of-complexity}:
	\begin{equation}
		\Kk \leq \Nn^2 - \Nn + 1
	\end{equation}

	We always initiate our algorithm with a Hermitian operator $|\Oo_0)$. The Lanczos coefficients allow us to represent the Liouvillian operator $\Ll$ as a tri-diagonal matrix where the off-diagonal elements are the Lanczos coefficients $\{ b_n\}$ while the diagonal elements are identically zero due to the structure of the Liouvillian operator that implies $(\Oo_n|\Ll|\Oo_n) = 0$ $\forall n$. To explicitly construct the matrix, let's evaluate the arbitrary matrix element $(\Oo_m| \Ll|\Oo_n)$ as follows:
	\begin{widetext}
		\begin{equation}
			\begin{aligned}
				(\Oo_m| \Ll|\Oo_n) = (\Oo_m| \left[|A_{n+1}) + b_n |\Oo_{n-1}) \right] =b_{n+1} \underbrace{(\Oo_m| \Oo_{n+1})}_{=\delta_{m, n+1}} + b_n \underbrace{(\Oo_m | \Oo_{n-1})}_{=\delta_{m, n-1}}  = \left\{ \begin{array}{ll}
					b_{n+1}, & m=n+1  \\
					b_n, & m=n-1
				\end{array} \right. 
			\end{aligned}
			\label{matrix element of Ll}
		\end{equation}
	\end{widetext}
	Therefore the general structure of the Liouvillian matrix in terms of Krylov operator states is given by
	\begin{widetext}
		\begin{equation}
			(\Oo_m | \Ll | \Oo_n) = \Ll_{mn}=\left(\begin{array}{cccccc}
				0 & b_1 & 0 & 0 & \ldots & 0 \\
				b_1 & 0 & b_2 & 0 & \ldots & 0 \\
				0 & b_2 & 0 & b_3 & \ldots & 0 \\
				\vdots & & \ddots & & \ddots & \vdots \\
				& & & & \ddots & b_{\Kk-1} \\
				0 & 0 & 0 & \ldots & b_{\Kk-1} & 0
			\end{array}\right)
			\label{Ll operator matrix form}
		\end{equation}
	\end{widetext}
	
	\subsection{Stability of the Lanczos algorithm}
	\label{stability subsection}
	
	The Lanczos algorithm, introduced in the previous section, is known to suffer from numerical instabilities when constructing the Krylov space.
	Numerical round-off error can lead to the degradation of the orthogonality between Krylov basis vectors and subsequently to unstable sequences of Lanczos coefficients $\{b_n\}$.
	
	In order to reduce the numerical instability and ensure that the Krylov basis stays orthogonal, we implement the Lanczos method with full orthogonalization (FO) throughout the whole work. FO means that on every Lanczos iteration step, we orthogonalize $A_n$ against all operator basis states already in the Krylov basis (after step 2 in the algorithm in the previous section). Since we use FO, we can also modify step 2 of the Lanczos algorithm to reduce computational time and still ensure that we get a proper ortho-normalized set of Krylov operator basis states. We now present the algorithm we used throughout this work (also see \cite{journey-to-edge, bose-hubbard}):
	\begin{enumerate}[label=(\alph*)]
		\item Set $b_0 \equiv 0$ and $\Oo_{-1} = 0$.
		\item For $n \geq 1: |A_n)= | \mathcal{L}  \mathcal{O}_{n-1} ) $ where $|A_n)$ is enforced to be Hermitian since we always initiate in this work with a Hermitian operator $|\Oo_0)$ (see the text below Eq. \eqref{Ll preserves hermiticity}). 
		\item Full orthogonalization (FO) of $|A_n)$ against all previous $|\Oo_i)$ with $i<n$: $|A_n) \to |A_n) -\sum_{i=0}^{n-1} | \mathcal{O}_i )(\mathcal{O}_i | A_n) $.
		\item Set $b_n = \sqrt{(A_n | A_n)}$.
		\item Stop if $b_n = 0$, else set $| \Oo_n) = \frac{|A_n)}{b_n}$ and jump to step 2.
	\end{enumerate}
	
	This again leads to a properly ortho-normalized Krylov basis as in Eq. \eqref{krylov basis}. Modifying step 2 in previous subsection to step (b) above does lead to a change of structure of the Liouvillian matrix in Eq. \eqref{Ll operator matrix form}. Using this algorithm, we get as matrix elements the following:
	\begin{widetext}
		\begin{equation}
			\begin{aligned}
				(\Oo_m| \Ll|\Oo_n) = (\Oo_m| A_{n+1}) \xrightarrow[]{FO} & (\Oo_m| \left[ |A_{n+1}) -\sum_{i=0}^{n} | \mathcal{O}_i )(\mathcal{O}_i | A_{n+1})  \right] \\
				&=  (\Oo_m| A_{n+1}) - \sum_{i=0}^{n} (\Oo_m| \mathcal{O}_i )(\mathcal{O}_i | A_{n+1})
				= \left\{ \begin{array}{ll}
					b_{n+1}, & m=n+1  \\
					0, & \forall m\neq n+1
				\end{array} \right. 
			\end{aligned}
			\label{lower-triangular Ll form}
		\end{equation}
	\end{widetext}
	Therefore we obtain a lower-triangular matrix for the Liouvillian matrix $\Ll_{mn} = (\Oo_m| \Ll|\Oo_n)$ where indices $m$ and $n$ label rows and columns, respectively. But we obtain the same set of Lanczos coefficients $\{ b_n\}$ as for the full Lanczos algorithm and this modified algorithm serves to make the numerical implementation more stable \cite{journey-to-edge}. 
	
	Next, we exploit the special structure of the Liouvillian operator $\Ll \equiv \i [H, \sbullet[0.75]]$ in the sense that this form of $\Ll$ preserves Hermiticity when applied on any arbitrary Hermitian operator $\Oo_h$:
	\begin{equation}
		\begin{aligned}
			(\Ll \Oo_h)^\dagger 
			= 
			\left( \i\left[\Hh, \Oo_h\right]\right)^\dagger 
			&= 
			-\i \left(\Hh \Oo_h - \Oo_h \Hh\right)^\dagger \\
			&= 
			\i( -\Oo_h \Hh - (-) \Hh \Oo_h)
			= 
			\Ll \Oo_h 
		\end{aligned}
		\label{Ll preserves hermiticity}
	\end{equation}
	
	Throughout this work, we always start with a Hermitian operator $|\Oo_0)$. 
	Therefore, all the Krylov operator basis states obtained through the Lanczos algorithm in Eq.~\eqref{krylov basis} are---by construction---also Hermitian. 
	Therefore all the intermediate $\{ |A_n) \}$ are all Hermitian (obtained in step (b) above). 
	We find that imposing the redundant step of re-enforcing the Hermiticity condition on the $|A_n)$ after applying the Lanczos step (step (b)) significantly improves the stability of the algorithm. Even though applying the FO on $|A_n)$ in step (c) can lead to loss of Hermiticity, we find that we \emph{never} have to impose Hermiticity condition after the (FO) step (c).
	The error manifested in the overlap of the Lanczos basis continues to be negligibly small. 
	
	Next, we quantify the error analysis that we will present below.
	To show that all of the Krylov basis states $\{| \Oo_n)\}$ in Eq. \eqref{krylov basis} are orthogonal within acceptable accuracy, we calculate 
	\begin{equation}
		\epsilon_n = \max\limits_{i < n}\,\, (\Oo_i | \Oo_n)\,,
		\label{error equation}
	\end{equation}
	to asses the accuracy of the sequence $\{b_n\}$.
	
\begin{figure}[htbp]
		\centering
		
		\begin{subfigure}{0.4848\linewidth}
			\includegraphics[width=\linewidth]{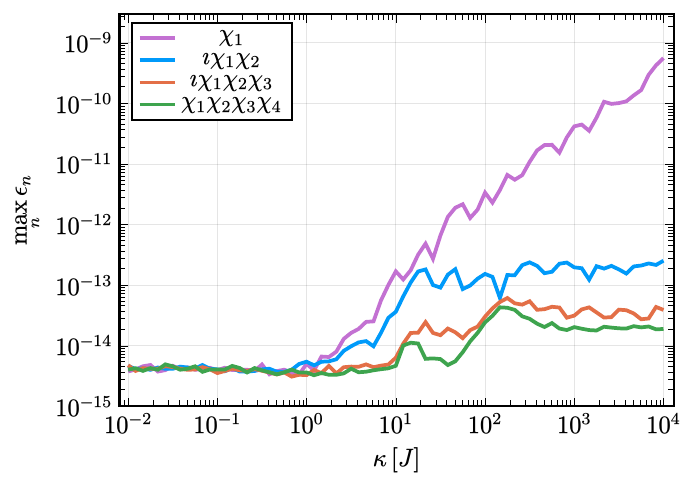}
			{\phantomsubcaption\label{fig:syk4_syk2_maxoverlaps}}
		\end{subfigure}%
		\begin{subfigure}{0.5\linewidth}
			\includegraphics[width=\linewidth]{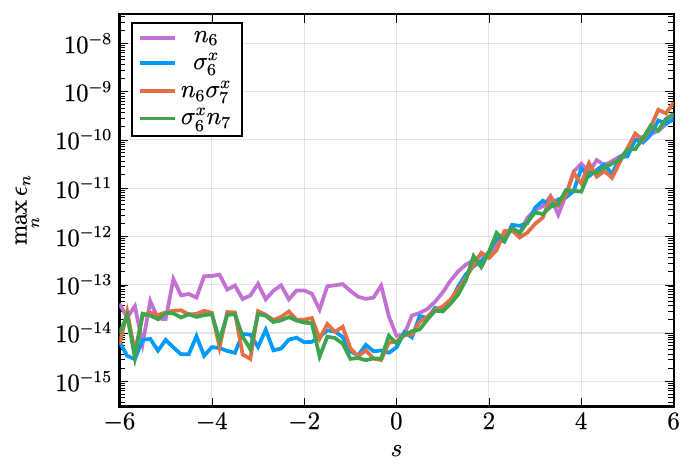}
			{\phantomsubcaption\label{fig:quantum_east_maxoverlaps}}
		\end{subfigure}
		
		\caption{
			Graphs of the maximal overlap (maximum of Eq. \eqref{error equation}) in the Lanczos basis as a function of system parameter for the coupled SYK model (left) and the quantum East model (right) for all operators shown in this work for system size $N=26$ and $L=13$, respectively.
			In the case of the coupled SYK model we not only take the maximum over all $n$ but also for all realizations.
			The numerical stability of the operator Lanczos algorithm shown here is representative of all other plots considered in this work that we have naturally checked for.
		}
		\label{error plots}
	\end{figure}

	We now present the accuracy of our data both for the coupled SYK model and the quantum East model by showing the maximum value of $\epsilon_n$. 
	As can be seen from the provided plots in Fig.~\ref{error plots}, $\max_n(\epsilon_n)$ is always negligibly small.  
	Naturally, we have also checked for other cases for all system sizes and all parameter values that we have considered in this work (including the appendices) and $\epsilon_n$ remains negligibly small. 
	From this we conclude that the Lanczos coefficients can be calculated robustly for arbitrary systems without numerical instabilities.

	\onecolumngrid
	
	\twocolumngrid

	\subsection{Krylov chain}
	\label{subsec. krylov chain}
	
	As argued in \cite{suppression-of-complexity}, the integrability of a system suppresses the $K$-complexity and this can be mapped onto the phenomenology of Anderson localization on the Krylov chain. The reason for this mapping to work, as we will also show below, is that there is a larger variance of the Lanczos coefficients $\{b_n\}$ which implies a stronger disorder in hopping amplitudes on the Krylov chain. Therefore integrability enhances localization on the Krylov chain where the localization length is given by \cite{Fleishman1977Mar, suppression-of-complexity} $l_{\text{loc}} \propto \sqrt{\Kk}/\sigma$ where $\Kk$ is the length of the Krylov chain and $\sigma$ is defined in Eq. \eqref{variance formula0} which measures
	the localization length on the Krylov chain. In the next section, we propose $\sigma^2$ as our suggested physical entity in this work for studying the delocalizing/localizing transition on the Krylov chain across ergodicity breaking after the scrambling time (this is the stage 2 of the $K$-complexity growth out of a total of 3 stages as explained in Appendix \ref{app. summary}). 
	
	We argue that due to the vast size of the operator Hilbert space $\mathcal{N}^2 - \mathcal{N} -1$, $\sigma^2$ will always be dominated by the dynamics after the ramp of the Lanczos coefficients $\{b_n\}$.
	Therefore, we ignore the initial ramp of $\{b_n\}$ and focus on the large $j$ statistics of $\{x_j\}$.
	However, the short-time dynamics on the Krylov chain are still relevant for the chaotic behavior of the system which is also captured on the Krylov chain as the scrambling time has yet not been reached. This is also the domain of the ``universal operator growth hypothesis'' \cite{Parker2019Oct}.
	Therefore, we show the analysis of $\sigma^2$ also for the entire sequence of $\{b_n\}$ below for the two models considered in this work (Fig.~\ref{fig:syk-variance_full} and Fig.~\ref{fig:east-variance_full}).
	However, these plots are only qualitative in nature, the acute sensitivity across ergodicity breaking can be deduced from the plots where the Lanczos coefficients are considered only after the scrambling time. 
	The reason is grounded in loss of notion of locality as explained and emphasized throughout this work after the scrambling time. This is what we will utilize in the next section to propose the probe \emph{after scrambling time} for studying the phenomenology of delocalization/localization transition on the Krylov chain across ergodicity breaking.
	
	\phantomsection\label{sec:main_results}
	\addcontentsline{toc}{section}{Main Results}
	
	\section{Main Results}
	The following picture emerges from the aforementioned discussions for chaotic systems with $f$ degrees of freedom between the dual behavior of Lanczos coefficients $\{b_n\}$ and Krylov complexity $K(t)$ \cite{Barbon2019Oct, integrability-to-chaos}: (1) initially a linear growth of $\{b_n\}$ for $1 \ll n<\Oo(f)$ implies an exponential growth in time of $K(t)$ for $0\lesssim t<\Oo(\log(f))$ as captured by the UOGH \cite{Parker2019Oct}, (2) a saturation after the linear growth happens for $\{b_n\}$ for $n \gg \Oo(f)$ that implies a linear-in-time growth of $K(t)$ for $t\gtrsim \Oo(\log(f))$, and (3) finally the descent of $\{b_n\}$ to zero for $n\sim \Oo(e^{2f})$ implying a saturation of $K(t)$ for $t\sim \Oo(e^{2f})$. 
	The last stage happens at an extremely late time scale that is beyond the scope of this work. 
	
	We only consider local operators having zero overlap with other conserved quantities of the system. 
	Stage (1) of the complexity growth still has notion of locality from the point of view of operators. 
	For ergodic systems, a random matrix behavior sets in during stage (2) of the evolution, allowing for a universal description of operators because the notion of locality is lost. 
	Therefore, our analysis starts \emph{after} stage (1) of complexity growth.

	Since the Krylov chain is a tight-binding model with disorder in the hopping elements, it is natural to expect phenomenology of Anderson localization (see, e.g., \cite{suppression-of-complexity, Fleishman1977Mar} for an integrable system).
	For a fixed length of the Krylov chain, we  propose the inverse of localization length given by the square root of
	\begin{equation}
		\sigma^2 =\text{Var}(x_j) \qquad \text{where} \quad x_j \equiv \ln\left(\frac{b_{2j-1}}{b_{2j}}\right) \,,
		\label{variance formula0}
	\end{equation}
	as a tool to study delocalization/localization transition on the Krylov chain as the physical system undergoes ergodicity breaking. Recall from Section \ref{subsec. krylov chain} that $\sigma$ is related to the localization length on the Krylov chain as $l_{\text{loc}} \propto \sqrt{\Kk}/\sigma$.
	We will refer to $\sigma^2$ as defined in Eq. \eqref{variance formula0} as the \emph{Krylov variance} or simply \emph{variance} for brevity from this point on. 
	We note the differences from the two related works \cite{suppression-of-complexity, Hashimoto2023Nov}. 
	Krylov variance has been studied in the context of integrable and weak-integrability broken systems in \cite{suppression-of-complexity} where they studied the localization on the Krylov chain in a qualitative sense. 
	Furthermore, the variance was used in \cite{Hashimoto2023Nov} to find correlations with quantum chaos where they included the entire spectrum of Lanczos coefficients including the stage 1 of the Krylov complexity growth (see Appendix \ref{app. summary}) where the notion of operator locality is still present. 
	We fundamentally differ from their analyses in the following ways: 
	(a) we propose the Krylov variance as a physical entity for ergodicity/weak ergodicity-breaking transition (instead of integrability breaking transition) by studying the phenomenology of delocalization/localization transition on the Krylov chain, 
	(b) we only include the Lanczos coefficients after the scrambling time (stage 2) where a sensible universal description of operators becomes viable, and 
	(c) we provide a tool to showcase a collapse for different local operators in the ergodic regime after the scrambling time.

	We wish to clarify that the Krylov variance can still be employed when keeping the entire sequence of $\{b_n\}$ (as we show for the case of the large-$q$ SYK model below, also see Fig. \ref{fig:syk-variance_full}) but the collapse and the universality across different operators can only be understood in qualitative terms. 
	The phenomenology of delocalization/localization transition on the Krylov chain across ergodicity breaking can only be analyzed after the scrambling time.
	Conceptually, the localization length is enhanced for integrable systems as shown in \cite{suppression-of-complexity} and we show via the analytical case of the large-$q$ SYK model that for the ergodic regime, $\sigma^2 \to 0$ for $n \to \infty$. Therefore the ergodicity/weak ergodicity-breaking transition is mapped to a delocalization/localization phenomenology on the Krylov chain. 
	
	An alternative analysis to probe quantum thermalization has been done in \cite{NandyScarProbe, Alishahiha2024Mar} where they use the Lanczos algorithm for eigenstates (unlike the operator Lanczos algorithm we use in this work) which generates a completely different Krylov basis than ours, therefore fundamentally differing from our approach.

	\phantomsection\label{sec:syk_model}
	\addcontentsline{toc}{section}{SYK Model}
	
	\begin{figure}[htbp]
		\centering
		\includegraphics[width=\linewidth]{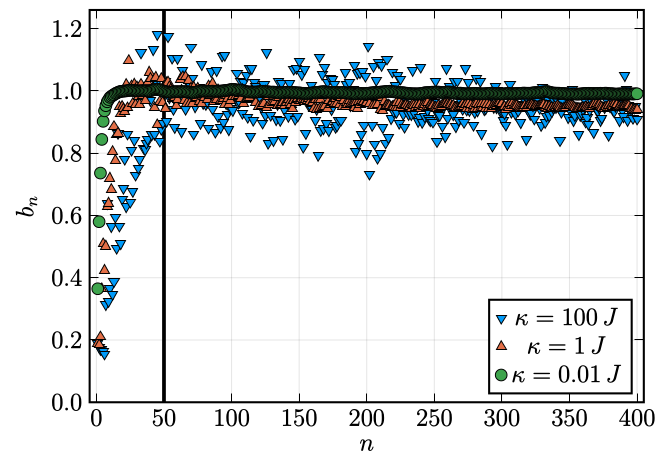}
		\caption{
			Lanczos coefficients $\{b_n\}$ at $\beta=0$ in a single realization of the Majorana SYK model (Eq.~\eqref{coupled syk model}) where $N=26$, the initial operator is $|\mathcal{O}_0)=\chi_1$. The vertical solid line at $n=50$ indicates where the initial ramp of the $b_n$ has ceased for all parameter values.
		}
		\label{fig:syk_bns}
	\end{figure}
	
	\begin{figure}[htbp]
		\centering
		\includegraphics[width=\linewidth]{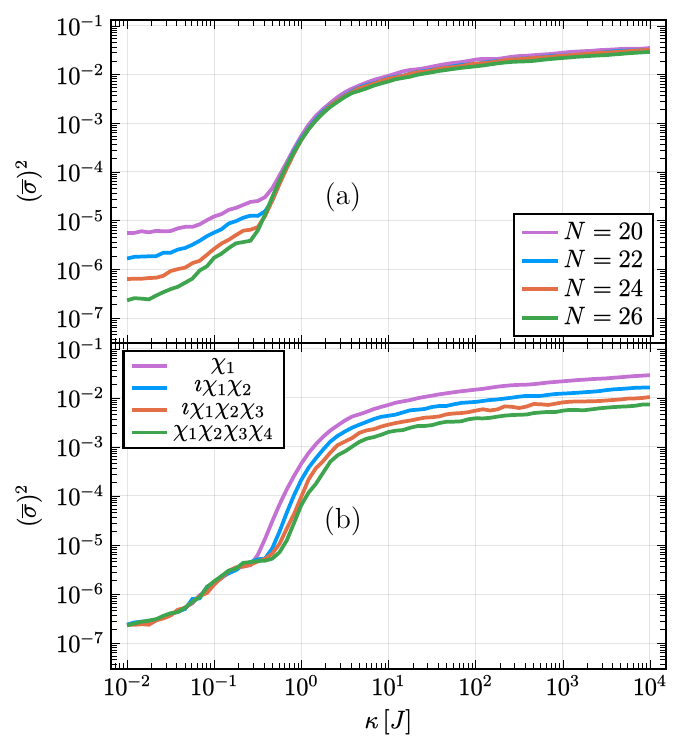}
		{\phantomsubcaption\label{fig:syk-variance_averaged_Ns}}
		{\phantomsubcaption\label{fig:syk-variance_averaged_operators}}
		\caption{
			Krylov variance $\sigma^2$ at $\beta=0$ against the system parameter $\kappa$ for the coupled SYK model (Eq.~\eqref{coupled syk model}) ignoring the first 50 Lanczos coefficients $\{b_n\}$ as justified through Fig.~\ref{fig:syk_bns} in the text. We have calculated the average $\overline{\sigma}$ from a total of 5 realizations before calculating $(\overline{\sigma})^2$. In \subref{fig:syk-variance_averaged_Ns}, we use $|\mathcal{O}_0)=\chi_1$ where the collapse of different system sizes captures the point of weak ergodicity-breaking.
			In \subref{fig:syk-variance_averaged_operators}, we show universal behavior across different operators for $N=26$ where there is a qualitative change in behavior around the transition point.
		}
		\label{fig:syk-variance_averaged}
	\end{figure}
	
	\section{Two Models}
	
	\subsection{SYK Model}
	\label{subsection syk model}
	Quantum ergodicity of the Majorana SYK model has been studied in \cite{Garcia-Garcia2016Dec, Altland2018May} where in the limit of large numbers of particles, the system is ergodic. Here we start by considering a general case of the large-$q$ SYK model whose Hamiltonian is given by $\Hh_q =\i^{q / 2} \sum\limits_{1 \leq i_1<\cdots<i_q \leq N} J_{i_1 \cdots i_q} \chi_{i_1} \cdots \chi_{i_q}$ (subscript $q$ denotes $q$-body interaction). Here $\chi_i$ are the Majorana fermions and $J_{i_1 \cdots i_q}$ are random variables derived from a Gaussian ensemble with zero mean and variance $\frac{(q-1) !\Jj^2}{2q N^{q-1}}  $ where we have introduced a re-scaled interaction strength $\Jj \equiv 2^{1-q} q J$ for some constant $J$. The dimension of the Hilbert space is $\Nn = 2^{N/2}$. We consider the large-$N$ semi-classical limit where the system is ergodic as studied in \cite{Garcia-Garcia2016Dec}. Using the large-$q$ expansion \cite{Maldacena-syk}, we can calculate the auto-correlation function corresponding to an initial operator, say $|\Oo_o) = \chi_1(t=0)$ that in turn leads to analytical evaluation of (non-rescaled) Lanczos coefficients $\{b_n\}$ for all $n$ \cite{Parker2019Oct}. In the infinite temperature limit, the Lanczos coefficients are given by $b_n = \Jj \sqrt{n(n-1)} + \Oo(1/q)$ for $n>1$ \footnote{$b_1$ is of the order $\Oo(\frac{1}{\sqrt{q}})$.} \cite{Parker2019Oct}. Therefore we can calculate the variance as in Eq.~\eqref{variance formula0} and find by including the entire sequence of $\{b_n\}$ that $\sigma^2 \to 0$ as $n \to \infty$. Since we know that large-$q$ SYK model instantaneously thermalizes with respect to local Green's function \cite{Eberlein2017Nov, Louw2022Feb}, the large-$q$ SYK model is ergodic. This sets the analytical expectation for ergodic systems that the Krylov variance vanishes. Therefore, on the level of the Krylov chain, we find a delocalization in contrast to the localization that happens in the weakly ergodicity-broken regime where $\sigma^2$ is enhanced. 
	Using this as a motivation, we study a system of coupled SYK models whose Hamiltonian is given by
	\begin{equation}
		\Hh = \frac{2}{\sqrt{N}}\sum\limits_{1 \leq i<j<k<l\leq N} J_{i j k l} \chi_i \chi_j \chi_k \chi_l 
		+ \i \sum\limits_{1 \leq i<j \leq N} \kappa_{i j}\chi_i \chi_j \,.
		\label{coupled syk model}
	\end{equation}
	The random couplings $J_{ijkl}$ and $\kappa_{ij}$ are drawn from a Gaussian distribution with variance $\frac{6J^2}{N^{3}}$ and $\frac{\kappa^2}{N}$ respectively, where $J$ and $\kappa$ are system parameters. We measure $\kappa$ in units of $J$ and we fix the unit system on the energy scale $J = 1$. 
	By tuning $\kappa$, we can study the ergodic and weakly ergodicity-broken regimes. 
	This model has also been considered in \cite{Nandy2022Dec} where they study thermalization dynamics of this model. 
	They employ the adiabatic gauge potential (AGP) and analyze the Thouless time via the spectral form factor \cite{Pandey2020Oct}. In their analysis using AGP, they find a weak ergodicity-breaking transition defined in terms of deviation from the conventional ETH ansataz (Eq. \eqref{convention ETH}). Finding this deviation from the ETH prediction leads to a delayed thermalization as can also be seen from the spectral form factor (Fig. 2 in Ref. \cite{Nandy2022Dec}). The conventional ETH scaling is violated as can be read from Fig. 1 in Ref. \cite{Nandy2022Dec} which they dub as $II \to III$ transition following the terminology of Ref. \cite{Monteiro2021Jan} where such a transition away from the ETH scaling was first proposed. 
	Here we consider the same model with same parameter values as considered in Ref. \cite{Nandy2022Dec} and study the same weak ergodicity-breaking transition using \emph{Krylov variance} where we find an agreement with their results.
	
	We observe a linear ramp in the sequence of $\{b_n\}$ in Fig.~\ref{fig:syk_bns} for the ergodic regime ($\kappa = 0.01 J$) as expected from UOGH \cite{Parker2019Oct}. 
	The vertical solid line at $n=50$ shows the end of this initial ramp and the onset of random matrix behavior beyond which we argue that all operators attain universal behavior due to the lost notion of locality.
	We note the large spread in the sequence of $\{b_n\}$ in the ergodicity-broken regime ($\kappa = 100\,J$) which we quantify by studying the Krylov variance $\sigma^2$ (Eq.~\eqref{variance formula0}) beyond the scrambling point ($n\ge 50$). 
	Note that the coupled SYK system needs to be averaged over disorder realizations due to the presence of random couplings.
	We have averaged $\sigma$ over multiple realizations and then plot the Krylov variance denoted by $(\overline{\sigma})^2$. 
	The reason we average over $\sigma$ instead of $\sigma^2$ is that the standard deviation $\sigma$ directly measures the inverse localization length on the Krylov chain, given by $l_{\text{loc}} \propto \sqrt{\Kk}/\sigma$. Here $\Kk$ is the length of the Krylov chain. Therefore, conceptually speaking, $\sigma$ is a quantity of interest on the Krylov chain.
	We plot the Krylov variance in Fig.~\ref{fig:syk-variance_averaged} against $\kappa$. 
	In Fig.~\ref{fig:syk-variance_averaged_Ns}, we find a near perfect collapse across different system sizes for the initial operator $|\Oo_0) = \chi_1$, capturing the weak ergodicity-breaking transition.
	This transition is also captured in \cite{Nandy2022Dec} (first proposed in \cite{Monteiro2021Jan}) where they observe a deviation from the ETH scaling of the AGP that leads to delayed thermalization (strong dependency on initial condition; see introduction for the definition of weak ergodicity-breaking) as analyzed via the spectral form factor. 
	We find a match with Ref. \cite{Nandy2022Dec} where the weak ergodicity-transition happens at $\kappa \sim 0.5$ for the given system size which keeps drifting with increasing system size.
	As can be seen in Fig.~\ref{fig:syk-variance_averaged_Ns} the value of $\sigma^2$ diminishes in the ergodic regime with increasing system size, while it is enhanced in the weakly ergodicity-broken regime.
	This signals delocalization and localization on the Krylov chain, respectively. 
	In order to show universal behavior across different operators, we plot the Krylov variance for different operators at $N=26$ in Fig.~\ref{fig:syk-variance_averaged_operators}.
	We do observe a qualitatively universal change of behavior across the transition point. 
	Again, we observe a near perfect collapse in the ergodic regime signaling \emph{universality}. 
	
\begin{figure}[htb]
		\centering
		\begin{subfigure}{0.5\linewidth}
			\includegraphics[width=\linewidth]{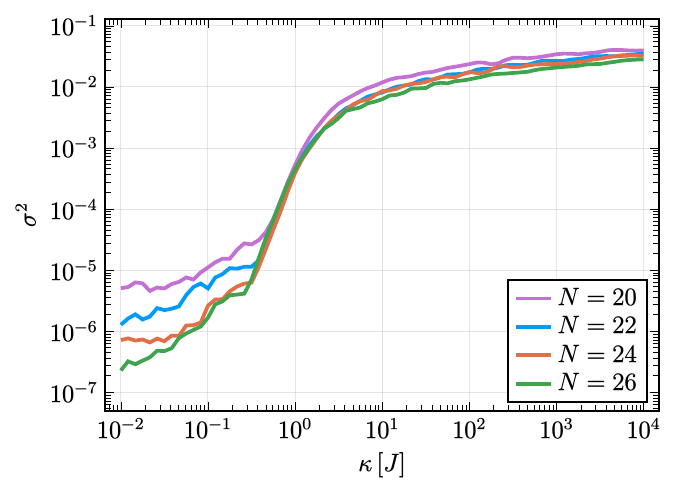}
			{\phantomsubcaption
				\label{fig:syk-variance_single}}
		\end{subfigure}%
		\begin{subfigure}{0.5\linewidth}
			\includegraphics[width=\linewidth]{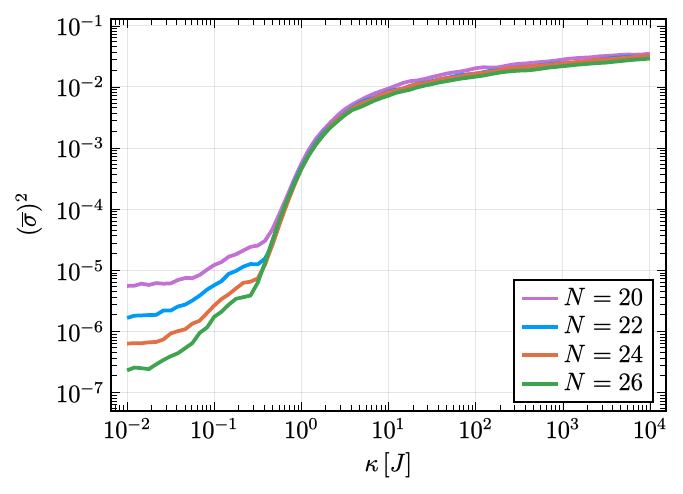}
			\phantomsubcaption
			\label{fig:syk-variance_averaged_supp}
		\end{subfigure}
		\caption{
			Plotting the Krylov variance $\sigma^2$ against $\kappa$ at $\beta=0$ for the coupled SYK model (Eq.~\eqref{coupled syk model}) ignoring the first 50 Lanczos coefficients $\{b_n\}$ where $|\mathcal{O}_0)=\chi_1$ is used.
			On the left, we show $\sigma^2$ for a single realization of the SYK model while on the right, we have calculated the average $\overline{\sigma}$ from a total of 5 realizations before calculating $(\overline{\sigma})^2$. The plot on the right is the same plot shown in Fig.~\ref{fig:syk-variance_averaged_Ns} which we again show here for comparison.
		}
		\label{fig:syk-variance}
	\end{figure}

	A natural question to ask is about the importance of the disorder realizations. We show the Krylov variance plotted against $\kappa$ in Fig.~\ref{fig:syk-variance} for a single realization as well as for multiple realizations of the SYK model.
	We can observe that a single realization shows similar features as the averaged-over-multiple-realization case which legitimizes our averaging procedure over $\sigma$ to calculate $(\overline\sigma)^2$. In Fig.~\ref{fig:syk-variance_operators}, we show the universality across different operators for a given system size by considering the Lanczos coefficients after the scrambling time. We again compare the results of a single realization against multiple realizations and we find a decent agreement between the two. In general we have found that the single realization case is already quite a good approximation to the case where we have averaged over multiple realizations.
	
\begin{figure}[htb]
		\centering
		\begin{subfigure}{0.5\linewidth}
			\includegraphics[width=\linewidth]{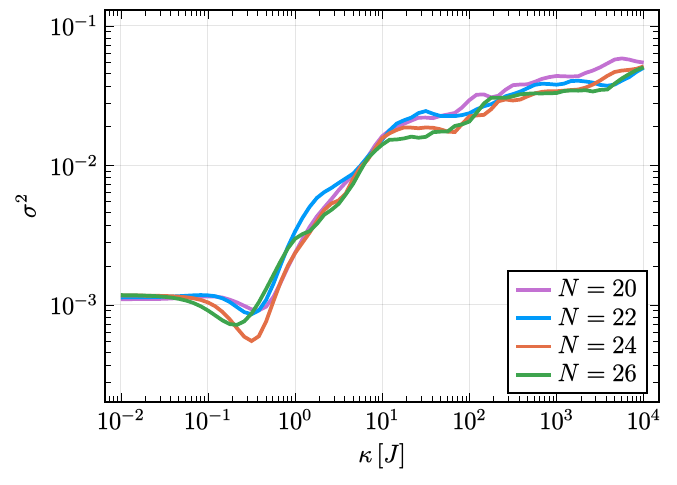}
			\phantomsubcaption
			\label{fig:syk-variance_full_single}
		\end{subfigure}%
		\begin{subfigure}{0.5\linewidth}
			\includegraphics[width=\linewidth]{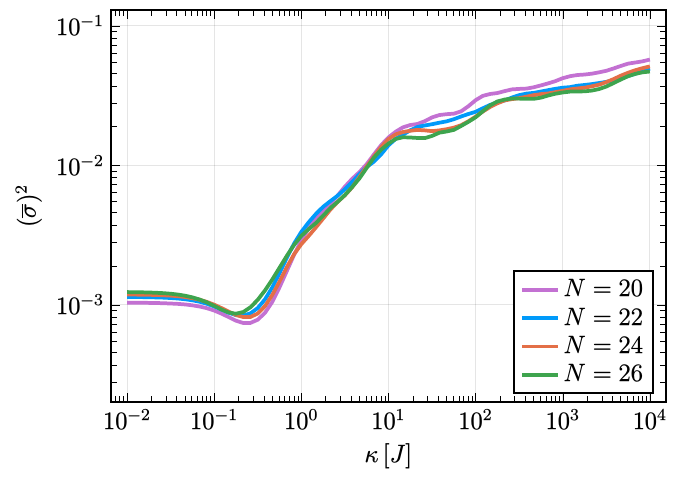}
			\phantomsubcaption
			\label{fig:syk-variance_full_averaged}
		\end{subfigure}
		\caption{
			Plotting the Krylov variance $\sigma^2$ against $\kappa$ at $\beta=0$ for the coupled SYK model (Eq.~\eqref{coupled syk model}) including \emph{all} the Lanczos coefficients $\{b_n\}$ where $|\mathcal{O}_0)=\chi_1$ is used.
			On the left, we show $\sigma^2$ for a single realization of the SYK model while on the right, we have calculated the average $\overline{\sigma}$ from a total of 5 realizations before calculating $(\overline{\sigma})^2$.
			We still observe a qualitative change in behavior around the point of weak ergodicity-breaking as observed in Fig.~\ref{fig:syk-variance_averaged} above by using our prescription. 
		}
		\label{fig:syk-variance_full}
	\end{figure}

\begin{figure}[htb]
		\centering
		\begin{subfigure}{0.5\linewidth}
			\includegraphics[width=\linewidth]{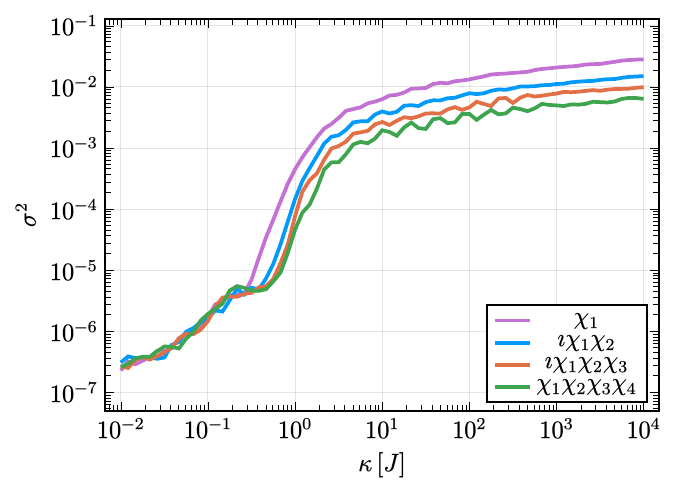}
			\phantomsubcaption
			\label{fig:syk-variance_operators_single}
		\end{subfigure}%
		\begin{subfigure}{0.5\linewidth}
			\includegraphics[width=\linewidth]{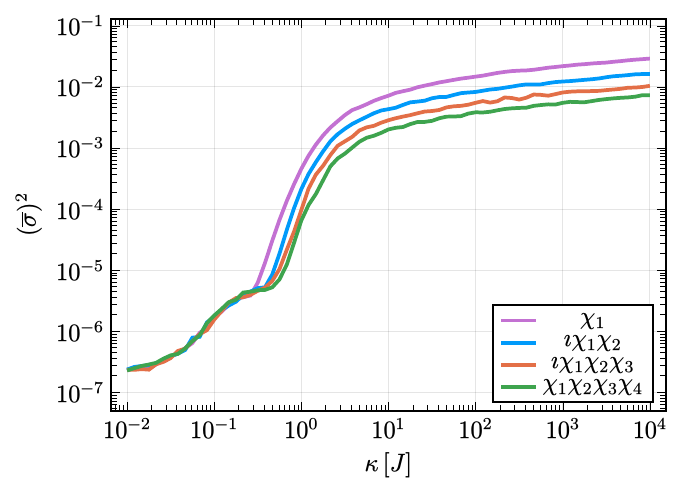}
			\phantomsubcaption
			\label{fig:syk-variance_operators_averaged}
		\end{subfigure}
		\caption{
			Plotting the Krylov variance $\sigma^2$ against $\kappa$ at $\beta=0$ for the coupled SYK model (Eq.~\eqref{coupled syk model}) for $N=26$ ignoring the first 50 Lanczos coefficients $b_n$ for different initial operators $|\Oo_0)$.
			On the left, we show $\sigma^2$ for a single realization of the SYK model while on the right, we have calculated the average $\overline{\sigma}$ from a total of 5 realizations before calculating $(\overline{\sigma})^2$. 
			The plot on the right is the same plot shown in Fig.~\ref{fig:syk-variance_averaged_operators} which we again show here for comparison.
		}
		\label{fig:syk-variance_operators}
	\end{figure}

	Moving forward, we show what happens to the Krylov variance in Eq. \eqref{variance formula0} if we consider the entire sequence of the Lanczos coefficients $\{b_n\}$ rather than just considering the sequence after the scrambling time as we have been doing until this point.  This question is pertinent also because the early Lanczos coefficients $\{b_n\}$ in the initial ramp control the early time dynamics on the Krylov chain.
	Therefore, even though we expect universality across operators for a given system size only after the scrambling has happened,
	the full Krylov variance (calculated by considering the full sequence of the Lanczos coefficients) may still give insight into the short-time dynamics of the initial state on the Krylov chain.
	Fig.~\ref{fig:syk-variance_full} shows $(\overline{\sigma})^2$ calculated using the \emph{full} sequence of $\{b_n\}$ analogous to how we detect the phenomenology of delocalization/localization transition on the Krylov chain while the physical system undergoes ergodicity breaking in Fig. \ref{fig:syk-variance_averaged} where we have considered Lanczos coefficients only after the scrambling point.
	We again see a qualitative change in behavior around the transition point but---in contrast to the results in Fig.~\ref{fig:syk-variance_averaged}---we can see that in the ergodic regime, there is already a difference of $\sim 10^3$ order of magnitude when comparing Fig.~\ref{fig:syk-variance_full} with Fig.~\ref{fig:syk-variance_operators}  for the largest system size $N=26$. 
	This reinforces our claim that a universal probe for analyzing the phenomenology of delocalization/localization transition on the Krylov chain across ergodicity breaking should be realized after the scrambling in a system has already happened.
	
	On a concluding note, we would like to remark that the Majorana SYK model possesses a particle-hole symmetry as well as charge-parity symmetry. This is discussed in detail in Appendix \ref{app. syk model}. The operator chosen in Fig. \ref{fig:syk_bns}, namely $|\Oo_0) = \chi_1$ has support in both charge parity sectors and we show in Appendix \ref{app. syk model}, the same analysis (Fig. \ref{fig:syk_bns_operators}) for an operator that has support in a single charge parity sector, namely $|\Oo_0) = \i \chi_1 \chi_2$. We find consistency in our analyses throughout and we refer the readers to Appendix \ref{app. syk model} for further details. We also compare the plot considered in Fig. \ref{fig:syk_bns} across different system sizes in Appendix \ref{app. syk model} (Fig. \ref{fig:syk_-bn-Ns}) for both the ergodic and ergodic-broken regimes.
	
	\onecolumngrid
	
	\twocolumngrid

	\FloatBarrier
	
	\onecolumngrid
	
	\twocolumngrid
	\phantomsection\label{sec:quantum_east_model}
	\addcontentsline{toc}{section}{Quantum East Model}
	
	\subsection{Quantum East Model}
	The Hamiltonian of the quantum East model is defined on a 1D lattice of size $L$
	\begin{equation}
		\Hh = - \frac{1}{2}\sum\limits_{i=1}^{L-1} n_i (e^{-s} \sigma^x_{i+1} - \mathbb{1}),
		\label{eq:quantum_east}
	\end{equation} 
	where $n_i$ is the projection on the spin-up state at lattice site $i$, $\sigma^x_i$ is the $x$-Pauli operator acting on that respective lattice site $i$ and $s$ is a system parameter. As $n_i$ projects on the spin-up state we can effectively consider any spin-down lattice site as a kinetic constraint for dynamics on the next lattice site. 
	
	\begin{figure}[htbp]
		\centering
		\includegraphics[width=\linewidth]{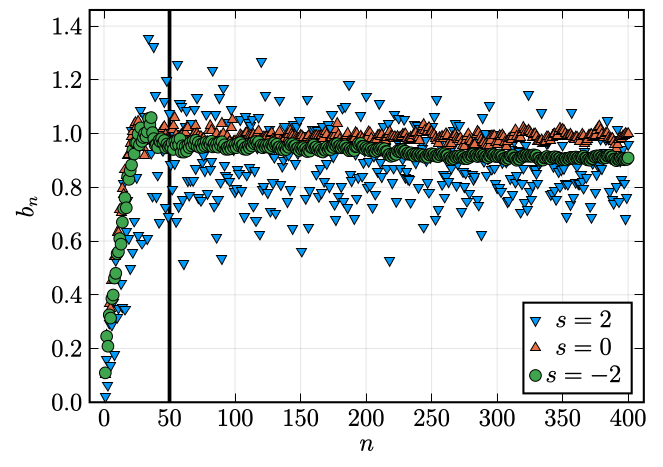}
		\caption{
			Lanczos coefficients $\{b_n\}$ at $\beta=0$ in the quantum East model (Eq.~\eqref{eq:quantum_east}) for $L=13$ where the initial operator is $|\mathcal{O}_0) =n_6$. The vertical solid line ($n=50$) shows the value of $n$ where the initial ramp of the Lanczos coefficients ends for all considered parameter values. 
		}
		\label{fig:quantum_east_bns}
	\end{figure}
	
	For appropriate boundary conditions, the ground-state of the quantum East model undergoes a sharp delocalization ($s<0$)/localization ($s>0$) transition at $s=0$ \cite{Pancotti2020Jun} (see Appendix \ref{app. quantum east model} for further details). Moreover in the regime $s \ge 0$, localized eigenstates can be constructed at arbitrary energy density. 
	As has been discussed in details in Ref. \cite{Pancotti2020Jun, VanHorssen2015Sept}, the regime for $s<0$ is ergodic in compliance with the conventional ETH (Eq. \eqref{convention ETH}) and therefore there remains no dependency whatsoever on initial conditions. 
	In contrast, for the regime $s>0$, there is a strong dependency on initial conditions, signaling the presence of some eigenstates that fail to satisfy the conventional ETH ansatz (Eq. \eqref{convention ETH}). Both of these behaviors are captured in Fig. 1 in Ref. \cite{Pancotti2020Jun}. 
	In line with considering the coupled SYK system exactly the same as in Ref. \cite{Nandy2022Dec} to match against their weak-ergodicity breaking transition in Section \ref{subsection syk model}, we use the same model for the quantum East as described in Ref. \cite{Pancotti2020Jun}. 
	
	In fact, we find below a perfect quantitative agreement using our proposed entity of Krylov variance that there exists at least a weak ergodicity-breaking transition point around $s=0$, the same as found in Ref. \cite{Pancotti2020Jun}, across which we observe a delocalization/localization transition on the Krylov chain.
	
\begin{figure}[htbp]
		\centering		\includegraphics[width=\linewidth]{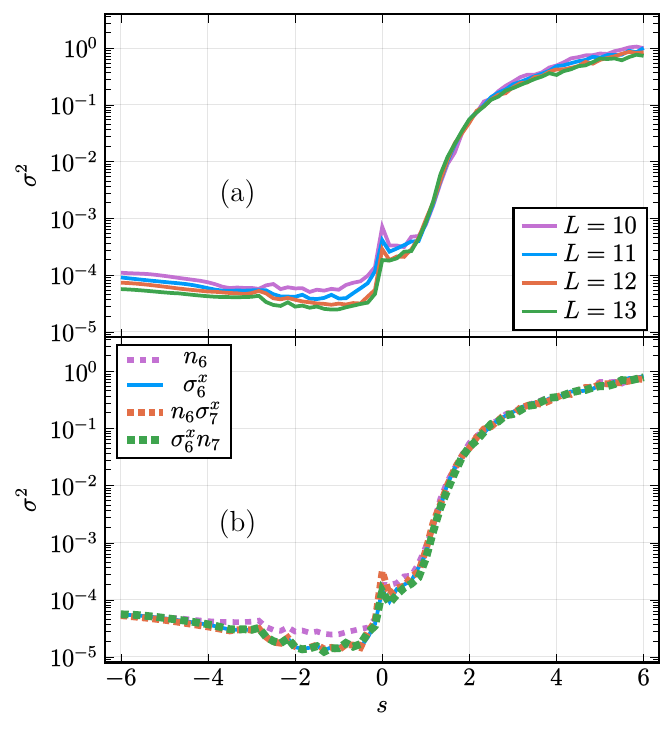}
		{\phantomsubcaption\label{fig:quantum_east_variance_Ls}}{\phantomsubcaption\label{fig:quantum_east_variance_operators}}
		\caption{
			Krylov variance $\sigma^2$ at $\beta=0$ against the system parameter $s$ in the quantum East model (Eq.~\eqref{eq:quantum_east}) for different system sizes. 
			We ignore the first 50 Lanczos coefficients $\{b_n\}$ as justified through Fig.~\ref{fig:quantum_east_bns} in the text. 
			In \subref{fig:quantum_east_variance_Ls}, we use $|\Oo_0) = n_{\lfloor L/2\rfloor}$ as an operator with support on the middle of the lattice. 
			A qualitative change in $\sigma^2$ around $s=0$ and subsequent collapse across different system sizes capture the transition point that leads to weak ergodicity-breaking. We note that the peak at $s=0$ decreases with the system size.
			In \subref{fig:quantum_east_variance_operators}, we plot for $L=13$ where the purpose is to show the qualitatively universal behavior of different operators across the transition point. 
		}
		\label{fig:quantum_east_variance}
	\end{figure}

	Similar to the SYK case, we first plot the Lanczos coefficients $\{b_n\}$ in Fig.~\ref{fig:quantum_east_bns}, where we observe a qualitatively different spreading in the sequence of $\{b_n\}$ between the ergodic and non-ergodic regime.
	In particular, we again observe a linear ramp in the ergodic regime $s=-2$ as expected from the UOGH \cite{Parker2019Oct} followed by a saturation as denoted by the vertical solid line, reminiscent of random matrix type behavior.
	To quantify the spread of $\{b_n\}$, we plot the Krylov variance $\sigma^2$ against $s$ in Fig.~\ref{fig:quantum_east_variance}, ignoring the $\{b_n\}$ of the initial ramp.
	Around the transition point $s=0$ we observe a qualitative change across different system sizes in Fig.~\ref{fig:quantum_east_variance_Ls}, signifying weak ergodicity-breaking with a subsequent collapse of the curves around $s > 1$. 
	The quantum East model is already known to have a first order phase transition at~$s=0$~\cite{Garrahan2007May, Banuls2019Nov}. 
	At the transition point~$s=0$, some eigenstates of the system undergo a sharp delocalization/localization transition which leads to non-thermal behavior and delayed relaxation to thermalization \cite{Pancotti2020Jun}.
	This delayed thermalization is manifested as weak ergodicity-breaking as already discussed in the introduction and this is indeed the point also captured by the Krylov variance in Fig.~\ref{fig:quantum_east_variance_Ls}.
	We leave the discussion of the collapse around $s=1$ for future work.
	Fig.~\ref{fig:quantum_east_variance_operators} captures how operators in the bulk of the system show universal behavior in the ergodic regime and across the transition, similar to the behavior in the coupled SYK model (Fig.~\ref{fig:syk-variance_averaged_operators}). We plot the Lanczos coefficients corresponding to the operator $|\Oo_0) = n_6$ (as used in Fig.~\ref{fig:quantum_east_bns}) for different system sizes in Appendix \ref{app. quantum east model} (Fig.~\ref{fig:quantum_east-bn-Ls}). We also provide data for another operator, namely $|\Oo_0) = \sigma^x_6$ in Appendix \ref{app. quantum east model} (Fig.~\ref{fig:quantum_east_bns_operators}) where we again find the robustness of our argument.
	
\begin{figure}[htbp]
		\centering
		\includegraphics[width=\linewidth]{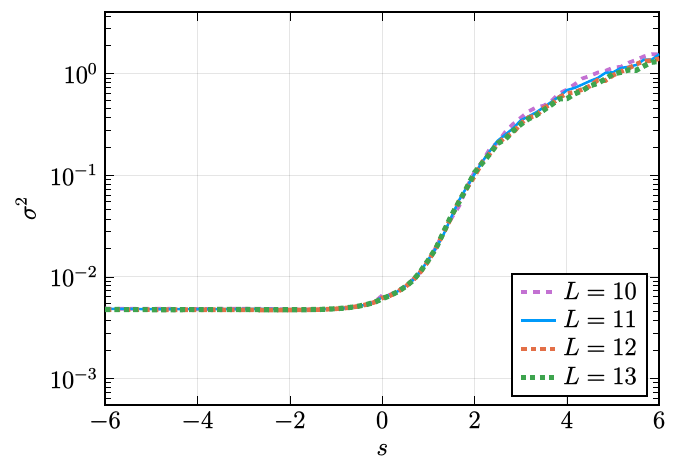}
		\caption{
			Plotting the Krylov variance $\sigma^2$ at $\beta=0$ for the quantum East model (Eq.~\eqref{eq:quantum_east}) for different system sizes where we have considered \emph{all} $b_n$ for the calculation of $\sigma^2$.
			The initial operator in the Lanczos iteration is $|\mathcal{O}_0) = n_{\lfloor L/2\rfloor}$. We can still observe a small qualitative change in behavior around $s=0$ which is the point of weak ergodicity-breaking (compare with Fig.~\ref{fig:quantum_east_variance}).
		}
		\label{fig:east-variance_full}
	\end{figure}

	We conclude this section by again comparing our analysis with the calculation of Krylov variance using \emph{all} Lanczos coefficients, both before and after scrambling. In Fig. \ref{fig:quantum_east_variance}, we showed the Krylov variance $\sigma^2$ by considering the Lanczos coefficients after the scrambling cut-off. 
	Since the early ramp of $\{b_n\}$ does capture the early-time dynamics on the Krylov chain, we also show the Krylov variance including all the Lanczos coefficients $\{b_n\}$ starting from $n=1$ in Fig.~\ref{fig:east-variance_full}. We insist that although there seems to be a perfect collapse, conceptually it only makes sense to consider the Lanczos coefficients after scrambling has happened  because only then the operators admit a universal description by loosing their respective notion of local behavior.
	Note also that in the ergodic regime there is a factor of $\sim 10^2$ difference between $\sigma^2$ with and without using the full $\{b_n\}$ while the saturation for ergodicity-broken regime remains the same.
	Therefore, there is an acute sensitivity between the two regimes when we only consider the Lanczos coefficients after the scrambling time. 

	\section{Conclusions and Discussion} 
	
	The UOGH \cite{Parker2019Oct} mapped the question of quantum chaos and operator complexity to a semi-infinite tight-binding Krylov chain. Here we argued that a universal scaling for all local operators is viable only after the scrambling time where the behavior is reminiscent of a random matrix theory. 
	We propose the measure of the inverse localization length (Eq.~\eqref{variance formula0}) on the Krylov chain as a probe for analyzing the phenomenology of delocalization/localization transition across ergodicity breaking by showing the collapse across different system sizes, where we also showed a universal behavior for different local operators in the ergodic regime. 
	Conceptually, we argued and then showed that considerations after the scrambling time allows for the Krylov variance in Eq. \eqref{variance formula0} to be used as a tool to find Anderson type localization and delocalization on the Krylov chain across ergodic-broken and ergodic regimes, respectively. Therefore we boiled down the question of ergodicity breaking to delocalization/localization phenomenology on the Krylov chain.
	The next logical extension is to generalize the findings to finite temperatures, where one has to use the Wightman inner-product as discussed above. The natural question to ask is about the universality of the Krylov variance can be answered by matching against other approaches for studying ergodicity breaking, in particular the adiabatic gauge potential (AGP) \cite{Pandey2020Oct, Lim2024Jan} following the analytical efforts made in \cite{Bhattacharjee2023Feb}. This means further inquiring into the nature of integrable systems and associated weak-integrability breaking. In this work we already compared the results for the coupled SYK model where we captured the same point of weak ergodicity-breaking where the ETH scaling is violated for the AGP leading to delayed thermalization \cite{Nandy2022Dec}. We believe that the ``raw'' Lanczos coefficients shown in Figs. \ref{fig:syk_bns} and \ref{fig:quantum_east_bns} capture a variety of physics, e.g., ergodicity breaking, phase transitions, etc. Different constructions of the Lanczos coefficients, such as the Krylov variance proposed in this work, \emph{may} capture different aspects of the underlying physics. Another playground for exploration can be open quantum systems as studied in \cite{dissipative-open-quantum-systems, open-quantum-systems, bi-lanczos, Liu2023Aug}.
	
	As an outlook, we would like to mention the possible new avenues that our findings might help exploring. Quantum chaos deals with finding the quantum signatures in the quantized model of a classically chaotic system. There are many standard probes proposed in the literature to study and characterize chaotic and ergodic behaviors in quantum systems, such as spectral statistics (see Ref. \cite{D'Alessio2016May} for a detailed review). The earliest results on which the current notion of quantum chaos and quantum ergodicity are based, came from the theory of Random Matrix Theory (RMT) which we have heavily used in our work as well. The theory of RMT came from systematic studies of the spectra of complex atomic nuclei by Wigner \cite{Wigner1, Wigner2, Wigner3} followed by Dyson \cite{Dyson1, Dyson2, Dyson3}. The other approaches to quantum chaos and quantum ergodicity are based on the study of different types of correlation functions \cite{correlation-outlook-1, correlation-outlook-2, correlation-outlook-3} and their relaxation dynamics which help characterize the system as ergodic, fractal or integrable \cite{correlation-outlook-4}. Understanding those correlation functions and their time averages in the context of the Krylov chain and using our result of mapping ergodic/ergodic-broken transition of a system to the delocalization/localization transition on the Krylov chain may shed light on what possibly can be the necessary and sufficient conditions for quantum chaos and ergodicity while addressing the difference between the two that has been recently addressed using fidelity susceptibility in Ref.~\cite{Lim2024Jan}. In particular, the correlation function and its associated properties proposed in Ref.~\cite{correlation-outlook-1} and generalized in Ref.~\cite{correlation-outlook-4} captures all three phases, namely ergodic, fractal and integrable phases~\footnote{These three phases are also captured independently using the adiabatic gauge potential (AGP), for example, as studied in the context of the coupled SYK model which was considered in this work as well \cite{Nandy2022Dec}.}. Identifying what this correlation function implies on the Krylov chain would open directions of research for characterizing integrability/integrability-transition on the Krylov chain which remains an open problem. As such, the relation between the variance of the correlation function studied in these references and the variance we have proposed in this work, namely the Krylov variance (Eq.~\eqref{variance formula0}), might lead to a deeper understanding of ergodic, fractal and localized/integrable phases including detection of such transitions, as shown in Ref.~\cite{correlation-outlook-4}. The probe suggested in this work, namely the Krylov variance using the Lanczos coefficients after the scrambling point, is sensitive to detecting deviations from the RMT behavior and the associated emerging localization but a full generalization to the integrability-breaking transitions remains obscure and the aforementioned way might serve to naturally generalize the current work. 
	Finding modifications so as to make Krylov variance a quantitative tool to detect weak ergodicity-breaking also remains an open problem that needs to be refined in future studies. 
	
	Furthermore, the scrambling time plays a crucial role in our analysis as we have argued that only after the scrambling point, the ergodic systems behave as a random matrix and therefore, the Krylov variance (Eq.~\eqref{variance formula0}) is helpful in detecting delocalization/localization transition on the Krylov chain across ergodicity-breaking transition by considering the Lanczos coefficients after the scrambling time \footnote{We also compared the Krylov variance where full sequence of Lanczos coefficients are used (both before and after the scrambling time) in Figs.~\ref{fig:syk-variance_full} and~\ref{fig:east-variance_full}. See the text in the manuscript for more details}. The Krylov chain seems to serve as a useful place in detecting the point where scrambling ends as the Lanczos coefficients saturate. This allows for a systematic study of scrambling time that can be compared against other useful time scales in the system, such as the Thouless time which is the longest relaxation time scale in a many-body system. To our best knowledge, such an analysis of Thouless time in scrambling systems has not been made in the literature. The natural question to ask is what Thouless time translates to on the Krylov chain and how the Krylov variance will behave before and after the Thouless time. There are models in the literature where precise ergodicity-breaking transition point is known. One such model that can serve as a useful playground to study such questions is the so called ``avalanche model'' or equivalently, the ``quantum Sun model'' \cite{original-quantum-sun-prl, quantum-sun-mobility-edge}. The reason for suggesting this model is that to the best of our knowledge, this is the only rigorously understood model in the many-body physics' literature that exhibits a sharp ergodic/non-ergodic transition, thereby making it a perfect playground for investigation.
	
	\section*{Data and Code Availability}
	All data and simulation codes are available on Zenodo on reasonable request \cite{menzler_2024_10975283}.

	\section*{Acknowledgments} 
	We are grateful to Salvatore R. Manmana, Fabian Heidrich-Meisner, Stefan Kehrein and Miroslav Hopjan for helpful insights and discussion.
	This work was supported by Deutsche Forschungsgemeinschaft (DFG, German Research Foundation) – 217133147, 499180199, 436382789, 493420525;  via SFB 1073 (project B03), FOR 5522, and large equipment grants (GOEGrid).	
	
	\bibliography{krylov.bib}	

\begin{thebibliography}{92}%
\makeatletter
\providecommand \@ifxundefined [1]{%
 \@ifx{#1\undefined}
}%
\providecommand \@ifnum [1]{%
 \ifnum #1\expandafter \@firstoftwo
 \else \expandafter \@secondoftwo
 \fi
}%
\providecommand \@ifx [1]{%
 \ifx #1\expandafter \@firstoftwo
 \else \expandafter \@secondoftwo
 \fi
}%
\providecommand \natexlab [1]{#1}%
\providecommand \enquote  [1]{``#1''}%
\providecommand \bibnamefont  [1]{#1}%
\providecommand \bibfnamefont [1]{#1}%
\providecommand \citenamefont [1]{#1}%
\providecommand \href@noop [0]{\@secondoftwo}%
\providecommand \href [0]{\begingroup \@sanitize@url \@href}%
\providecommand \@href[1]{\@@startlink{#1}\@@href}%
\providecommand \@@href[1]{\endgroup#1\@@endlink}%
\providecommand \@sanitize@url [0]{\catcode `\\12\catcode `\$12\catcode
  `\&12\catcode `\#12\catcode `\^12\catcode `\_12\catcode `\%12\relax}%
\providecommand \@@startlink[1]{}%
\providecommand \@@endlink[0]{}%
\providecommand \url  [0]{\begingroup\@sanitize@url \@url }%
\providecommand \@url [1]{\endgroup\@href {#1}{\urlprefix }}%
\providecommand \urlprefix  [0]{URL }%
\providecommand \Eprint [0]{\href }%
\providecommand \doibase [0]{https://doi.org/}%
\providecommand \selectlanguage [0]{\@gobble}%
\providecommand \bibinfo  [0]{\@secondoftwo}%
\providecommand \bibfield  [0]{\@secondoftwo}%
\providecommand \translation [1]{[#1]}%
\providecommand \BibitemOpen [0]{}%
\providecommand \bibitemStop [0]{}%
\providecommand \bibitemNoStop [0]{.\EOS\space}%
\providecommand \EOS [0]{\spacefactor3000\relax}%
\providecommand \BibitemShut  [1]{\csname bibitem#1\endcsname}%
\let\auto@bib@innerbib\@empty
\bibitem [{\citenamefont {Landau}\ and\ \citenamefont
  {Lifshitz}(2014)}]{LandauE.M.Lifshitz2014Jan}%
  \BibitemOpen
  \bibfield  {author} {\bibinfo {author} {\bibfnamefont {L.~D.}\ \bibnamefont
  {Landau}}\ and\ \bibinfo {author} {\bibfnamefont {E.~M.}\ \bibnamefont
  {Lifshitz}},\ }\href
  {https://shop.elsevier.com/books/statistical-physics/lifshitz/978-0-08-050350-9}
  {\emph {\bibinfo {title} {Course of Theoretical Physics, Vol. 9 Statistical
  Physics, Part-2}}}\ (\bibinfo  {publisher} {Elsevier India},\ \bibinfo {year}
  {2014})\BibitemShut {NoStop}%
\bibitem [{\citenamefont {Victor}\ and\ \citenamefont
  {Tabor}(1977)}]{Victor1977Sep}%
  \BibitemOpen
  \bibfield  {author} {\bibinfo {author} {\bibfnamefont {B.~M.}\ \bibnamefont
  {Victor}}\ and\ \bibinfo {author} {\bibfnamefont {M.}~\bibnamefont {Tabor}},\
  }\bibfield  {title} {\bibinfo {title} {{Level clustering in the regular
  spectrum}},\ }\href {https://doi.org/10.1098/rspa.1977.0140} {\bibfield
  {journal} {\bibinfo  {journal} {Proc. R. Soc. Lond. A.}\ }\textbf {\bibinfo
  {volume} {356}},\ \bibinfo {pages} {375} (\bibinfo {year}
  {1977})}\BibitemShut {NoStop}%
\bibitem [{\citenamefont {Berry}(1977)}]{Berry1977Dec}%
  \BibitemOpen
  \bibfield  {author} {\bibinfo {author} {\bibfnamefont {M.~V.}\ \bibnamefont
  {Berry}},\ }\bibfield  {title} {\bibinfo {title} {{Regular and irregular
  semiclassical wavefunctions}},\ }\href
  {https://doi.org/10.1088/0305-4470/10/12/016} {\bibfield  {journal} {\bibinfo
   {journal} {J. Phys. A: Math. Gen.}\ }\textbf {\bibinfo {volume} {10}},\
  \bibinfo {pages} {2083} (\bibinfo {year} {1977})}\BibitemShut {NoStop}%
\bibitem [{\citenamefont {Deutsch}(1991)}]{Deutsch1991Feb}%
  \BibitemOpen
  \bibfield  {author} {\bibinfo {author} {\bibfnamefont {J.~M.}\ \bibnamefont
  {Deutsch}},\ }\bibfield  {title} {\bibinfo {title} {{Quantum statistical
  mechanics in a closed system}},\ }\href
  {https://doi.org/10.1103/PhysRevA.43.2046} {\bibfield  {journal} {\bibinfo
  {journal} {Phys. Rev. A}\ }\textbf {\bibinfo {volume} {43}},\ \bibinfo
  {pages} {2046} (\bibinfo {year} {1991})}\BibitemShut {NoStop}%
\bibitem [{\citenamefont {Srednicki}(1994)}]{Srednicki1994Aug}%
  \BibitemOpen
  \bibfield  {author} {\bibinfo {author} {\bibfnamefont {M.}~\bibnamefont
  {Srednicki}},\ }\bibfield  {title} {\bibinfo {title} {{Chaos and quantum
  thermalization}},\ }\href {https://doi.org/10.1103/PhysRevE.50.888}
  {\bibfield  {journal} {\bibinfo  {journal} {Phys. Rev. E}\ }\textbf {\bibinfo
  {volume} {50}},\ \bibinfo {pages} {888} (\bibinfo {year} {1994})}\BibitemShut
  {NoStop}%
\bibitem [{\citenamefont {Srednicki}(1996)}]{Srednicki1996Feb}%
  \BibitemOpen
  \bibfield  {author} {\bibinfo {author} {\bibfnamefont {M.}~\bibnamefont
  {Srednicki}},\ }\bibfield  {title} {\bibinfo {title} {{Thermal fluctuations
  in quantized chaotic systems}},\ }\href
  {https://doi.org/10.1088/0305-4470/29/4/003} {\bibfield  {journal} {\bibinfo
  {journal} {J. Phys. A: Math. Gen.}\ }\textbf {\bibinfo {volume} {29}},\
  \bibinfo {pages} {L75} (\bibinfo {year} {1996})}\BibitemShut {NoStop}%
\bibitem [{\citenamefont {Srednicki}(1999)}]{Srednicki1999Feb}%
  \BibitemOpen
  \bibfield  {author} {\bibinfo {author} {\bibfnamefont {M.}~\bibnamefont
  {Srednicki}},\ }\bibfield  {title} {\bibinfo {title} {{The approach to
  thermal equilibrium in quantized chaotic systems}},\ }\href
  {https://doi.org/10.1088/0305-4470/32/7/007} {\bibfield  {journal} {\bibinfo
  {journal} {J. Phys. A: Math. Gen.}\ }\textbf {\bibinfo {volume} {32}},\
  \bibinfo {pages} {1163} (\bibinfo {year} {1999})}\BibitemShut {NoStop}%
\bibitem [{\citenamefont {Rigol}\ \emph {et~al.}(2008)\citenamefont {Rigol},
  \citenamefont {Dunjko},\ and\ \citenamefont {Olshanii}}]{Rigol2008Apr}%
  \BibitemOpen
  \bibfield  {author} {\bibinfo {author} {\bibfnamefont {M.}~\bibnamefont
  {Rigol}}, \bibinfo {author} {\bibfnamefont {V.}~\bibnamefont {Dunjko}},\ and\
  \bibinfo {author} {\bibfnamefont {M.}~\bibnamefont {Olshanii}},\ }\bibfield
  {title} {\bibinfo {title} {Thermalization and its mechanism for generic
  isolated quantum systems},\ }\href {https://doi.org/10.1038/nature06838}
  {\bibfield  {journal} {\bibinfo  {journal} {Nature}\ }\textbf {\bibinfo
  {volume} {452}},\ \bibinfo {pages} {854} (\bibinfo {year}
  {2008})}\BibitemShut {NoStop}%
\bibitem [{\citenamefont {D'Alessio}\ \emph {et~al.}(2016)\citenamefont
  {D'Alessio}, \citenamefont {Kafri}, \citenamefont {Polkovnikov},\ and\
  \citenamefont {Rigol}}]{D'Alessio2016May}%
  \BibitemOpen
  \bibfield  {author} {\bibinfo {author} {\bibfnamefont {L.}~\bibnamefont
  {D'Alessio}}, \bibinfo {author} {\bibfnamefont {Y.}~\bibnamefont {Kafri}},
  \bibinfo {author} {\bibfnamefont {A.}~\bibnamefont {Polkovnikov}},\ and\
  \bibinfo {author} {\bibfnamefont {M.}~\bibnamefont {Rigol}},\ }\bibfield
  {title} {\bibinfo {title} {{From quantum chaos and eigenstate thermalization
  to statistical mechanics and thermodynamics}},\ }\href
  {https://www.tandfonline.com/doi/full/10.1080/00018732.2016.1198134}
  {\bibfield  {journal} {\bibinfo  {journal} {Adv. Phys.}\ } (\bibinfo {year}
  {2016})}\BibitemShut {NoStop}%
\bibitem [{\citenamefont {Deutsch}(2018)}]{Deutsch2018Jul}%
  \BibitemOpen
  \bibfield  {author} {\bibinfo {author} {\bibfnamefont {J.~M.}\ \bibnamefont
  {Deutsch}},\ }\bibfield  {title} {\bibinfo {title} {{Eigenstate
  thermalization hypothesis}},\ }\href
  {https://doi.org/10.1088/1361-6633/aac9f1} {\bibfield  {journal} {\bibinfo
  {journal} {Rep. Prog. Phys.}\ }\textbf {\bibinfo {volume} {81}},\ \bibinfo
  {pages} {082001} (\bibinfo {year} {2018})}\BibitemShut {NoStop}%
\bibitem [{\citenamefont {Abanin}\ \emph {et~al.}(2019)\citenamefont {Abanin},
  \citenamefont {Altman}, \citenamefont {Bloch},\ and\ \citenamefont
  {Serbyn}}]{Abanin2019May}%
  \BibitemOpen
  \bibfield  {author} {\bibinfo {author} {\bibfnamefont {D.~A.}\ \bibnamefont
  {Abanin}}, \bibinfo {author} {\bibfnamefont {E.}~\bibnamefont {Altman}},
  \bibinfo {author} {\bibfnamefont {I.}~\bibnamefont {Bloch}},\ and\ \bibinfo
  {author} {\bibfnamefont {M.}~\bibnamefont {Serbyn}},\ }\bibfield  {title}
  {\bibinfo {title} {{Colloquium: Many-body localization, thermalization, and
  entanglement}},\ }\href {https://doi.org/10.1103/RevModPhys.91.021001}
  {\bibfield  {journal} {\bibinfo  {journal} {Rev. Mod. Phys.}\ }\textbf
  {\bibinfo {volume} {91}},\ \bibinfo {pages} {021001} (\bibinfo {year}
  {2019})}\BibitemShut {NoStop}%
\bibitem [{\citenamefont {Lim}\ \emph {et~al.}(2024)\citenamefont {Lim},
  \citenamefont {Matirko}, \citenamefont {Polkovnikov},\ and\ \citenamefont
  {Flynn}}]{Lim2024Jan}%
  \BibitemOpen
  \bibfield  {author} {\bibinfo {author} {\bibfnamefont {C.}~\bibnamefont
  {Lim}}, \bibinfo {author} {\bibfnamefont {K.}~\bibnamefont {Matirko}},
  \bibinfo {author} {\bibfnamefont {A.}~\bibnamefont {Polkovnikov}},\ and\
  \bibinfo {author} {\bibfnamefont {M.~O.}\ \bibnamefont {Flynn}},\ }\href
  {https://arxiv.org/abs/2401.01927} {\bibinfo {title} {Defining classical and
  quantum chaos through adiabatic transformations}} (\bibinfo {year} {2024}),\
  \Eprint {https://arxiv.org/abs/2401.01927} {arXiv:2401.01927
  [cond-mat.stat-mech]} \BibitemShut {NoStop}%
\bibitem [{\citenamefont {Parker}\ \emph {et~al.}(2019)\citenamefont {Parker},
  \citenamefont {Cao}, \citenamefont {Avdoshkin}, \citenamefont {Scaffidi},\
  and\ \citenamefont {Altman}}]{Parker2019Oct}%
  \BibitemOpen
  \bibfield  {author} {\bibinfo {author} {\bibfnamefont {D.~E.}\ \bibnamefont
  {Parker}}, \bibinfo {author} {\bibfnamefont {X.}~\bibnamefont {Cao}},
  \bibinfo {author} {\bibfnamefont {A.}~\bibnamefont {Avdoshkin}}, \bibinfo
  {author} {\bibfnamefont {T.}~\bibnamefont {Scaffidi}},\ and\ \bibinfo
  {author} {\bibfnamefont {E.}~\bibnamefont {Altman}},\ }\bibfield  {title}
  {\bibinfo {title} {{A Universal Operator Growth Hypothesis}},\ }\href
  {https://doi.org/10.1103/PhysRevX.9.041017} {\bibfield  {journal} {\bibinfo
  {journal} {Phys. Rev. X}\ }\textbf {\bibinfo {volume} {9}},\ \bibinfo {pages}
  {041017} (\bibinfo {year} {2019})}\BibitemShut {NoStop}%
\bibitem [{\citenamefont {Rabinovici}\ \emph {et~al.}(2021)\citenamefont
  {Rabinovici}, \citenamefont {S{\ifmmode\acute{a}\else\'{a}\fi}nchez-Garrido},
  \citenamefont {Shir},\ and\ \citenamefont {Sonner}}]{journey-to-edge}%
  \BibitemOpen
  \bibfield  {author} {\bibinfo {author} {\bibfnamefont {E.}~\bibnamefont
  {Rabinovici}}, \bibinfo {author} {\bibfnamefont {A.}~\bibnamefont
  {S{\ifmmode\acute{a}\else\'{a}\fi}nchez-Garrido}}, \bibinfo {author}
  {\bibfnamefont {R.}~\bibnamefont {Shir}},\ and\ \bibinfo {author}
  {\bibfnamefont {J.}~\bibnamefont {Sonner}},\ }\bibfield  {title} {\bibinfo
  {title} {{Operator complexity: a journey to the edge of Krylov space}},\
  }\href {https://doi.org/10.1007/JHEP06(2021)062} {\bibfield  {journal}
  {\bibinfo  {journal} {J. High Energy Phys.}\ }\textbf {\bibinfo {volume}
  {2021}}\bibinfo  {number} { (6)},\ \bibinfo {pages} {1}}\BibitemShut
  {NoStop}%
\bibitem [{\citenamefont {Caputa}\ \emph {et~al.}(2022)\citenamefont {Caputa},
  \citenamefont {Magan},\ and\ \citenamefont {Patramanis}}]{geometry}%
  \BibitemOpen
\bibfield  {number} {  }\bibfield  {author} {\bibinfo {author} {\bibfnamefont
  {P.}~\bibnamefont {Caputa}}, \bibinfo {author} {\bibfnamefont {J.~M.}\
  \bibnamefont {Magan}},\ and\ \bibinfo {author} {\bibfnamefont
  {D.}~\bibnamefont {Patramanis}},\ }\bibfield  {title} {\bibinfo {title}
  {{Geometry of Krylov complexity}},\ }\href
  {https://doi.org/10.1103/PhysRevResearch.4.013041} {\bibfield  {journal}
  {\bibinfo  {journal} {Phys. Rev. Res.}\ }\textbf {\bibinfo {volume} {4}},\
  \bibinfo {pages} {013041} (\bibinfo {year} {2022})}\BibitemShut {NoStop}%
\bibitem [{\citenamefont {Ballar~Trigueros}\ and\ \citenamefont
  {Lin}(2022)}]{mbl}%
  \BibitemOpen
  \bibfield  {author} {\bibinfo {author} {\bibfnamefont {F.}~\bibnamefont
  {Ballar~Trigueros}}\ and\ \bibinfo {author} {\bibfnamefont {C.-J.}\
  \bibnamefont {Lin}},\ }\bibfield  {title} {\bibinfo {title} {{Krylov
  complexity of many-body localization: Operator localization in Krylov
  basis}},\ }\href {https://doi.org/10.21468/SciPostPhys.13.2.037} {\bibfield
  {journal} {\bibinfo  {journal} {SciPost Phys.}\ }\textbf {\bibinfo {volume}
  {13}},\ \bibinfo {pages} {037} (\bibinfo {year} {2022})}\BibitemShut
  {NoStop}%
\bibitem [{\citenamefont {Rabinovici}\ \emph
  {et~al.}(2022{\natexlab{a}})\citenamefont {Rabinovici}, \citenamefont
  {S{\ifmmode\acute{a}\else\'{a}\fi}nchez-Garrido}, \citenamefont {Shir},\ and\
  \citenamefont {Sonner}}]{integrability-to-chaos}%
  \BibitemOpen
  \bibfield  {author} {\bibinfo {author} {\bibfnamefont {E.}~\bibnamefont
  {Rabinovici}}, \bibinfo {author} {\bibfnamefont {A.}~\bibnamefont
  {S{\ifmmode\acute{a}\else\'{a}\fi}nchez-Garrido}}, \bibinfo {author}
  {\bibfnamefont {R.}~\bibnamefont {Shir}},\ and\ \bibinfo {author}
  {\bibfnamefont {J.}~\bibnamefont {Sonner}},\ }\bibfield  {title} {\bibinfo
  {title} {{Krylov complexity from integrability to chaos}},\ }\href
  {https://doi.org/10.1007/JHEP07(2022)151} {\bibfield  {journal} {\bibinfo
  {journal} {J. High Energy Phys.}\ }\textbf {\bibinfo {volume} {2022}}\bibinfo
   {number} { (7)},\ \bibinfo {pages} {1}}\BibitemShut {NoStop}%
\bibitem [{\citenamefont {Rabinovici}\ \emph
  {et~al.}(2022{\natexlab{b}})\citenamefont {Rabinovici}, \citenamefont
  {S{\ifmmode\acute{a}\else\'{a}\fi}nchez-Garrido}, \citenamefont {Shir},\ and\
  \citenamefont {Sonner}}]{suppression-of-complexity}%
  \BibitemOpen
\bibfield  {number} {  }\bibfield  {author} {\bibinfo {author} {\bibfnamefont
  {E.}~\bibnamefont {Rabinovici}}, \bibinfo {author} {\bibfnamefont
  {A.}~\bibnamefont {S{\ifmmode\acute{a}\else\'{a}\fi}nchez-Garrido}}, \bibinfo
  {author} {\bibfnamefont {R.}~\bibnamefont {Shir}},\ and\ \bibinfo {author}
  {\bibfnamefont {J.}~\bibnamefont {Sonner}},\ }\bibfield  {title} {\bibinfo
  {title} {{Krylov localization and suppression of complexity}},\ }\href
  {https://doi.org/10.1007/JHEP03(2022)211} {\bibfield  {journal} {\bibinfo
  {journal} {J. High Energy Phys.}\ }\textbf {\bibinfo {volume} {2022}}\bibinfo
   {number} { (3)},\ \bibinfo {pages} {1}}\BibitemShut {NoStop}%
\bibitem [{\citenamefont {Bhattacharjee}\ \emph
  {et~al.}(2022{\natexlab{a}})\citenamefont {Bhattacharjee}, \citenamefont
  {Cao}, \citenamefont {Nandy},\ and\ \citenamefont
  {Pathak}}]{saddle-dominated-scrambling}%
  \BibitemOpen
\bibfield  {number} {  }\bibfield  {author} {\bibinfo {author} {\bibfnamefont
  {B.}~\bibnamefont {Bhattacharjee}}, \bibinfo {author} {\bibfnamefont
  {X.}~\bibnamefont {Cao}}, \bibinfo {author} {\bibfnamefont {P.}~\bibnamefont
  {Nandy}},\ and\ \bibinfo {author} {\bibfnamefont {T.}~\bibnamefont
  {Pathak}},\ }\bibfield  {title} {\bibinfo {title} {{Krylov complexity in
  saddle-dominated scrambling}},\ }\href
  {https://doi.org/10.1007/JHEP05(2022)174} {\bibfield  {journal} {\bibinfo
  {journal} {J. High Energy Phys.}\ }\textbf {\bibinfo {volume} {2022}}\bibinfo
   {number} { (5)},\ \bibinfo {pages} {1}}\BibitemShut {NoStop}%
\bibitem [{\citenamefont {Bhattacharjee}\ \emph
  {et~al.}(2023{\natexlab{a}})\citenamefont {Bhattacharjee}, \citenamefont
  {Nandy},\ and\ \citenamefont {Pathak}}]{double-scaled-syk}%
  \BibitemOpen
\bibfield  {number} {  }\bibfield  {author} {\bibinfo {author} {\bibfnamefont
  {B.}~\bibnamefont {Bhattacharjee}}, \bibinfo {author} {\bibfnamefont
  {P.}~\bibnamefont {Nandy}},\ and\ \bibinfo {author} {\bibfnamefont
  {T.}~\bibnamefont {Pathak}},\ }\bibfield  {title} {\bibinfo {title} {{Krylov
  complexity in large q and double-scaled SYK model}},\ }\href
  {https://doi.org/10.1007/JHEP08(2023)099} {\bibfield  {journal} {\bibinfo
  {journal} {J. High Energy Phys.}\ }\textbf {\bibinfo {volume} {2023}}\bibinfo
   {number} { (8)},\ \bibinfo {pages} {1}}\BibitemShut {NoStop}%
\bibitem [{\citenamefont {Bhattacharya}\ \emph {et~al.}(2024)\citenamefont
  {Bhattacharya}, \citenamefont {Nath},\ and\ \citenamefont
  {Sahu}}]{non-local-spin-chains}%
  \BibitemOpen
\bibfield  {number} {  }\bibfield  {author} {\bibinfo {author} {\bibfnamefont
  {A.}~\bibnamefont {Bhattacharya}}, \bibinfo {author} {\bibfnamefont {P.~P.}\
  \bibnamefont {Nath}},\ and\ \bibinfo {author} {\bibfnamefont
  {H.}~\bibnamefont {Sahu}},\ }\bibfield  {title} {\bibinfo {title} {{Krylov
  complexity for nonlocal spin chains}},\ }\href
  {https://doi.org/10.1103/PhysRevD.109.066010} {\bibfield  {journal} {\bibinfo
   {journal} {Phys. Rev. D}\ }\textbf {\bibinfo {volume} {109}},\ \bibinfo
  {pages} {066010} (\bibinfo {year} {2024})}\BibitemShut {NoStop}%
\bibitem [{\citenamefont {Bhattacharyya}\ \emph {et~al.}(2023)\citenamefont
  {Bhattacharyya}, \citenamefont {Ghosh},\ and\ \citenamefont
  {Nandi}}]{bose-hubbard}%
  \BibitemOpen
  \bibfield  {author} {\bibinfo {author} {\bibfnamefont {A.}~\bibnamefont
  {Bhattacharyya}}, \bibinfo {author} {\bibfnamefont {D.}~\bibnamefont
  {Ghosh}},\ and\ \bibinfo {author} {\bibfnamefont {P.}~\bibnamefont {Nandi}},\
  }\bibfield  {title} {\bibinfo {title} {{Operator growth and Krylov complexity
  in Bose-Hubbard model}},\ }\href {https://doi.org/10.1007/JHEP12(2023)112}
  {\bibfield  {journal} {\bibinfo  {journal} {J. High Energy Phys.}\ }\textbf
  {\bibinfo {volume} {2023}}\bibinfo  {number} { (12)},\ \bibinfo {pages}
  {1}}\BibitemShut {NoStop}%
\bibitem [{\citenamefont {Tang}(2023)}]{Tang2023Dec}%
  \BibitemOpen
\bibfield  {number} {  }\bibfield  {author} {\bibinfo {author} {\bibfnamefont
  {H.}~\bibnamefont {Tang}},\ }\href {https://arxiv.org/abs/2312.17416}
  {\bibinfo {title} {Operator krylov complexity in random matrix theory}}
  (\bibinfo {year} {2023}),\ \Eprint {https://arxiv.org/abs/2312.17416}
  {arXiv:2312.17416 [hep-th]} \BibitemShut {NoStop}%
\bibitem [{\citenamefont {Camargo}\ \emph {et~al.}(2024)\citenamefont
  {Camargo}, \citenamefont {Jahnke}, \citenamefont {Jeong}, \citenamefont
  {Kim},\ and\ \citenamefont {Nishida}}]{Jeong1}%
  \BibitemOpen
  \bibfield  {author} {\bibinfo {author} {\bibfnamefont {H.~A.}\ \bibnamefont
  {Camargo}}, \bibinfo {author} {\bibfnamefont {V.}~\bibnamefont {Jahnke}},
  \bibinfo {author} {\bibfnamefont {H.-S.}\ \bibnamefont {Jeong}}, \bibinfo
  {author} {\bibfnamefont {K.-Y.}\ \bibnamefont {Kim}},\ and\ \bibinfo {author}
  {\bibfnamefont {M.}~\bibnamefont {Nishida}},\ }\bibfield  {title} {\bibinfo
  {title} {{Spectral and Krylov complexity in billiard systems}},\ }\href
  {https://doi.org/10.1103/PhysRevD.109.046017} {\bibfield  {journal} {\bibinfo
   {journal} {Phys. Rev. D}\ }\textbf {\bibinfo {volume} {109}},\ \bibinfo
  {pages} {046017} (\bibinfo {year} {2024})}\BibitemShut {NoStop}%
\bibitem [{\citenamefont {Caputa}\ \emph {et~al.}(2024)\citenamefont {Caputa},
  \citenamefont {Jeong}, \citenamefont {Liu}, \citenamefont {Pedraza},\ and\
  \citenamefont {Qu}}]{Jeong2}%
  \BibitemOpen
  \bibfield  {author} {\bibinfo {author} {\bibfnamefont {P.}~\bibnamefont
  {Caputa}}, \bibinfo {author} {\bibfnamefont {H.-S.}\ \bibnamefont {Jeong}},
  \bibinfo {author} {\bibfnamefont {S.}~\bibnamefont {Liu}}, \bibinfo {author}
  {\bibfnamefont {J.~F.}\ \bibnamefont {Pedraza}},\ and\ \bibinfo {author}
  {\bibfnamefont {L.-C.}\ \bibnamefont {Qu}},\ }\bibfield  {title} {\bibinfo
  {title} {{Krylov complexity of density matrix operators}},\ }\href
  {https://doi.org/10.1007/JHEP05(2024)337} {\bibfield  {journal} {\bibinfo
  {journal} {J. High Energy Phys.}\ }\textbf {\bibinfo {volume} {2024}}\bibinfo
   {number} { (5)},\ \bibinfo {pages} {1}}\BibitemShut {NoStop}%
\bibitem [{\citenamefont {Hashimoto}\ \emph {et~al.}(2023)\citenamefont
  {Hashimoto}, \citenamefont {Murata}, \citenamefont {Tanahashi},\ and\
  \citenamefont {Watanabe}}]{Hashimoto2023Nov}%
  \BibitemOpen
\bibfield  {number} {  }\bibfield  {author} {\bibinfo {author} {\bibfnamefont
  {K.}~\bibnamefont {Hashimoto}}, \bibinfo {author} {\bibfnamefont
  {K.}~\bibnamefont {Murata}}, \bibinfo {author} {\bibfnamefont
  {N.}~\bibnamefont {Tanahashi}},\ and\ \bibinfo {author} {\bibfnamefont
  {R.}~\bibnamefont {Watanabe}},\ }\bibfield  {title} {\bibinfo {title}
  {{Krylov complexity and chaos in quantum mechanics}},\ }\href
  {https://doi.org/10.1007/JHEP11(2023)040} {\bibfield  {journal} {\bibinfo
  {journal} {J. High Energy Phys.}\ }\textbf {\bibinfo {volume} {2023}}\bibinfo
   {number} { (11)},\ \bibinfo {pages} {1}}\BibitemShut {NoStop}%
\bibitem [{\citenamefont {Bhattacharya}\ \emph {et~al.}(2022)\citenamefont
  {Bhattacharya}, \citenamefont {Nandy}, \citenamefont {Nath},\ and\
  \citenamefont {Sahu}}]{dissipative-open-quantum-systems}%
  \BibitemOpen
\bibfield  {number} {  }\bibfield  {author} {\bibinfo {author} {\bibfnamefont
  {A.}~\bibnamefont {Bhattacharya}}, \bibinfo {author} {\bibfnamefont
  {P.}~\bibnamefont {Nandy}}, \bibinfo {author} {\bibfnamefont {P.~P.}\
  \bibnamefont {Nath}},\ and\ \bibinfo {author} {\bibfnamefont
  {H.}~\bibnamefont {Sahu}},\ }\bibfield  {title} {\bibinfo {title} {{Operator
  growth and Krylov construction in dissipative open quantum systems}},\ }\href
  {https://doi.org/10.1007/JHEP12(2022)081} {\bibfield  {journal} {\bibinfo
  {journal} {J. High Energy Phys.}\ }\textbf {\bibinfo {volume} {2022}}\bibinfo
   {number} { (12)},\ \bibinfo {pages} {1}}\BibitemShut {NoStop}%
\bibitem [{\citenamefont {Bhattacharjee}\ \emph
  {et~al.}(2023{\natexlab{b}})\citenamefont {Bhattacharjee}, \citenamefont
  {Cao}, \citenamefont {Nandy},\ and\ \citenamefont
  {Pathak}}]{open-quantum-systems}%
  \BibitemOpen
\bibfield  {number} {  }\bibfield  {author} {\bibinfo {author} {\bibfnamefont
  {B.}~\bibnamefont {Bhattacharjee}}, \bibinfo {author} {\bibfnamefont
  {X.}~\bibnamefont {Cao}}, \bibinfo {author} {\bibfnamefont {P.}~\bibnamefont
  {Nandy}},\ and\ \bibinfo {author} {\bibfnamefont {T.}~\bibnamefont
  {Pathak}},\ }\bibfield  {title} {\bibinfo {title} {{Operator growth in open
  quantum systems: lessons from the dissipative SYK}},\ }\href
  {https://doi.org/10.1007/JHEP03(2023)054} {\bibfield  {journal} {\bibinfo
  {journal} {J. High Energy Phys.}\ }\textbf {\bibinfo {volume} {2023}}\bibinfo
   {number} { (3)},\ \bibinfo {pages} {1}}\BibitemShut {NoStop}%
\bibitem [{\citenamefont {Bhattacharya}\ \emph {et~al.}(2023)\citenamefont
  {Bhattacharya}, \citenamefont {Nandy}, \citenamefont {Nath},\ and\
  \citenamefont {Sahu}}]{bi-lanczos}%
  \BibitemOpen
\bibfield  {number} {  }\bibfield  {author} {\bibinfo {author} {\bibfnamefont
  {A.}~\bibnamefont {Bhattacharya}}, \bibinfo {author} {\bibfnamefont
  {P.}~\bibnamefont {Nandy}}, \bibinfo {author} {\bibfnamefont {P.~P.}\
  \bibnamefont {Nath}},\ and\ \bibinfo {author} {\bibfnamefont
  {H.}~\bibnamefont {Sahu}},\ }\bibfield  {title} {\bibinfo {title} {{On Krylov
  complexity in open systems: an approach via bi-Lanczos algorithm}},\ }\href
  {https://doi.org/10.1007/JHEP12(2023)066} {\bibfield  {journal} {\bibinfo
  {journal} {J. High Energy Phys.}\ }\textbf {\bibinfo {volume} {2023}}\bibinfo
   {number} { (12)},\ \bibinfo {pages} {1}}\BibitemShut {NoStop}%
\bibitem [{\citenamefont {Liu}\ \emph {et~al.}(2023)\citenamefont {Liu},
  \citenamefont {Tang},\ and\ \citenamefont {Zhai}}]{Liu2023Aug}%
  \BibitemOpen
\bibfield  {number} {  }\bibfield  {author} {\bibinfo {author} {\bibfnamefont
  {C.}~\bibnamefont {Liu}}, \bibinfo {author} {\bibfnamefont {H.}~\bibnamefont
  {Tang}},\ and\ \bibinfo {author} {\bibfnamefont {H.}~\bibnamefont {Zhai}},\
  }\bibfield  {title} {\bibinfo {title} {{Krylov complexity in open quantum
  systems}},\ }\href {https://doi.org/10.1103/PhysRevResearch.5.033085}
  {\bibfield  {journal} {\bibinfo  {journal} {Phys. Rev. Res.}\ }\textbf
  {\bibinfo {volume} {5}},\ \bibinfo {pages} {033085} (\bibinfo {year}
  {2023})}\BibitemShut {NoStop}%
\bibitem [{\citenamefont {Dymarsky}\ and\ \citenamefont
  {Smolkin}(2021)}]{Dymarsky2021Oct}%
  \BibitemOpen
  \bibfield  {author} {\bibinfo {author} {\bibfnamefont {A.}~\bibnamefont
  {Dymarsky}}\ and\ \bibinfo {author} {\bibfnamefont {M.}~\bibnamefont
  {Smolkin}},\ }\bibfield  {title} {\bibinfo {title} {{Krylov complexity in
  conformal field theory}},\ }\href
  {https://doi.org/10.1103/PhysRevD.104.L081702} {\bibfield  {journal}
  {\bibinfo  {journal} {Phys. Rev. D}\ }\textbf {\bibinfo {volume} {104}},\
  \bibinfo {pages} {L081702} (\bibinfo {year} {2021})}\BibitemShut {NoStop}%
\bibitem [{\citenamefont {Heveling}\ \emph {et~al.}(2022)\citenamefont
  {Heveling}, \citenamefont {Wang},\ and\ \citenamefont
  {Gemmer}}]{Heveling2022Jul}%
  \BibitemOpen
  \bibfield  {author} {\bibinfo {author} {\bibfnamefont {R.}~\bibnamefont
  {Heveling}}, \bibinfo {author} {\bibfnamefont {J.}~\bibnamefont {Wang}},\
  and\ \bibinfo {author} {\bibfnamefont {J.}~\bibnamefont {Gemmer}},\
  }\bibfield  {title} {\bibinfo {title} {{Numerically probing the universal
  operator growth hypothesis}},\ }\href
  {https://doi.org/10.1103/PhysRevE.106.014152} {\bibfield  {journal} {\bibinfo
   {journal} {Phys. Rev. E}\ }\textbf {\bibinfo {volume} {106}},\ \bibinfo
  {pages} {014152} (\bibinfo {year} {2022})}\BibitemShut {NoStop}%
\bibitem [{\citenamefont {Avdoshkin}\ \emph {et~al.}(2024)\citenamefont
  {Avdoshkin}, \citenamefont {Dymarsky},\ and\ \citenamefont
  {Smolkin}}]{Dymarsky2022}%
  \BibitemOpen
  \bibfield  {author} {\bibinfo {author} {\bibfnamefont {A.}~\bibnamefont
  {Avdoshkin}}, \bibinfo {author} {\bibfnamefont {A.}~\bibnamefont
  {Dymarsky}},\ and\ \bibinfo {author} {\bibfnamefont {M.}~\bibnamefont
  {Smolkin}},\ }\bibfield  {title} {\bibinfo {title} {{Krylov complexity in
  quantum field theory, and beyond}},\ }\href
  {https://doi.org/10.1007/JHEP06(2024)066} {\bibfield  {journal} {\bibinfo
  {journal} {J. High Energy Phys.}\ }\textbf {\bibinfo {volume} {2024}}\bibinfo
   {number} { (6)},\ \bibinfo {pages} {1}}\BibitemShut {NoStop}%
\bibitem [{\citenamefont {Adhikari}\ \emph {et~al.}(2023)\citenamefont
  {Adhikari}, \citenamefont {Choudhury},\ and\ \citenamefont
  {Roy}}]{Adhikari2023Aug}%
  \BibitemOpen
\bibfield  {number} {  }\bibfield  {author} {\bibinfo {author} {\bibfnamefont
  {K.}~\bibnamefont {Adhikari}}, \bibinfo {author} {\bibfnamefont
  {S.}~\bibnamefont {Choudhury}},\ and\ \bibinfo {author} {\bibfnamefont
  {A.}~\bibnamefont {Roy}},\ }\bibfield  {title} {\bibinfo {title} {{Krylov
  Complexity in Quantum Field Theory}},\ }\href
  {https://doi.org/10.1016/j.nuclphysb.2023.116263} {\bibfield  {journal}
  {\bibinfo  {journal} {Nucl. Phys. B}\ }\textbf {\bibinfo {volume} {993}},\
  \bibinfo {pages} {116263} (\bibinfo {year} {2023})}\BibitemShut {NoStop}%
\bibitem [{\citenamefont {Xu}\ \emph {et~al.}(2020)\citenamefont {Xu},
  \citenamefont {Scaffidi},\ and\ \citenamefont {Cao}}]{Xu2020Apr}%
  \BibitemOpen
  \bibfield  {author} {\bibinfo {author} {\bibfnamefont {T.}~\bibnamefont
  {Xu}}, \bibinfo {author} {\bibfnamefont {T.}~\bibnamefont {Scaffidi}},\ and\
  \bibinfo {author} {\bibfnamefont {X.}~\bibnamefont {Cao}},\ }\bibfield
  {title} {\bibinfo {title} {{Does Scrambling Equal Chaos?}},\ }\href
  {https://doi.org/10.1103/PhysRevLett.124.140602} {\bibfield  {journal}
  {\bibinfo  {journal} {Phys. Rev. Lett.}\ }\textbf {\bibinfo {volume} {124}},\
  \bibinfo {pages} {140602} (\bibinfo {year} {2020})}\BibitemShut {NoStop}%
\bibitem [{\citenamefont {Dowling}\ \emph {et~al.}(2023)\citenamefont
  {Dowling}, \citenamefont {Kos},\ and\ \citenamefont {Modi}}]{Dowling2023Nov}%
  \BibitemOpen
  \bibfield  {author} {\bibinfo {author} {\bibfnamefont {N.}~\bibnamefont
  {Dowling}}, \bibinfo {author} {\bibfnamefont {P.}~\bibnamefont {Kos}},\ and\
  \bibinfo {author} {\bibfnamefont {K.}~\bibnamefont {Modi}},\ }\bibfield
  {title} {\bibinfo {title} {{Scrambling Is Necessary but Not Sufficient for
  Chaos}},\ }\href {https://doi.org/10.1103/PhysRevLett.131.180403} {\bibfield
  {journal} {\bibinfo  {journal} {Phys. Rev. Lett.}\ }\textbf {\bibinfo
  {volume} {131}},\ \bibinfo {pages} {180403} (\bibinfo {year}
  {2023})}\BibitemShut {NoStop}%
\bibitem [{\citenamefont {Turner}\ \emph {et~al.}(2018)\citenamefont {Turner},
  \citenamefont {Michailidis}, \citenamefont {Abanin}, \citenamefont {Serbyn},\
  and\ \citenamefont {Papi{\ifmmode\acute{c}\else\'{c}\fi}}}]{Turner2018Jul}%
  \BibitemOpen
  \bibfield  {author} {\bibinfo {author} {\bibfnamefont {C.~J.}\ \bibnamefont
  {Turner}}, \bibinfo {author} {\bibfnamefont {A.~A.}\ \bibnamefont
  {Michailidis}}, \bibinfo {author} {\bibfnamefont {D.~A.}\ \bibnamefont
  {Abanin}}, \bibinfo {author} {\bibfnamefont {M.}~\bibnamefont {Serbyn}},\
  and\ \bibinfo {author} {\bibfnamefont {Z.}~\bibnamefont
  {Papi{\ifmmode\acute{c}\else\'{c}\fi}}},\ }\bibfield  {title} {\bibinfo
  {title} {{Weak ergodicity breaking from quantum many-body scars}},\ }\href
  {https://doi.org/10.1038/s41567-018-0137-5} {\bibfield  {journal} {\bibinfo
  {journal} {Nat. Phys.}\ }\textbf {\bibinfo {volume} {14}},\ \bibinfo {pages}
  {745} (\bibinfo {year} {2018})}\BibitemShut {NoStop}%
\bibitem [{\citenamefont {Schecter}\ and\ \citenamefont
  {Iadecola}(2019)}]{Schecter2019Oct}%
  \BibitemOpen
  \bibfield  {author} {\bibinfo {author} {\bibfnamefont {M.}~\bibnamefont
  {Schecter}}\ and\ \bibinfo {author} {\bibfnamefont {T.}~\bibnamefont
  {Iadecola}},\ }\bibfield  {title} {\bibinfo {title} {{Weak Ergodicity
  Breaking and Quantum Many-Body Scars in Spin-1 $XY$ Magnets}},\ }\href
  {https://doi.org/10.1103/PhysRevLett.123.147201} {\bibfield  {journal}
  {\bibinfo  {journal} {Phys. Rev. Lett.}\ }\textbf {\bibinfo {volume} {123}},\
  \bibinfo {pages} {147201} (\bibinfo {year} {2019})}\BibitemShut {NoStop}%
\bibitem [{\citenamefont {Kliczkowski}\ \emph {et~al.}(2024)\citenamefont
  {Kliczkowski}, \citenamefont {Swietek}, \citenamefont {Hopjan},\ and\
  \citenamefont {Vidmar}}]{fading-ergodicity}%
  \BibitemOpen
  \bibfield  {author} {\bibinfo {author} {\bibfnamefont {M.}~\bibnamefont
  {Kliczkowski}}, \bibinfo {author} {\bibfnamefont {R.}~\bibnamefont
  {Swietek}}, \bibinfo {author} {\bibfnamefont {M.}~\bibnamefont {Hopjan}},\
  and\ \bibinfo {author} {\bibfnamefont {L.}~\bibnamefont {Vidmar}},\ }\href
  {https://arxiv.org/abs/2407.16773} {\bibinfo {title} {Fading ergodicity}}
  (\bibinfo {year} {2024}),\ \Eprint {https://arxiv.org/abs/2407.16773}
  {arXiv:2407.16773 [cond-mat.stat-mech]} \BibitemShut {NoStop}%
\bibitem [{\citenamefont {Pandey}\ \emph {et~al.}(2020)\citenamefont {Pandey},
  \citenamefont {Claeys}, \citenamefont {Campbell}, \citenamefont
  {Polkovnikov},\ and\ \citenamefont {Sels}}]{Pandey2020Oct}%
  \BibitemOpen
  \bibfield  {author} {\bibinfo {author} {\bibfnamefont {M.}~\bibnamefont
  {Pandey}}, \bibinfo {author} {\bibfnamefont {P.~W.}\ \bibnamefont {Claeys}},
  \bibinfo {author} {\bibfnamefont {D.~K.}\ \bibnamefont {Campbell}}, \bibinfo
  {author} {\bibfnamefont {A.}~\bibnamefont {Polkovnikov}},\ and\ \bibinfo
  {author} {\bibfnamefont {D.}~\bibnamefont {Sels}},\ }\bibfield  {title}
  {\bibinfo {title} {{Adiabatic Eigenstate Deformations as a Sensitive Probe
  for Quantum Chaos}},\ }\href {https://doi.org/10.1103/PhysRevX.10.041017}
  {\bibfield  {journal} {\bibinfo  {journal} {Phys. Rev. X}\ }\textbf {\bibinfo
  {volume} {10}},\ \bibinfo {pages} {041017} (\bibinfo {year}
  {2020})}\BibitemShut {NoStop}%
\bibitem [{\citenamefont {Nandy}\ \emph {et~al.}(2022)\citenamefont {Nandy},
  \citenamefont
  {{\ifmmode\check{C}\else\v{C}\fi}ade{\ifmmode\check{z}\else\v{z}\fi}},
  \citenamefont {Dietz}, \citenamefont {Andreanov},\ and\ \citenamefont
  {Rosa}}]{Nandy2022Dec}%
  \BibitemOpen
  \bibfield  {author} {\bibinfo {author} {\bibfnamefont {D.~K.}\ \bibnamefont
  {Nandy}}, \bibinfo {author} {\bibfnamefont {T.}~\bibnamefont
  {{\ifmmode\check{C}\else\v{C}\fi}ade{\ifmmode\check{z}\else\v{z}\fi}}},
  \bibinfo {author} {\bibfnamefont {B.}~\bibnamefont {Dietz}}, \bibinfo
  {author} {\bibfnamefont {A.}~\bibnamefont {Andreanov}},\ and\ \bibinfo
  {author} {\bibfnamefont {D.}~\bibnamefont {Rosa}},\ }\bibfield  {title}
  {\bibinfo {title} {{Delayed thermalization in the mass-deformed
  Sachdev-Ye-Kitaev model}},\ }\href
  {https://doi.org/10.1103/PhysRevB.106.245147} {\bibfield  {journal} {\bibinfo
   {journal} {Phys. Rev. B}\ }\textbf {\bibinfo {volume} {106}},\ \bibinfo
  {pages} {245147} (\bibinfo {year} {2022})}\BibitemShut {NoStop}%
\bibitem [{\citenamefont {Sala}\ \emph {et~al.}(2020)\citenamefont {Sala},
  \citenamefont {Rakovszky}, \citenamefont {Verresen}, \citenamefont {Knap},\
  and\ \citenamefont {Pollmann}}]{Sala2020Feb}%
  \BibitemOpen
  \bibfield  {author} {\bibinfo {author} {\bibfnamefont {P.}~\bibnamefont
  {Sala}}, \bibinfo {author} {\bibfnamefont {T.}~\bibnamefont {Rakovszky}},
  \bibinfo {author} {\bibfnamefont {R.}~\bibnamefont {Verresen}}, \bibinfo
  {author} {\bibfnamefont {M.}~\bibnamefont {Knap}},\ and\ \bibinfo {author}
  {\bibfnamefont {F.}~\bibnamefont {Pollmann}},\ }\bibfield  {title} {\bibinfo
  {title} {{Ergodicity Breaking Arising from Hilbert Space Fragmentation in
  Dipole-Conserving Hamiltonians}},\ }\href
  {https://doi.org/10.1103/PhysRevX.10.011047} {\bibfield  {journal} {\bibinfo
  {journal} {Phys. Rev. X}\ }\textbf {\bibinfo {volume} {10}},\ \bibinfo
  {pages} {011047} (\bibinfo {year} {2020})}\BibitemShut {NoStop}%
\bibitem [{\citenamefont {Khemani}\ \emph {et~al.}(2020)\citenamefont
  {Khemani}, \citenamefont {Hermele},\ and\ \citenamefont
  {Nandkishore}}]{Khemani2020May}%
  \BibitemOpen
  \bibfield  {author} {\bibinfo {author} {\bibfnamefont {V.}~\bibnamefont
  {Khemani}}, \bibinfo {author} {\bibfnamefont {M.}~\bibnamefont {Hermele}},\
  and\ \bibinfo {author} {\bibfnamefont {R.}~\bibnamefont {Nandkishore}},\
  }\bibfield  {title} {\bibinfo {title} {{Localization from Hilbert space
  shattering: From theory to physical realizations}},\ }\href
  {https://doi.org/10.1103/PhysRevB.101.174204} {\bibfield  {journal} {\bibinfo
   {journal} {Phys. Rev. B}\ }\textbf {\bibinfo {volume} {101}},\ \bibinfo
  {pages} {174204} (\bibinfo {year} {2020})}\BibitemShut {NoStop}%
\bibitem [{\citenamefont {Serbyn}\ \emph {et~al.}(2021)\citenamefont {Serbyn},
  \citenamefont {Abanin},\ and\ \citenamefont
  {Papi{\ifmmode\acute{c}\else\'{c}\fi}}}]{Serbyn2021Jun}%
  \BibitemOpen
  \bibfield  {author} {\bibinfo {author} {\bibfnamefont {M.}~\bibnamefont
  {Serbyn}}, \bibinfo {author} {\bibfnamefont {D.~A.}\ \bibnamefont {Abanin}},\
  and\ \bibinfo {author} {\bibfnamefont {Z.}~\bibnamefont
  {Papi{\ifmmode\acute{c}\else\'{c}\fi}}},\ }\bibfield  {title} {\bibinfo
  {title} {{Quantum many-body scars and weak breaking of ergodicity}},\ }\href
  {https://doi.org/10.1038/s41567-021-01230-2} {\bibfield  {journal} {\bibinfo
  {journal} {Nat. Phys.}\ }\textbf {\bibinfo {volume} {17}},\ \bibinfo {pages}
  {675} (\bibinfo {year} {2021})}\BibitemShut {NoStop}%
\bibitem [{\citenamefont {Su}\ \emph {et~al.}(2023)\citenamefont {Su},
  \citenamefont {Sun}, \citenamefont {Hudomal}, \citenamefont {Desaules},
  \citenamefont {Zhou}, \citenamefont {Yang}, \citenamefont {Halimeh},
  \citenamefont {Yuan}, \citenamefont {Papic},\ and\ \citenamefont
  {Pan}}]{Halimeh1}%
  \BibitemOpen
  \bibfield  {author} {\bibinfo {author} {\bibfnamefont {G.-X.}\ \bibnamefont
  {Su}}, \bibinfo {author} {\bibfnamefont {H.}~\bibnamefont {Sun}}, \bibinfo
  {author} {\bibfnamefont {A.}~\bibnamefont {Hudomal}}, \bibinfo {author}
  {\bibfnamefont {J.-Y.}\ \bibnamefont {Desaules}}, \bibinfo {author}
  {\bibfnamefont {Z.-Y.}\ \bibnamefont {Zhou}}, \bibinfo {author}
  {\bibfnamefont {B.}~\bibnamefont {Yang}}, \bibinfo {author} {\bibfnamefont
  {J.~C.}\ \bibnamefont {Halimeh}}, \bibinfo {author} {\bibfnamefont {Z.-S.}\
  \bibnamefont {Yuan}}, \bibinfo {author} {\bibfnamefont {Z.}~\bibnamefont
  {Papic}},\ and\ \bibinfo {author} {\bibfnamefont {J.-W.}\ \bibnamefont
  {Pan}},\ }\bibfield  {title} {\bibinfo {title} {{Observation of many-body
  scarring in a Bose-Hubbard quantum simulator}},\ }\href
  {https://doi.org/10.1103/PhysRevResearch.5.023010} {\bibfield  {journal}
  {\bibinfo  {journal} {Phys. Rev. Res.}\ }\textbf {\bibinfo {volume} {5}},\
  \bibinfo {pages} {023010} (\bibinfo {year} {2023})}\BibitemShut {NoStop}%
\bibitem [{\citenamefont {Desaules}\ \emph {et~al.}(2023)\citenamefont
  {Desaules}, \citenamefont {Hudomal}, \citenamefont {Banerjee}, \citenamefont
  {Sen}, \citenamefont {Papic},\ and\ \citenamefont {Halimeh}}]{Halimeh2}%
  \BibitemOpen
  \bibfield  {author} {\bibinfo {author} {\bibfnamefont {J.-Y.}\ \bibnamefont
  {Desaules}}, \bibinfo {author} {\bibfnamefont {A.}~\bibnamefont {Hudomal}},
  \bibinfo {author} {\bibfnamefont {D.}~\bibnamefont {Banerjee}}, \bibinfo
  {author} {\bibfnamefont {A.}~\bibnamefont {Sen}}, \bibinfo {author}
  {\bibfnamefont {Z.}~\bibnamefont {Papic}},\ and\ \bibinfo {author}
  {\bibfnamefont {J.~C.}\ \bibnamefont {Halimeh}},\ }\bibfield  {title}
  {\bibinfo {title} {{Prominent quantum many-body scars in a truncated
  Schwinger model}},\ }\href {https://doi.org/10.1103/PhysRevB.107.205112}
  {\bibfield  {journal} {\bibinfo  {journal} {Phys. Rev. B}\ }\textbf {\bibinfo
  {volume} {107}},\ \bibinfo {pages} {205112} (\bibinfo {year}
  {2023})}\BibitemShut {NoStop}%
\bibitem [{\citenamefont {Halimeh}\ \emph {et~al.}(2023)\citenamefont
  {Halimeh}, \citenamefont {Barbiero}, \citenamefont {Hauke}, \citenamefont
  {Grusdt},\ and\ \citenamefont {Bohrdt}}]{Halimeh3}%
  \BibitemOpen
  \bibfield  {author} {\bibinfo {author} {\bibfnamefont {J.~C.}\ \bibnamefont
  {Halimeh}}, \bibinfo {author} {\bibfnamefont {L.}~\bibnamefont {Barbiero}},
  \bibinfo {author} {\bibfnamefont {P.}~\bibnamefont {Hauke}}, \bibinfo
  {author} {\bibfnamefont {F.}~\bibnamefont {Grusdt}},\ and\ \bibinfo {author}
  {\bibfnamefont {A.}~\bibnamefont {Bohrdt}},\ }\bibfield  {title} {\bibinfo
  {title} {{Robust quantum many-body scars in lattice gauge theories}},\ }\href
  {https://doi.org/10.22331/q-2023-05-15-1004} {\bibfield  {journal} {\bibinfo
  {journal} {Quantum}\ }\textbf {\bibinfo {volume} {7}},\ \bibinfo {pages}
  {1004} (\bibinfo {year} {2023})},\ \Eprint
  {https://arxiv.org/abs/2203.08828v5} {2203.08828v5} \BibitemShut {NoStop}%
\bibitem [{\citenamefont {Sachdev}\ and\ \citenamefont
  {Ye}(1993)}]{Sachdev1993May}%
  \BibitemOpen
  \bibfield  {author} {\bibinfo {author} {\bibfnamefont {S.}~\bibnamefont
  {Sachdev}}\ and\ \bibinfo {author} {\bibfnamefont {J.}~\bibnamefont {Ye}},\
  }\bibfield  {title} {\bibinfo {title} {{Gapless spin-fluid ground state in a
  random quantum Heisenberg magnet}},\ }\href
  {https://doi.org/10.1103/PhysRevLett.70.3339} {\bibfield  {journal} {\bibinfo
   {journal} {Phys. Rev. Lett.}\ }\textbf {\bibinfo {volume} {70}},\ \bibinfo
  {pages} {3339} (\bibinfo {year} {1993})}\BibitemShut {NoStop}%
\bibitem [{\citenamefont {Kitaev}(2015)}]{Kitaev2015}%
  \BibitemOpen
  \bibfield  {author} {\bibinfo {author} {\bibfnamefont {A.}~\bibnamefont
  {Kitaev}},\ }\bibfield  {title} {\bibinfo {title} {A simple model of quantum
  holography}} (\bibinfo {year} {2015}),\ \bibinfo {note} {{ Talks given at
  ``KITP: Entanglement in Strongly-Correlated Quantum Matter'',
  \href{https://online.kitp.ucsb.edu/online/entangled15/kitaev/}{(Part 1,}
  \href{https://online.kitp.ucsb.edu/online/entangled15/kitaev2/}{ Part
  2)}}}\BibitemShut {NoStop}%
\bibitem [{\citenamefont {Maldacena}\ and\ \citenamefont
  {Stanford}(2016)}]{Maldacena-syk}%
  \BibitemOpen
  \bibfield  {author} {\bibinfo {author} {\bibfnamefont {J.}~\bibnamefont
  {Maldacena}}\ and\ \bibinfo {author} {\bibfnamefont {D.}~\bibnamefont
  {Stanford}},\ }\bibfield  {title} {\bibinfo {title} {{Remarks on the
  Sachdev-Ye-Kitaev model}},\ }\href
  {https://doi.org/10.1103/PhysRevD.94.106002} {\bibfield  {journal} {\bibinfo
  {journal} {Phys. Rev. D}\ }\textbf {\bibinfo {volume} {94}},\ \bibinfo
  {pages} {106002} (\bibinfo {year} {2016})}\BibitemShut {NoStop}%
\bibitem [{\citenamefont {Chowdhury}\ \emph {et~al.}(2022)\citenamefont
  {Chowdhury}, \citenamefont {Georges}, \citenamefont {Parcollet},\ and\
  \citenamefont {Sachdev}}]{Chowdhury2022Sep}%
  \BibitemOpen
  \bibfield  {author} {\bibinfo {author} {\bibfnamefont {D.}~\bibnamefont
  {Chowdhury}}, \bibinfo {author} {\bibfnamefont {A.}~\bibnamefont {Georges}},
  \bibinfo {author} {\bibfnamefont {O.}~\bibnamefont {Parcollet}},\ and\
  \bibinfo {author} {\bibfnamefont {S.}~\bibnamefont {Sachdev}},\ }\bibfield
  {title} {\bibinfo {title} {{Sachdev-Ye-Kitaev models and beyond: Window into
  non-Fermi liquids}},\ }\href {https://doi.org/10.1103/RevModPhys.94.035004}
  {\bibfield  {journal} {\bibinfo  {journal} {Rev. Mod. Phys.}\ }\textbf
  {\bibinfo {volume} {94}},\ \bibinfo {pages} {035004} (\bibinfo {year}
  {2022})}\BibitemShut {NoStop}%
\bibitem [{\citenamefont {van Horssen}\ \emph {et~al.}(2015)\citenamefont {van
  Horssen}, \citenamefont {Levi},\ and\ \citenamefont
  {Garrahan}}]{VanHorssen2015Sept}%
  \BibitemOpen
  \bibfield  {author} {\bibinfo {author} {\bibfnamefont {M.}~\bibnamefont {van
  Horssen}}, \bibinfo {author} {\bibfnamefont {E.}~\bibnamefont {Levi}},\ and\
  \bibinfo {author} {\bibfnamefont {J.~P.}\ \bibnamefont {Garrahan}},\
  }\bibfield  {title} {\bibinfo {title} {{Dynamics of many-body localization in
  a translation-invariant quantum glass model}},\ }\href
  {https://doi.org/10.1103/PhysRevB.92.100305} {\bibfield  {journal} {\bibinfo
  {journal} {Phys. Rev. B}\ }\textbf {\bibinfo {volume} {92}},\ \bibinfo
  {pages} {100305} (\bibinfo {year} {2015})}\BibitemShut {NoStop}%
\bibitem [{\citenamefont {Crowley}(2023)}]{Crowley2017}%
  \BibitemOpen
  \bibfield  {author} {\bibinfo {author} {\bibfnamefont {P.}~\bibnamefont
  {Crowley}},\ }\emph {\bibinfo {title} {Entanglement and Thermalization in
  Many Body Quantum Systems}},\ \href@noop {} {Ph.D. thesis},\ \bibinfo
  {school} {University College London} (\bibinfo {year} {2023}),\ \bibinfo
  {note}
  {\href{https://discovery.ucl.ac.uk/id/eprint/1553310}{https://discovery.ucl.ac.uk/id/eprint/1553310}}\BibitemShut
  {NoStop}%
\bibitem [{\citenamefont {Pancotti}\ \emph {et~al.}(2020)\citenamefont
  {Pancotti}, \citenamefont {Giudice}, \citenamefont {Cirac}, \citenamefont
  {Garrahan},\ and\ \citenamefont
  {Ba{\ifmmode\tilde{n}\else\~{n}\fi}uls}}]{Pancotti2020Jun}%
  \BibitemOpen
  \bibfield  {author} {\bibinfo {author} {\bibfnamefont {N.}~\bibnamefont
  {Pancotti}}, \bibinfo {author} {\bibfnamefont {G.}~\bibnamefont {Giudice}},
  \bibinfo {author} {\bibfnamefont {J.~I.}\ \bibnamefont {Cirac}}, \bibinfo
  {author} {\bibfnamefont {J.~P.}\ \bibnamefont {Garrahan}},\ and\ \bibinfo
  {author} {\bibfnamefont {M.~C.}\ \bibnamefont
  {Ba{\ifmmode\tilde{n}\else\~{n}\fi}uls}},\ }\bibfield  {title} {\bibinfo
  {title} {{Quantum East Model: Localization, Nonthermal Eigenstates, and Slow
  Dynamics}},\ }\href {https://doi.org/10.1103/PhysRevX.10.021051} {\bibfield
  {journal} {\bibinfo  {journal} {Phys. Rev. X}\ }\textbf {\bibinfo {volume}
  {10}},\ \bibinfo {pages} {021051} (\bibinfo {year} {2020})}\BibitemShut
  {NoStop}%
\bibitem [{\citenamefont {Bertini}\ \emph {et~al.}(2024)\citenamefont
  {Bertini}, \citenamefont {Kos},\ and\ \citenamefont
  {Prosen}}]{Bertini2024Feb}%
  \BibitemOpen
  \bibfield  {author} {\bibinfo {author} {\bibfnamefont {B.}~\bibnamefont
  {Bertini}}, \bibinfo {author} {\bibfnamefont {P.}~\bibnamefont {Kos}},\ and\
  \bibinfo {author} {\bibfnamefont {T.}~\bibnamefont {Prosen}},\ }\bibfield
  {title} {\bibinfo {title} {Localized dynamics in the floquet quantum east
  model},\ }\bibfield  {journal} {\bibinfo  {journal} {Phys. Rev. Lett.}\
  }\textbf {\bibinfo {volume} {132}},\ \href
  {https://doi.org/10.1103/physrevlett.132.080401}
  {10.1103/physrevlett.132.080401} (\bibinfo {year} {2024})\BibitemShut
  {NoStop}%
\bibitem [{\citenamefont {Nandkishore}\ and\ \citenamefont
  {Huse}(2015)}]{Nandkishore2015Mar}%
  \BibitemOpen
  \bibfield  {author} {\bibinfo {author} {\bibfnamefont {R.}~\bibnamefont
  {Nandkishore}}\ and\ \bibinfo {author} {\bibfnamefont {D.~A.}\ \bibnamefont
  {Huse}},\ }\bibfield  {title} {\bibinfo {title} {{Many-Body Localization and
  Thermalization in Quantum Statistical Mechanics}},\ }\href
  {https://doi.org/10.1146/annurev-conmatphys-031214-014726} {\bibfield
  {journal} {\bibinfo  {journal} {Annu. Rev. Condens. Matter Phys.}\ }\textbf
  {\bibinfo {volume} {6}},\ \bibinfo {pages} {15} (\bibinfo {year}
  {2015})}\BibitemShut {NoStop}%
\bibitem [{\citenamefont {Royen}\ \emph {et~al.}(2024)\citenamefont {Royen},
  \citenamefont {Mondal}, \citenamefont {Pollmann},\ and\ \citenamefont
  {Heidrich-Meisner}}]{Royen2024Feb}%
  \BibitemOpen
  \bibfield  {author} {\bibinfo {author} {\bibfnamefont {K.}~\bibnamefont
  {Royen}}, \bibinfo {author} {\bibfnamefont {S.}~\bibnamefont {Mondal}},
  \bibinfo {author} {\bibfnamefont {F.}~\bibnamefont {Pollmann}},\ and\
  \bibinfo {author} {\bibfnamefont {F.}~\bibnamefont {Heidrich-Meisner}},\
  }\bibfield  {title} {\bibinfo {title} {Enhanced many-body localization in a
  kinetically constrained model},\ }\bibfield  {journal} {\bibinfo  {journal}
  {Phys. Rev. E}\ }\textbf {\bibinfo {volume} {109}},\ \href
  {https://doi.org/10.1103/physreve.109.024136} {10.1103/physreve.109.024136}
  (\bibinfo {year} {2024})\BibitemShut {NoStop}%
\bibitem [{\citenamefont {Chandran}\ \emph {et~al.}(2023)\citenamefont
  {Chandran}, \citenamefont {Iadecola}, \citenamefont {Khemani},\ and\
  \citenamefont {Moessner}}]{Chandran2023Mar}%
  \BibitemOpen
  \bibfield  {author} {\bibinfo {author} {\bibfnamefont {A.}~\bibnamefont
  {Chandran}}, \bibinfo {author} {\bibfnamefont {T.}~\bibnamefont {Iadecola}},
  \bibinfo {author} {\bibfnamefont {V.}~\bibnamefont {Khemani}},\ and\ \bibinfo
  {author} {\bibfnamefont {R.}~\bibnamefont {Moessner}},\ }\bibfield  {title}
  {\bibinfo {title} {Quantum many-body scars: A quasiparticle perspective},\
  }\href {https://doi.org/10.1146/annurev-conmatphys-031620-101617} {\bibfield
  {journal} {\bibinfo  {journal} {Annual Review of Condensed Matter Physics}\
  }\textbf {\bibinfo {volume} {14}},\ \bibinfo {pages} {443} (\bibinfo {year}
  {2023})}\BibitemShut {NoStop}%
\bibitem [{\citenamefont {Brighi}\ \emph {et~al.}(2023)\citenamefont {Brighi},
  \citenamefont {Ljubotina},\ and\ \citenamefont {Serbyn}}]{Brighi2023Sep}%
  \BibitemOpen
  \bibfield  {author} {\bibinfo {author} {\bibfnamefont {P.}~\bibnamefont
  {Brighi}}, \bibinfo {author} {\bibfnamefont {M.}~\bibnamefont {Ljubotina}},\
  and\ \bibinfo {author} {\bibfnamefont {M.}~\bibnamefont {Serbyn}},\
  }\bibfield  {title} {\bibinfo {title} {Hilbert space fragmentation and slow
  dynamics in particle-conserving quantum east models},\ }\bibfield  {journal}
  {\bibinfo  {journal} {SciPost Physics}\ }\textbf {\bibinfo {volume} {15}},\
  \href {https://doi.org/10.21468/scipostphys.15.3.093}
  {10.21468/scipostphys.15.3.093} (\bibinfo {year} {2023})\BibitemShut
  {NoStop}%
\bibitem [{\citenamefont {Viswanath}\ and\ \citenamefont
  {M{\ifmmode\ddot{u}\else\"{u}\fi}ller}(1994)}]{Viswanath1994}%
  \BibitemOpen
  \bibfield  {author} {\bibinfo {author} {\bibfnamefont {V.~S.}\ \bibnamefont
  {Viswanath}}\ and\ \bibinfo {author} {\bibfnamefont {G.}~\bibnamefont
  {M{\ifmmode\ddot{u}\else\"{u}\fi}ller}},\ }\href
  {https://link.springer.com/book/10.1007/978-3-540-48651-0} {\emph {\bibinfo
  {title} {{The Recursion Method}}}}\ (\bibinfo  {publisher} {Springer},\
  \bibinfo {address} {Berlin, Germany},\ \bibinfo {year} {1994})\BibitemShut
  {NoStop}%
\bibitem [{Note1()}]{Note1}%
  \BibitemOpen
  \bibinfo {note} {As derived in Appendix \ref {app. krylov complexity}, if we
  would use a different expansion coefficient for the operator such as
  $|\protect \mathcal {O}(t) )=\DOTSB \sum@ \slimits@ _{n=0}^{\protect
  \ensuremath {{\protect \cal K}}-1} \protect \ensuremath {{\imath }}^n \psi
  _n(t) | \protect \mathcal {O}_n)$, then we obtain a ``Schro{\"o}dinger-type''
  structure, namely $\protect \ensuremath {{\imath }}\protect \ensuremath
  {{\partial }}_t \psi _n(t) = b_n \psi _{n-1}(t) - b_{n+1}\psi _{n+1}(t) $.
  The coefficients $\phi _n(t)$ and $\psi _n(t)$ are related through $ \phi
  _n(t) = \protect \ensuremath {{\imath }}^n \psi _n(t)$.}\BibitemShut {Stop}%
\bibitem [{\citenamefont {Tan}\ \emph {et~al.}(2024)\citenamefont {Tan},
  \citenamefont {Wei},\ and\ \citenamefont
  {Zhang}}]{wightman-vs-standard-inner-product}%
  \BibitemOpen
  \bibfield  {author} {\bibinfo {author} {\bibfnamefont {C.}~\bibnamefont
  {Tan}}, \bibinfo {author} {\bibfnamefont {Z.}~\bibnamefont {Wei}},\ and\
  \bibinfo {author} {\bibfnamefont {R.}~\bibnamefont {Zhang}},\ }\href
  {https://arxiv.org/abs/2401.10499} {\bibinfo {title} {Scaling relations of
  spectrum form factor and krylov complexity at finite temperature}} (\bibinfo
  {year} {2024}),\ \Eprint {https://arxiv.org/abs/2401.10499} {arXiv:2401.10499
  [cond-mat.stat-mech]} \BibitemShut {NoStop}%
\bibitem [{\citenamefont {Fleishman}\ and\ \citenamefont
  {Licciardello}(1977)}]{Fleishman1977Mar}%
  \BibitemOpen
  \bibfield  {author} {\bibinfo {author} {\bibfnamefont {L.}~\bibnamefont
  {Fleishman}}\ and\ \bibinfo {author} {\bibfnamefont {D.~C.}\ \bibnamefont
  {Licciardello}},\ }\bibfield  {title} {\bibinfo {title} {{Fluctuations and
  localization in one dimension}},\ }\href
  {https://doi.org/10.1088/0022-3719/10/6/003} {\bibfield  {journal} {\bibinfo
  {journal} {J. Phys. C: Solid State Phys.}\ }\textbf {\bibinfo {volume}
  {10}},\ \bibinfo {pages} {L125} (\bibinfo {year} {1977})}\BibitemShut
  {NoStop}%
\bibitem [{\citenamefont {Barb{\ifmmode\acute{o}\else\'{o}\fi}n}\ \emph
  {et~al.}(2019)\citenamefont {Barb{\ifmmode\acute{o}\else\'{o}\fi}n},
  \citenamefont {Rabinovici}, \citenamefont {Shir},\ and\ \citenamefont
  {Sinha}}]{Barbon2019Oct}%
  \BibitemOpen
  \bibfield  {author} {\bibinfo {author} {\bibfnamefont {J.~L.~F.}\
  \bibnamefont {Barb{\ifmmode\acute{o}\else\'{o}\fi}n}}, \bibinfo {author}
  {\bibfnamefont {E.}~\bibnamefont {Rabinovici}}, \bibinfo {author}
  {\bibfnamefont {R.}~\bibnamefont {Shir}},\ and\ \bibinfo {author}
  {\bibfnamefont {R.}~\bibnamefont {Sinha}},\ }\bibfield  {title} {\bibinfo
  {title} {{On the evolution of operator complexity beyond scrambling}},\
  }\href {https://doi.org/10.1007/JHEP10(2019)264} {\bibfield  {journal}
  {\bibinfo  {journal} {J. High Energy Phys.}\ }\textbf {\bibinfo {volume}
  {2019}}\bibinfo  {number} { (10)},\ \bibinfo {pages} {1}}\BibitemShut
  {NoStop}%
\bibitem [{\citenamefont {Bhattacharjee}\ \emph
  {et~al.}(2022{\natexlab{b}})\citenamefont {Bhattacharjee}, \citenamefont
  {Sur},\ and\ \citenamefont {Nandy}}]{NandyScarProbe}%
  \BibitemOpen
\bibfield  {number} {  }\bibfield  {author} {\bibinfo {author} {\bibfnamefont
  {B.}~\bibnamefont {Bhattacharjee}}, \bibinfo {author} {\bibfnamefont
  {S.}~\bibnamefont {Sur}},\ and\ \bibinfo {author} {\bibfnamefont
  {P.}~\bibnamefont {Nandy}},\ }\bibfield  {title} {\bibinfo {title} {{Probing
  quantum scars and weak ergodicity breaking through quantum complexity}},\
  }\href {https://doi.org/10.1103/PhysRevB.106.205150} {\bibfield  {journal}
  {\bibinfo  {journal} {Phys. Rev. B}\ }\textbf {\bibinfo {volume} {106}},\
  \bibinfo {pages} {205150} (\bibinfo {year} {2022}{\natexlab{b}})}\BibitemShut
  {NoStop}%
\bibitem [{\citenamefont {Alishahiha}\ and\ \citenamefont
  {Vasli}(2024)}]{Alishahiha2024Mar}%
  \BibitemOpen
  \bibfield  {author} {\bibinfo {author} {\bibfnamefont {M.}~\bibnamefont
  {Alishahiha}}\ and\ \bibinfo {author} {\bibfnamefont {M.~J.}\ \bibnamefont
  {Vasli}},\ }\href {https://arxiv.org/abs/2403.06655} {\bibinfo {title}
  {Thermalization in krylov basis}} (\bibinfo {year} {2024}),\ \Eprint
  {https://arxiv.org/abs/2403.06655} {arXiv:2403.06655 [quant-ph]} \BibitemShut
  {NoStop}%
\bibitem [{\citenamefont {Garcia-Garcia}\ and\ \citenamefont
  {Verbaarschot}(2016)}]{Garcia-Garcia2016Dec}%
  \BibitemOpen
  \bibfield  {author} {\bibinfo {author} {\bibfnamefont {A.~M.}\ \bibnamefont
  {Garcia-Garcia}}\ and\ \bibinfo {author} {\bibfnamefont {J.~J.~M.}\
  \bibnamefont {Verbaarschot}},\ }\bibfield  {title} {\bibinfo {title}
  {{Spectral and thermodynamic properties of the Sachdev-Ye-Kitaev model}},\
  }\href {https://doi.org/10.1103/PhysRevD.94.126010} {\bibfield  {journal}
  {\bibinfo  {journal} {Phys. Rev. D}\ }\textbf {\bibinfo {volume} {94}},\
  \bibinfo {pages} {126010} (\bibinfo {year} {2016})}\BibitemShut {NoStop}%
\bibitem [{\citenamefont {Altland}\ and\ \citenamefont
  {Bagrets}(2018)}]{Altland2018May}%
  \BibitemOpen
  \bibfield  {author} {\bibinfo {author} {\bibfnamefont {A.}~\bibnamefont
  {Altland}}\ and\ \bibinfo {author} {\bibfnamefont {D.}~\bibnamefont
  {Bagrets}},\ }\bibfield  {title} {\bibinfo {title} {{Quantum ergodicity in
  the SYK model}},\ }\href {https://doi.org/10.1016/j.nuclphysb.2018.02.015}
  {\bibfield  {journal} {\bibinfo  {journal} {Nucl. Phys. B}\ }\textbf
  {\bibinfo {volume} {930}},\ \bibinfo {pages} {45} (\bibinfo {year}
  {2018})}\BibitemShut {NoStop}%
\bibitem [{Note2()}]{Note2}%
  \BibitemOpen
  \bibinfo {note} {$b_1$ is of the order $\protect \ensuremath {{\protect \cal
  O}}(\protect \frac {1}{\protect \sqrt {q}})$.}\BibitemShut {Stop}%
\bibitem [{\citenamefont {Eberlein}\ \emph {et~al.}(2017)\citenamefont
  {Eberlein}, \citenamefont {Kasper}, \citenamefont {Sachdev},\ and\
  \citenamefont {Steinberg}}]{Eberlein2017Nov}%
  \BibitemOpen
  \bibfield  {author} {\bibinfo {author} {\bibfnamefont {A.}~\bibnamefont
  {Eberlein}}, \bibinfo {author} {\bibfnamefont {V.}~\bibnamefont {Kasper}},
  \bibinfo {author} {\bibfnamefont {S.}~\bibnamefont {Sachdev}},\ and\ \bibinfo
  {author} {\bibfnamefont {J.}~\bibnamefont {Steinberg}},\ }\bibfield  {title}
  {\bibinfo {title} {{Quantum quench of the Sachdev-Ye-Kitaev model}},\ }\href
  {https://doi.org/10.1103/PhysRevB.96.205123} {\bibfield  {journal} {\bibinfo
  {journal} {Phys. Rev. B}\ }\textbf {\bibinfo {volume} {96}},\ \bibinfo
  {pages} {205123} (\bibinfo {year} {2017})}\BibitemShut {NoStop}%
\bibitem [{\citenamefont {Louw}\ and\ \citenamefont
  {Kehrein}(2022)}]{Louw2022Feb}%
  \BibitemOpen
  \bibfield  {author} {\bibinfo {author} {\bibfnamefont {J.~C.}\ \bibnamefont
  {Louw}}\ and\ \bibinfo {author} {\bibfnamefont {S.}~\bibnamefont {Kehrein}},\
  }\bibfield  {title} {\bibinfo {title} {{Thermalization of many many-body
  interacting Sachdev-Ye-Kitaev models}},\ }\href
  {https://doi.org/10.1103/PhysRevB.105.075117} {\bibfield  {journal} {\bibinfo
   {journal} {Phys. Rev. B}\ }\textbf {\bibinfo {volume} {105}},\ \bibinfo
  {pages} {075117} (\bibinfo {year} {2022})}\BibitemShut {NoStop}%
\bibitem [{\citenamefont {Monteiro}\ \emph {et~al.}(2021)\citenamefont
  {Monteiro}, \citenamefont {Micklitz}, \citenamefont {Tezuka},\ and\
  \citenamefont {Altland}}]{Monteiro2021Jan}%
  \BibitemOpen
  \bibfield  {author} {\bibinfo {author} {\bibfnamefont {F.}~\bibnamefont
  {Monteiro}}, \bibinfo {author} {\bibfnamefont {T.}~\bibnamefont {Micklitz}},
  \bibinfo {author} {\bibfnamefont {M.}~\bibnamefont {Tezuka}},\ and\ \bibinfo
  {author} {\bibfnamefont {A.}~\bibnamefont {Altland}},\ }\bibfield  {title}
  {\bibinfo {title} {{Minimal model of many-body localization}},\ }\href
  {https://doi.org/10.1103/PhysRevResearch.3.013023} {\bibfield  {journal}
  {\bibinfo  {journal} {Phys. Rev. Res.}\ }\textbf {\bibinfo {volume} {3}},\
  \bibinfo {pages} {013023} (\bibinfo {year} {2021})}\BibitemShut {NoStop}%
\bibitem [{\citenamefont {Garrahan}\ \emph {et~al.}(2007)\citenamefont
  {Garrahan}, \citenamefont {Jack}, \citenamefont {Lecomte}, \citenamefont
  {Pitard}, \citenamefont {van Duijvendijk},\ and\ \citenamefont {van
  Wijland}}]{Garrahan2007May}%
  \BibitemOpen
  \bibfield  {author} {\bibinfo {author} {\bibfnamefont {J.~P.}\ \bibnamefont
  {Garrahan}}, \bibinfo {author} {\bibfnamefont {R.~L.}\ \bibnamefont {Jack}},
  \bibinfo {author} {\bibfnamefont {V.}~\bibnamefont {Lecomte}}, \bibinfo
  {author} {\bibfnamefont {E.}~\bibnamefont {Pitard}}, \bibinfo {author}
  {\bibfnamefont {K.}~\bibnamefont {van Duijvendijk}},\ and\ \bibinfo {author}
  {\bibfnamefont {F.}~\bibnamefont {van Wijland}},\ }\bibfield  {title}
  {\bibinfo {title} {{Dynamical First-Order Phase Transition in Kinetically
  Constrained Models of Glasses}},\ }\href
  {https://doi.org/10.1103/PhysRevLett.98.195702} {\bibfield  {journal}
  {\bibinfo  {journal} {Phys. Rev. Lett.}\ }\textbf {\bibinfo {volume} {98}},\
  \bibinfo {pages} {195702} (\bibinfo {year} {2007})}\BibitemShut {NoStop}%
\bibitem [{\citenamefont {Ba{\ifmmode\tilde{n}\else\~{n}\fi}uls}\ and\
  \citenamefont {Garrahan}(2019)}]{Banuls2019Nov}%
  \BibitemOpen
  \bibfield  {author} {\bibinfo {author} {\bibfnamefont {M.~C.}\ \bibnamefont
  {Ba{\ifmmode\tilde{n}\else\~{n}\fi}uls}}\ and\ \bibinfo {author}
  {\bibfnamefont {J.~P.}\ \bibnamefont {Garrahan}},\ }\bibfield  {title}
  {\bibinfo {title} {Using matrix product states to study the dynamical large
  deviations of kinetically constrained models},\ }\href
  {https://doi.org/10.1103/PhysRevLett.123.200601} {\bibfield  {journal}
  {\bibinfo  {journal} {Phys. Rev. Lett.}\ }\textbf {\bibinfo {volume} {123}},\
  \bibinfo {pages} {200601} (\bibinfo {year} {2019})}\BibitemShut {NoStop}%
\bibitem [{\citenamefont {Bhattacharjee}(2023)}]{Bhattacharjee2023Feb}%
  \BibitemOpen
  \bibfield  {author} {\bibinfo {author} {\bibfnamefont {B.}~\bibnamefont
  {Bhattacharjee}},\ }\href {https://arxiv.org/abs/2302.07228} {\bibinfo
  {title} {A lanczos approach to the adiabatic gauge potential}} (\bibinfo
  {year} {2023}),\ \Eprint {https://arxiv.org/abs/2302.07228} {arXiv:2302.07228
  [quant-ph]} \BibitemShut {NoStop}%
\bibitem [{\citenamefont {Wigner}(1955)}]{Wigner1}%
  \BibitemOpen
  \bibfield  {author} {\bibinfo {author} {\bibfnamefont {E.~P.}\ \bibnamefont
  {Wigner}},\ }\bibfield  {title} {\bibinfo {title} {{Characteristic Vectors of
  Bordered Matrices With Infinite Dimensions on JSTOR}},\ }\href
  {https://www.jstor.org/stable/1970079} {\bibfield  {journal} {\bibinfo
  {journal} {Ann. Of Math.}\ }\textbf {\bibinfo {volume} {62}},\ \bibinfo
  {pages} {548} (\bibinfo {year} {1955})},\ \bibinfo {note} {[Online; accessed
  5. Aug. 2024]}\BibitemShut {NoStop}%
\bibitem [{\citenamefont {Wigner}(1957)}]{Wigner2}%
  \BibitemOpen
  \bibfield  {author} {\bibinfo {author} {\bibfnamefont {E.~P.}\ \bibnamefont
  {Wigner}},\ }\bibfield  {title} {\bibinfo {title} {{Characteristics Vectors
  of Bordered Matrices with Infinite Dimensions II on JSTOR}},\ }\href
  {https://www.jstor.org/stable/1969956} {\bibfield  {journal} {\bibinfo
  {journal} {Ann. Of Math.}\ }\textbf {\bibinfo {volume} {65}},\ \bibinfo
  {pages} {203} (\bibinfo {year} {1957})},\ \bibinfo {note} {[Online; accessed
  5. Aug. 2024]}\BibitemShut {NoStop}%
\bibitem [{\citenamefont {Wigner}(1958)}]{Wigner3}%
  \BibitemOpen
  \bibfield  {author} {\bibinfo {author} {\bibfnamefont {E.~P.}\ \bibnamefont
  {Wigner}},\ }\href {https://www.jstor.org/stable/1970008} {\bibinfo {title}
  {{On the Distribution of the Roots of Certain Symmetric Matrices on JSTOR}}}
  (\bibinfo {year} {1958}),\ \bibinfo {note} {[Online; accessed 5. Aug.
  2024]}\BibitemShut {NoStop}%
\bibitem [{\citenamefont {Dyson}(1962{\natexlab{a}})}]{Dyson1}%
  \BibitemOpen
  \bibfield  {author} {\bibinfo {author} {\bibfnamefont {F.~J.}\ \bibnamefont
  {Dyson}},\ }\bibfield  {title} {\bibinfo {title} {{Statistical Theory of the
  Energy Levels of Complex Systems. I}},\ }\bibfield  {journal} {\bibinfo
  {journal} {J. Math. Phys.}\ }\textbf {\bibinfo {volume} {3}},\ \href
  {https://doi.org/10.1063/1.1703773} {10.1063/1.1703773} (\bibinfo {year}
  {1962}{\natexlab{a}})\BibitemShut {NoStop}%
\bibitem [{\citenamefont {Dyson}(1962{\natexlab{b}})}]{Dyson2}%
  \BibitemOpen
  \bibfield  {author} {\bibinfo {author} {\bibfnamefont {F.~J.}\ \bibnamefont
  {Dyson}},\ }\bibfield  {title} {\bibinfo {title} {{Statistical Theory of the
  Energy Levels of Complex Systems. II}},\ }\bibfield  {journal} {\bibinfo
  {journal} {J. Math. Phys.}\ }\textbf {\bibinfo {volume} {3}},\ \href
  {https://doi.org/10.1063/1.1703774} {10.1063/1.1703774} (\bibinfo {year}
  {1962}{\natexlab{b}})\BibitemShut {NoStop}%
\bibitem [{\citenamefont {Dyson}(1962{\natexlab{c}})}]{Dyson3}%
  \BibitemOpen
  \bibfield  {author} {\bibinfo {author} {\bibfnamefont {F.~J.}\ \bibnamefont
  {Dyson}},\ }\bibfield  {title} {\bibinfo {title} {{Statistical Theory of the
  Energy Levels of Complex Systems. III}},\ }\href
  {https://doi.org/10.1063/1.1703775} {\bibfield  {journal} {\bibinfo
  {journal} {J. Math. Phys.}\ }\textbf {\bibinfo {volume} {3}},\ \bibinfo
  {pages} {166} (\bibinfo {year} {1962}{\natexlab{c}})}\BibitemShut {NoStop}%
\bibitem [{\citenamefont {Lakshminarayan}(1997)}]{correlation-outlook-1}%
  \BibitemOpen
  \bibfield  {author} {\bibinfo {author} {\bibfnamefont {A.}~\bibnamefont
  {Lakshminarayan}},\ }\bibfield  {title} {\bibinfo {title} {{Relaxation
  fluctuations about an equilibrium in quantum chaos}},\ }\href
  {https://doi.org/10.1103/PhysRevE.56.2540} {\bibfield  {journal} {\bibinfo
  {journal} {Phys. Rev. E}\ }\textbf {\bibinfo {volume} {56}},\ \bibinfo
  {pages} {2540} (\bibinfo {year} {1997})}\BibitemShut {NoStop}%
\bibitem [{\citenamefont {Lakshminarayan}(1998)}]{correlation-outlook-2}%
  \BibitemOpen
  \bibfield  {author} {\bibinfo {author} {\bibfnamefont {A.}~\bibnamefont
  {Lakshminarayan}},\ }\href {https://arxiv.org/abs/chao-dyn/9804015} {\bibinfo
  {title} {Relaxation fluctuations in quantum chaos}} (\bibinfo {year}
  {1998}),\ \Eprint {https://arxiv.org/abs/chao-dyn/9804015}
  {arXiv:chao-dyn/9804015 [chao-dyn]} \BibitemShut {NoStop}%
\bibitem [{\citenamefont {Lee}(2001)}]{correlation-outlook-3}%
  \BibitemOpen
  \bibfield  {author} {\bibinfo {author} {\bibfnamefont {M.~H.}\ \bibnamefont
  {Lee}},\ }\bibfield  {title} {\bibinfo {title} {{Ergodic Theory, Infinite
  Products, and Long Time Behavior in Hermitian Models}},\ }\href
  {https://doi.org/10.1103/PhysRevLett.87.250601} {\bibfield  {journal}
  {\bibinfo  {journal} {Phys. Rev. Lett.}\ }\textbf {\bibinfo {volume} {87}},\
  \bibinfo {pages} {250601} (\bibinfo {year} {2001})}\BibitemShut {NoStop}%
\bibitem [{\citenamefont {Pathak}(2024)}]{correlation-outlook-4}%
  \BibitemOpen
  \bibfield  {author} {\bibinfo {author} {\bibfnamefont {T.}~\bibnamefont
  {Pathak}},\ }\href {https://arxiv.org/abs/2407.21644} {\bibinfo {title}
  {Relaxation fluctuations of correlation functions: Spin and random matrix
  models}} (\bibinfo {year} {2024}),\ \Eprint
  {https://arxiv.org/abs/2407.21644} {arXiv:2407.21644 [quant-ph]} \BibitemShut
  {NoStop}%
\bibitem [{Note3()}]{Note3}%
  \BibitemOpen
  \bibinfo {note} {These three phases are also captured independently using the
  adiabatic gauge potential (AGP), for example, as studied in the context of
  the coupled SYK model which was considered in this work as well \cite
  {Nandy2022Dec}.}\BibitemShut {Stop}%
\bibitem [{Note4()}]{Note4}%
  \BibitemOpen
  \bibinfo {note} {We also compared the Krylov variance where full sequence of
  Lanczos coefficients are used (both before and after the scrambling time) in
  Figs.~\ref {fig:syk-variance_full} and~\ref {fig:east-variance_full}. See the
  text in the manuscript for more details}\BibitemShut {NoStop}%
\bibitem [{\citenamefont {{\ifmmode\check{S}\else\v{S}\fi}untajs}\ and\
  \citenamefont {Vidmar}(2022)}]{original-quantum-sun-prl}%
  \BibitemOpen
  \bibfield  {author} {\bibinfo {author} {\bibfnamefont {J.}~\bibnamefont
  {{\ifmmode\check{S}\else\v{S}\fi}untajs}}\ and\ \bibinfo {author}
  {\bibfnamefont {L.}~\bibnamefont {Vidmar}},\ }\bibfield  {title} {\bibinfo
  {title} {{Ergodicity Breaking Transition in Zero Dimensions}},\ }\href
  {https://doi.org/10.1103/PhysRevLett.129.060602} {\bibfield  {journal}
  {\bibinfo  {journal} {Phys. Rev. Lett.}\ }\textbf {\bibinfo {volume} {129}},\
  \bibinfo {pages} {060602} (\bibinfo {year} {2022})}\BibitemShut {NoStop}%
\bibitem [{\citenamefont {Pawlik}\ \emph {et~al.}(2024)\citenamefont {Pawlik},
  \citenamefont {Sierant}, \citenamefont {Vidmar},\ and\ \citenamefont
  {Zakrzewski}}]{quantum-sun-mobility-edge}%
  \BibitemOpen
  \bibfield  {author} {\bibinfo {author} {\bibfnamefont {K.}~\bibnamefont
  {Pawlik}}, \bibinfo {author} {\bibfnamefont {P.}~\bibnamefont {Sierant}},
  \bibinfo {author} {\bibfnamefont {L.}~\bibnamefont {Vidmar}},\ and\ \bibinfo
  {author} {\bibfnamefont {J.}~\bibnamefont {Zakrzewski}},\ }\bibfield  {title}
  {\bibinfo {title} {{Many-body mobility edge in quantum sun models}},\ }\href
  {https://doi.org/10.1103/PhysRevB.109.L180201} {\bibfield  {journal}
  {\bibinfo  {journal} {Phys. Rev. B}\ }\textbf {\bibinfo {volume} {109}},\
  \bibinfo {pages} {L180201} (\bibinfo {year} {2024})}\BibitemShut {NoStop}%
\bibitem [{\citenamefont {Menzler}\ and\ \citenamefont
  {Jha}(2024)}]{menzler_2024_10975283}%
  \BibitemOpen
  \bibfield  {author} {\bibinfo {author} {\bibfnamefont {H.~G.}\ \bibnamefont
  {Menzler}}\ and\ \bibinfo {author} {\bibfnamefont {R.}~\bibnamefont {Jha}},\
  }\bibfield  {title} {\bibinfo {title} {{Krylov Delocalization/Localization
  across Ergodicity Breaking}},\ }\href
  {https://doi.org/10.5281/zenodo.10975283} {10.5281/zenodo.10975283} (\bibinfo
  {year} {2024})\BibitemShut {NoStop}%
\bibitem [{\citenamefont {Maldacena}\ \emph {et~al.}(2016)\citenamefont
  {Maldacena}, \citenamefont {Shenker},\ and\ \citenamefont
  {Stanford}}]{Maldacena2016Aug}%
  \BibitemOpen
  \bibfield  {author} {\bibinfo {author} {\bibfnamefont {J.}~\bibnamefont
  {Maldacena}}, \bibinfo {author} {\bibfnamefont {S.~H.}\ \bibnamefont
  {Shenker}},\ and\ \bibinfo {author} {\bibfnamefont {D.}~\bibnamefont
  {Stanford}},\ }\bibfield  {title} {\bibinfo {title} {{A bound on chaos}},\
  }\href {https://doi.org/10.1007/JHEP08(2016)106} {\bibfield  {journal}
  {\bibinfo  {journal} {J. High Energy Phys.}\ }\textbf {\bibinfo {volume}
  {2016}}\bibinfo  {number} { (8)},\ \bibinfo {pages} {1}}\BibitemShut
  {NoStop}%
\bibitem [{\citenamefont {Cotler}\ \emph {et~al.}(2017)\citenamefont {Cotler},
  \citenamefont {Gur-Ari}, \citenamefont {Hanada}, \citenamefont {Polchinski},
  \citenamefont {Saad}, \citenamefont {Shenker}, \citenamefont {Stanford},
  \citenamefont {Streicher},\ and\ \citenamefont {Tezuka}}]{Cotler2017May}%
  \BibitemOpen
\bibfield  {number} {  }\bibfield  {author} {\bibinfo {author} {\bibfnamefont
  {J.~S.}\ \bibnamefont {Cotler}}, \bibinfo {author} {\bibfnamefont
  {G.}~\bibnamefont {Gur-Ari}}, \bibinfo {author} {\bibfnamefont
  {M.}~\bibnamefont {Hanada}}, \bibinfo {author} {\bibfnamefont
  {J.}~\bibnamefont {Polchinski}}, \bibinfo {author} {\bibfnamefont
  {P.}~\bibnamefont {Saad}}, \bibinfo {author} {\bibfnamefont {S.~H.}\
  \bibnamefont {Shenker}}, \bibinfo {author} {\bibfnamefont {D.}~\bibnamefont
  {Stanford}}, \bibinfo {author} {\bibfnamefont {A.}~\bibnamefont
  {Streicher}},\ and\ \bibinfo {author} {\bibfnamefont {M.}~\bibnamefont
  {Tezuka}},\ }\bibfield  {title} {\bibinfo {title} {{Black holes and random
  matrices}},\ }\href {https://doi.org/10.1007/JHEP05(2017)118} {\bibfield
  {journal} {\bibinfo  {journal} {J. High Energy Phys.}\ }\textbf {\bibinfo
  {volume} {2017}}\bibinfo  {number} { (5)},\ \bibinfo {pages} {1}}\BibitemShut
  {NoStop}%
\end{thebibliography}%
	
	\appendix

	\section{Krylov Complexity $K(t)$}
	\label{app. krylov complexity}
	
	\subsection{Definition}
	\label{krylov subsection}
	Since we have the Krylov basis as in Eq.~\eqref{krylov basis}, we can expand any operator $|\Oo(t) )$ in terms of these basis states as
	\begin{equation}
		|\mathcal{O}(t) )=\sum_{n=0}^{\Kk-1}  \phi_n(t) | \mathcal{O}_n)\,,
		\label{phi introduced}
	\end{equation}
	where we have $| \Oo(t=0) ) = | \Oo_0)$. Here $\phi_n(t)$ are time-dependent functions that capture the spread of the operator over different Krylov basis vectors. We now use the Heisenberg equation of motion
	\begin{equation}
		\partial_t \Oo(t) = \i [\Hh, \Oo(t)] = \Ll \Oo(t) \qquad (\Ll \equiv \i [\Hh, \sbullet[0.75]])\,,
		\label{heisenberg eom}
	\end{equation}
	where we substitute Eq. \eqref{phi introduced} for $|\Oo(t))$ to get the following differential equation for $\phi(t)$:
	\begin{equation}
		\p_t \phi_n(t)=b_n \phi_{n-1}(t)+b_{n+1} \phi_{n+1} (t)\,,
		\label{phi differential equation}
	\end{equation}
	with the initial conditions $\phi_n(t=0) = \delta_{n,0}$, $b_{n=0} = 0$, and $\phi_{n=-1}(t)= 0$. Therefore solving this ``real-wave-equation-type'' structure for $\phi_n(t)$ is equivalent to evolving the operator $|\Oo(t))$ (Eq.~\eqref{phi introduced}). Thus the Lanczos coefficients $\{ b_n\}_{n=1}^{\Kk-1}$ are physically interpreted as ``nearest-neighbor hopping amplitudes'' on this one-dimensional chain with $\phi_n(t)$ being the ``wavefunction'' of the moving ``particle'' along the chain. With this setup, the Krylov complexity (also sometimes referred to as $K$-complexity) is defined as
	\begin{equation}
		K(t)\equiv \sum\limits_{n=0}^{\Kk-1} n\left|\phi_n(t)\right|^2\,,
		\label{krylov complexity def}
	\end{equation}
	which can be physically interpreted as expectation value of ``position'' on the above Krylov chain. It can be easily verified that if we use a different expansion coefficient in Eq.~\eqref{phi introduced} such as 
	\begin{equation}
		|\mathcal{O}(t) )=\sum_{n=0}^{\Kk-1} \i^n \psi_n(t) | \mathcal{O}_n)
		\label{psi introduced}
	\end{equation}
	and use the Heisenberg's equation of motion for $|\Oo(t))$ in Eq. \eqref{heisenberg eom}, we get the following differential equation for $\psi_n(t)$:
	\begin{equation}
		\i \p_t \psi_n(t) = b_n \psi_{n-1}(t) - b_{n+1}\psi_{n+1}(t)\,,
		\label{psi differential equation}
	\end{equation}
	which has a ``Schr{\"o}dinger-type'' structure and the same initial conditions as before, namely $\psi_n(t=0) = \delta_{n,0}$, $b_{n=0} = 0$, and $\psi_{n=-1}(t)= 0$. The coefficients $\phi_n(t)$ and $\psi_n(t)$ are related through $ \phi_n(t) = \i^n \psi_n(t)$.
	
	\subsection{Growth of Krylov complexity}
	
	For a chaotic system, we now analyze the various stages of growth of Krylov complexity. We proceed in chronological dynamical order to form a complete picture.
	
	\subsubsection{Universal Operator Growth Hypothesis}
	
	\label{ogh subsection}
	A ``universal operator growth hypothesis'' was proposed in \cite{Parker2019Oct} whose fundamental tenet is that any chaotic system implies a linear growth of Lanczos coefficients $\{b_n\}$ with $n$, or in other words $b_n \sim  \alpha n$ (with logarithmic corrections in one dimension \cite{Parker2019Oct}) where the proportionality $\alpha$ is system dependent. It was shown \cite{Parker2019Oct} that $b_n \sim \alpha n$ implies an exponential growth of Krylov complexity defined in Eq.~\eqref{krylov complexity def} as $	K (t) \sim e^{2 \alpha t}$ and that $\alpha$ bounds the Lyapunov exponent as $\lambda_L \leq 2 \alpha$. This is an improvement over the famous Maldacena-Shenker-Stanford bound on chaos \cite{Maldacena2016Aug} (see Fig.~8 of \cite{Parker2019Oct} for the case of large-$q$ SYK model).
	
	\subsubsection{Random matrix theory}
	\label{rmt subsection}
	
	Operator growth hypothesis as mentioned above deals with the initial growth of operators where the Lanczos coefficients $b_n$ grows with $n$ \cite{Tang2023Dec}. In the context of random matrix theory (RMT), the asymptotic value of $b_n$ in large-$N$ theories saturates to $1$. In other words,
	\begin{equation}
		\lim_{n\to \infty} \lim_{N \to \infty} b_n = 1 \qquad (\text{RMT})
	\end{equation}
	
	We know that eigenvalues of quantum many-body chaotic systems are extensive in the degrees of freedom of the system while while the eigenvalues for a random matrix Hamiltonian is usually scaled to be of $\Oo(1)$. That's why in order to compare the growth of operators with the random matrix behavior, we need to properly scale our results using the fact that the energy scale is conjugate to the time scale. 
	
	The spectrum of a random matrix can be described by a semi-circle of width two and the non-re-scaled Lanczos coefficients are subject to the same scale as the eigenvalues.
	Therefore, we re-scale the original Lanczos coefficients $\tilde{b}_n$ with respect to the largest and smallest eigenvalues $E_\mathrm{max}$ and $E_\mathrm{min}$ by 
	
	\begin{equation}
		\label{eq:lanczos_coeff_rescale}
		b_n = \tilde{b}_n / r_\mathrm{spectrum}\,,
		\quad
		\text{where  }
		r_\mathrm{spectrum} = \frac{E_\mathrm{max} - E_\mathrm{min}}{2}
		\,.
	\end{equation}
	
	This ensures that the Lanczos coefficients $b_n$---as shown throughout this work---are all re-scaled in such a way that they are comparable with the results of RMT allowing for their proper comparison throughout the parameter space.
	Indeed, this process resembles re-scaling of the Hamiltonian to energy scales of the original Wigner semi-circle law for random matrices.
	
	\subsubsection{Decay at the edge of Krylov space}
	At the edge of the Krylov space \cite{journey-to-edge}, the Lanczos coefficients $\{ b_n\}$ descent back to zero where the algorithm stops and this signifies the saturation of the Krylov complexity. 
	
	\subsubsection{Summary}
	\label{app. summary}
	Therefore the picture that emerges for the growth of Krylov complexity and its mapping to the Krylov space in terms of the Lanczos coefficients $\{b_n\}$ has three stages where the third stage kicks in after a significantly long time as mentioned below. The three stages are \cite{Barbon2019Oct, journey-to-edge, integrability-to-chaos}
	\begin{enumerate}
		\item initially a linear growth of $\{b_n\}$ for $\ll n<\Oo(f)$ implies an exponential growth in time of $K(t)$ for $0\lesssim t<\Oo(\log(f))$,
		\item a saturation after the linear growth happens for $\{b_n\}$ for $n \gg \Oo(f)$ that implies a linear-in-time growth of $K(t)$ for $t\gtrsim \Oo(\log(f))$, and
		\item finally the descent of $\{b_n\}$ to zero for $n\sim \Oo(e^{2f})$ implying a saturation of $K(t)$ for $t\sim \Oo(e^{2f})$.
	\end{enumerate}
	We argued in this work that all local operators looses their sense of locality once the stage (2) of the growth kicks in and that's why this allows for a possibility to have universal behavior for all operators (with vanishing overlap with any other conserved quantity in the system).

	\section{Equivalence of Different Approaches}
	\label{app. equivalence}
	
	In the literature on Krylov complexity, there is another physically equivalent definition of Liouvillian is used which is given by
	\begin{equation}
		\Lltil \equiv [\Hh, \sbullet[0.75]]\,,
	\end{equation}
	in contrast to our definition of Liouvillian considered throughout this work, namely $\Ll \equiv \i [\Hh, \sbullet[0.75]]$. We now show equivalence between these two approaches. We always initialize the Lanczos algorithm with a Hermitian operator $|\Oo_0)$. As we showed in Eq. \eqref{Ll preserves hermiticity} that $\Ll$ preserves Hermiticity, this is not true for $\Lltil$ which acts on any arbitrary Hermitian operator $\Oo_h$ to give an anti-Hermitian operator $\Oo_{ah}$ and \textit{vice versa}:
	\begin{widetext}
		\begin{equation}
			\begin{aligned}
				(\Lltil \Oo_h)^\dagger &= \left( \left[\Hh, \Oo_h\right]\right)^\dagger =  \left(\Hh \Oo_h - \Oo_h \Hh\right)^\dagger =  ( \Oo_h \Hh -  \Hh \Oo_h) =-\Lltil \Oo_h    \\    
				(\Lltil \Oo_{ah})^\dagger &=\left( \left[\Hh, \Oo_{ah}\right]\right)^\dagger =  ( \Oo_{ah}^\dagger \Hh -  \Hh \Oo_{ah}^\dagger) =( -\Oo_{ah} \Hh + \Hh \Oo_{ah}) =\Lltil \Oo_{ah}   
			\end{aligned}
			\label{Lltil does not preserve hermiticity}
		\end{equation}
	\end{widetext}
	
	Therefore when the Lanczos algorithm, as explained in Sec. \ref{lanczos algo subsection}, is implemented using $\Lltil$ instead of $\Ll$ by starting from an initial Hermitian operator $|\Oo_0^\prime)$, we get the Krylov basis states as $\{ |\Oo_n^\prime )\}_{n=0}^{\Kk-1}$ which are alternating between Hermitian (for even values of $n$) and anti-Hermitian (for odd values of $n$) operator states. Therefore $\i^n \Oo_n^\prime$ is Hermitian for all values of $n$. Moreover, for both $\Ll$ and $\Lltil$, the diagonal elements in the Liouvillian matrix representation are identically zero. This is because $(\Oo_n^\prime | \Lltil |\Oo_n^\prime) = 0$ irrespective of whether $|\Oo_n^\prime)$ is Hermitian or anti-Hermitian. Recall the definition of inner-product: $(A|B) = \frac{1}{\Nn} \Tr[A^\dagger B]$ where $\Nn $ is the dimension of the Hilbert space. Let the off-diagonal Lanczos elements be $\{ b_n\}$ and $\{ b_n^\prime \}$ which are obtained through $\Ll$ and $\Lltil$, respectively. Both $\Ll$ and $\Lltil$ have a tri-diagonal matrix representation (Eq. \eqref{Ll operator matrix form}) in $\{b_n\}$ and $\{ b_n^\prime \}$, respectively which can be written as ($\Kk$ is the dimension of the Krylov space)
	\begin{widetext}
		\begin{equation}
			\Ll = \sum\limits_{n=0}^{\Kk-2} b_{n+1} \left( |\mathcal{O}_n)(\mathcal{O}_{n+1}|+| \mathcal{O}_{n+1})(\mathcal{O}_n |\right), \qquad \Lltil = \sum\limits_{n=0}^{\Kk-2} b_{n+1}^\prime \left( |\mathcal{O}_n^\prime)(\mathcal{O}_{n+1}^\prime|+| \mathcal{O}_{n+1}^\prime)(\mathcal{O}_n^\prime |\right)
			\,.
		\end{equation}
	\end{widetext}

	In order to understand the connection between the two approaches, we study the evolution of an arbitrary operator with respect to both $\Ll$ (already shown in Eq. \eqref{evolution of operator by Ll} and reproduced here for convenience) and $\Lltil$
	\begin{equation}
		\begin{aligned}
			| \Oo(t) )&=e^{ \mathcal{L} t} | \Oo_0) =\sum_{n=0}^{\infty} \frac{t^n}{n !} | \mathcal{L}^n \Oo_0), \\
			| \Oo^\prime(\tau) )&=e^{ \i \Lltil \tau} | \Oo_0^\prime) =\sum_{n=0}^{\infty} \frac{(\i \tau)^n}{n !} | \Lltil^n \Oo_0^\prime)\,,
		\end{aligned}
		\label{evolution through Ll and Lltil}
	\end{equation}
	where time is labeled by $t$ and $\tau$ to denote the time evolution by $\Ll$ and $\Lltil$ respectively. Since any operator matches at initial time $t=\tau=0$, namely $|\Oo(t=0)) = |\Oo^\prime(\tau=0))$, this implies for $n=0$ Krylov operator basis state that $|\Oo_0) = |\Oo_0^\prime)$ which, for instance, is used as an initial operator in the Lanczos algorithm either using $\Ll$ or $\Lltil$. Therefore, assuming analytic continuation, we find equivalence between the two definitions using the \emph{mapping} $\left\{t \leftrightarrow \i \tau, \Ll \leftrightarrow \Lltil \right\}$ which ensures that the physical content stays the same. For instance, the Heisenberg's equation of motion in Eq. \eqref{heisenberg eom} structurally remains the same under this transformation where we take $t \to \i \tau$ and $\Ll \to \Lltil$ as well as denote $\Oo(t\to \i \tau) = \Oo^\prime(\tau)$.

	We know that the evolution of operator in Eq. \eqref{evolution through Ll and Lltil} can be translated to a differential equation of ``wave-equations'' on a one-dimensional Krylov chain (e.g., Eq. \eqref{phi differential equation} corresponding to Eq. \eqref{phi introduced}, or equivalently Eq. \eqref{psi differential equation} corresponding to Eq. \eqref{psi introduced}). Similarly we expand an arbitrary operator in terms of Krylov basis operator states $\{ |\Oo_n^\prime )\}_{n=0}^{\Kk-1}$ corresponding to the evolution through $\Lltil$:
	\begin{equation}
		|\Oo^\prime(\tau)) = \sum\limits_{n=0}^{\Kk-1} \varphi_n(\tau) |\Oo_n^\prime)
	\end{equation}
	Then the Heisenberg equation of motion leads to the following ``Schr{\"o}dinger-type'' differential equation corresponding to operator growth with respect to $\Lltil$:
	\begin{equation}
		\i \p_\tau \varphi_n(\tau) = - b_{n+1}^\prime \varphi_{n+1}(\tau) - b_n^\prime \varphi_{n-1}(\tau)\,,
	\end{equation}
	with the initial conditions $\varphi_n(\tau=0) = \delta_{n,0} $, $b_{n=0}^\prime = 0$ and $\varphi_{n=-1}(\tau)=0$. Here we used the Lanczos step $|\Lltil \Oo_n^\prime) = b_{n+1}^\prime |\Oo_{n+1}^\prime) + b_n^\prime |\Oo_{n-1}^\prime)$. This is the same differential equation as found in \cite{Tang2023Dec} using $\Lltil$. If we translate this differential equation using the aforementioned mapping and denote $\varphi_n(\tau \to -\i t) = \phi(t)$, then we re-derive our ``real-wave-equation-type'' differential equation in Eq. \eqref{phi differential equation} along with the same initial conditions where we have used $\Ll$ for operator evolution. Therefore the mapping obtained here also ensures that the Lanczos coefficients corresponding to $\Ll$ and $\Lltil$ are mapped onto each other thereby proving the equivalence of the two approaches. As a redundant check, we express an arbitrary operator using another set of expansion coefficients (like Eq. \eqref{psi introduced}) as
	\begin{equation}
		|\Oo^\prime(\tau)) = \sum\limits_{n=0}^{\Kk-1}\i^n \Psi_n(\tau) |\Oo_n^\prime)\,,
	\end{equation}
	which leads to the following ``real-wave-equation-type'' differential equation
	\begin{equation}
		\p_\tau \Psi(\tau) = - b_{n+1}^\prime \Psi_{n+1}(\tau) + b_{n}^\prime \Psi_{n-1}(\tau)\,,
	\end{equation}
	with the initial conditions $\Psi_n(\tau=0) = \delta_{n,0} $, $b_{n=0}^\prime = 0$ and $\Psi_{n=-1}(\tau)=0$. This is the same as obtained in \cite{Parker2019Oct, Tang2023Dec} using $\Lltil$. We again re-derive ``Schr{\"o}dinger-type'' differential equation in Eq. \eqref{psi differential equation} when we use the aforementioned mapping where we denote $\Psi(\tau \to -\i t) = \psi(t)$. Therefore, we again found that the Lanczos coefficients corresponding to $\Ll$ and $\Lltil$ are mapped one-to-one using the mapping. Hence we have established that the physical content of both the approaches corresponding to $\Ll$ and $\Lltil$ are equivalent and can be mapped onto each other using the aforementioned mapping. 
	
	
	\section{Coupled Sachdev-Ye-Kitaev model}
	\label{app. syk model}

	The Hamiltonian considered in the main manuscript is
	\begin{equation}
		\Hh = \frac{2}{\sqrt{N}}\sum\limits_{1 \leq i<j<k<l\leq N} J_{i j k l} \chi_i \chi_j \chi_k \chi_l 
		+ i \sum\limits_{1 \leq i<j \leq N} \kappa_{i j}\chi_i \chi_j\,.
		\label{app. coupled syk model}
	\end{equation}
	The random couplings $J_{ijkl}$ and $\kappa_{ij}$ are drawn from a Gaussian distribution with variance $\frac{6J^2}{N^{3}}$ and $\frac{\kappa^2}{N}$ respectively, where $J$ and $\kappa$ are system parameters and we take the large-$N$ limit \cite{Maldacena-syk}. We measure $\kappa$ in units of $J$ where we have kept $J=1$ to fix the unit system.
	This is the same Hamiltonian as studied in \cite{Nandy2022Dec} where they studied the adiabatic gauge potential (AGP) and found a delayed thermalization due to the violation of ETH scaling in the AGP as confirmed by the spectral form factor analyses.

	\onecolumngrid
	
	\begin{figure}[htbp]
		\centering
		\includegraphics[width=0.9\linewidth]{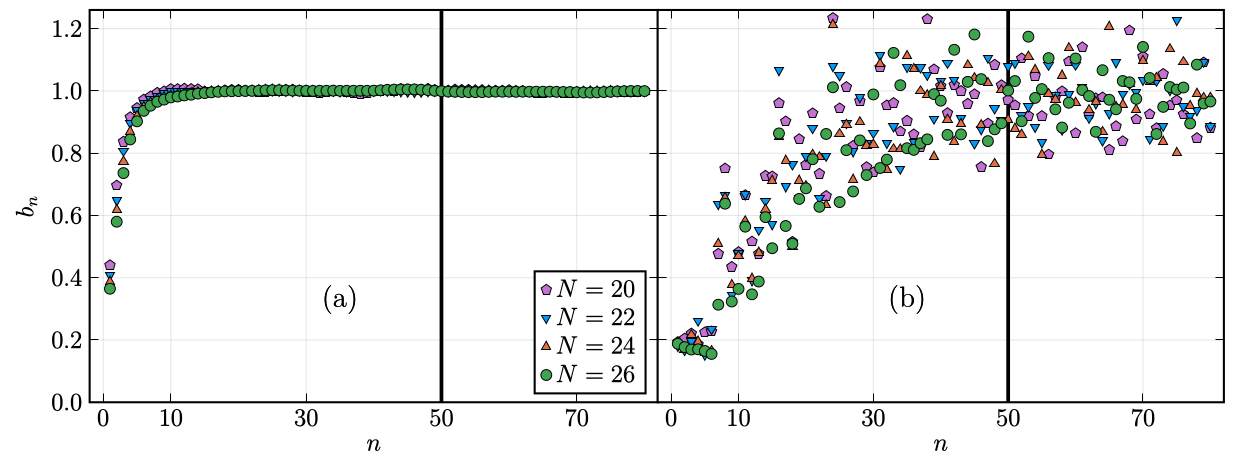}
		{\phantomsubcaption\label{fig:syk-bn-Ns_ergodic}}
		{\phantomsubcaption\label{fig:syk-bn-Ns_nonergodic}}
		\caption{
			Lanczos coefficients $\{b_n\}$ compared for different system sizes for the coupled SYK model (Eq.~\eqref{app. coupled syk model}) in \subref{fig:syk-bn-Ns_ergodic} the ergodic regime $\kappa=0.01\,J$  and in \subref{fig:syk-bn-Ns_nonergodic} the ergodicity-broken regime $\kappa = 100\,J$ with the initial operator $|\mathcal{O}_0) = \chi_1$ at $N=26$.
			We visualize the end of the ramp for all parameters considered with a vertical solid line at $n=50$.
		}
		\label{fig:syk_-bn-Ns}
	\end{figure}
	
	\twocolumngrid

	\begin{figure}[htbp]
		\centering
		\includegraphics[width=\linewidth]{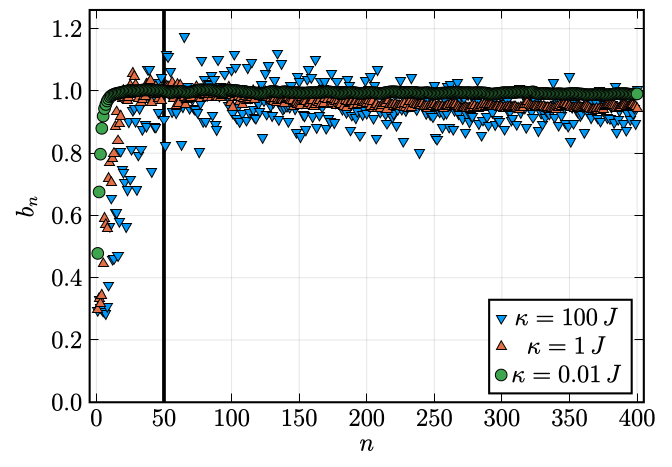}
		\label{fig:syk_bns_chi12}
		
		\caption{
			Lanczos coefficients $\{b_n\}$ at $\beta = 0$ for the coupled SYK model model \eqref{coupled syk model} for $N=26$. The initial operator is $|\mathcal{O}_0) = \i \chi_1 \chi_2$ which---in contrast to $\chi_1$---has support in only a single charge parity sector.
			Again, we mark the value of $n=50$ where the initial ramp in the Lanczos coefficients stops with a solid vertical line.
		}
		\label{fig:syk_bns_operators}
	\end{figure}

	The Majorana SYK has a particle-hole symmetry as well as charge-parity symmetry. We formalize this now. For $N$ Majorana fermions, the dimension of the Hilbert space is $2^{N/2}$ which is equivalent to having $L=N/2$ Dirac fermions. Often, it is convenient to use the Dirac basis to computationally handle Majorana fermions. Let the Dirac operators be denoted by $c_j$ where $j = 1, 2, \ldots , L$ where $L=N/2$. They satisfy the standard anti-commutation relations: $\{c_i, c_j^\dagger\} = \delta_{ij}$, $\{c_i, c_j \} = 0$ and $\{c^\dagger_i, c^\dagger_j \}=0$. Then the Majorana fermions can be written as \cite{Cotler2017May}
	\begin{equation}
		\chi_{2j} = \frac{c_j + c^\dagger_j}{\sqrt{2}}, \quad\chi_{2j-1} = \frac{\i (c_j - c^\dagger_j)}{\sqrt{2}}\,.
	\end{equation}
	Then, we can define the total charge of the system as $Q = \sum_{j=1}^{L=N/2} c^\dagger_j c_j$. 
	Clearly, $[\Hh_q, Q] \neq 0$ where $\Hh_q$ is the Hamiltonian of single arbitrary $q$ body Majorana SYK model where $q$ is always considered to be even. 
	It is given by $\Hh_q =\i^{q / 2} \sum_{1 \leq i_1<\cdots<i_q \leq N} J_{i_1 \cdots i_q} \chi_{i_1} \cdots \chi_{i_q}$ as also mentioned in the main manuscript. 
	Therefore the charge is not conserved but $(Q \mod (q/2))$ commutes with the Hamiltonian $\Hh_q$. 
	For our case $q=4$, therefore $(Q \mod 2)$ commutes with $\Hh_4$ and therefore, the Hamiltonian $\Hh_4$ is divided into two blocks: even sector and odd sector corresponding to even and odd values of $Q$. 
	Two choices of initial operators, namely $\chi_1$ and $\i \chi_1 \chi_2 \chi_3$, have support in both symmetry sectors and in order to be consistent, we have considered the full spectrum for all operators, to generate the plots presented in the main manuscript and here in the appendix.
	
	Finally, the symmetry between holes and particles are captured by the operator $\hat{P} = T \prod_{j=1}^{L=N/2} (c_j + c^\dagger_j)$ where $T: \mathbb{C} \to \mathbb{C}$ such that $T z = \overline{z}$ (complex conjugate) $\forall$ $z \in \mathbb{C}$. 
	For $q=4$, $\hat{P}^2 = +1$ for $L \mod 4 = 0, 1$ while $\hat{P}^2 = -1$ for $L \mod 4 = 2, 3$. 
	Using the relation $\hat{P}c_j \hat{P} = \alpha c^\dagger_j$ and $\hat{P}c^\dagger_j\hat{P} = \alpha c_j$, we get $\hat{P} \chi_j \hat{P} = \alpha \chi_j$ where $\alpha \equiv (-1)^{L-1}\hat{P}^2$.
	Using all these properties, one can show that $[\Hh_4, \hat{P}] = 0$ and hence is a symmetry of the Hamiltonian.
	Generalization to $q$-body case is straightforward.
	
	In Fig.~\ref{fig:syk_bns} of the main manuscript, we showed the Lanczos coefficients corresponding to the operator $\chi_1$ for $N=26$, here we show in Fig.~\ref{fig:syk_-bn-Ns} the coefficients for different system sizes in the ergodic and ergodic-broken regimes. We also show in Fig.~\ref{fig:syk_bns_operators} the Lanczos coefficients for another operator $|\Oo_0) = \i \chi_1 \chi_2$ for a given system size which has a support in both symmetry sectors of the Hamiltonian as explained above. Both the plots show that our cut-off $n=50$ for the initial ramp holds across all considered system sizes. We have also checked for all combinations of operators and system sizes to ensure the validity of the cut-off.


	\section{Quantum East Model}
	\label{app. quantum east model}

	The Hamiltonian of the Quantum East model as defined on a 1D lattice of size $L$ is \cite{Pancotti2020Jun}
	\begin{equation}
		\Hh = - \frac{1}{2}\sum\limits_{i=1}^{L-1} n_i (e^{-s} \sigma^x_{i+1} - \mathbb{1})\,,
		\label{app.eq:quantum_east}
	\end{equation} 
	where $n_i$ is the projection on the spin-up state at lattice site $i$, $\sigma^x_i$ is the $x$-Pauli operator acting on that respective lattice site $i$ and $s$ is a system parameter.
	As $n_i$ projects on the spin-up state, we can consider any spin-down lattice site as a kinetic constraint for dynamics on the next lattice site.
	
	\onecolumngrid
	
	\begin{figure}[htbp]
		\centering
		\includegraphics[width=0.9\linewidth]{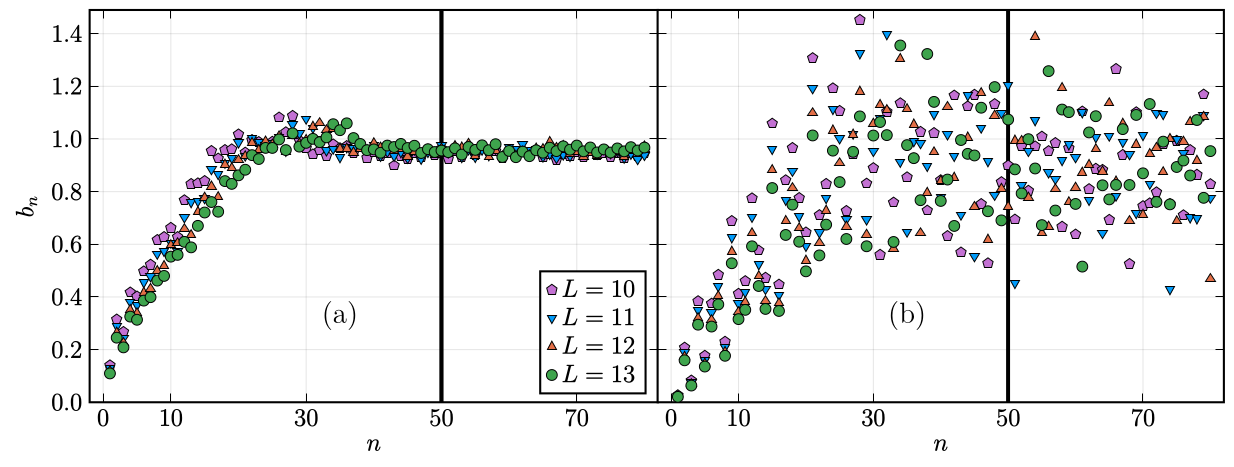}
		{\phantomsubcaption\label{fig:quantum_east-bn-Ls_ergodic}}
		{\phantomsubcaption\label{fig:quantum_east-bn-Ls_nonergodic}}
		\caption{
			Lanczos coefficients $\{b_n\}$ compared for different system sizes for the quantum East model (Eq.~\eqref{app.eq:quantum_east}) in \subref{fig:quantum_east-bn-Ls_ergodic} the ergodic regime $s=-2$  and in \subref{fig:quantum_east-bn-Ls_nonergodic} the ergodicity-broken regime $s=2$ for the initial operator $|\mathcal{O}_0) = n_6$ at $L=13$.
			We visualize the end of the ramp with a vertical solid line at $n=50$.
		}
		\label{fig:quantum_east-bn-Ls}
	\end{figure}
	
	\twocolumngrid
	
	\begin{figure}[htbp]
		\centering
		\includegraphics[width=\linewidth]{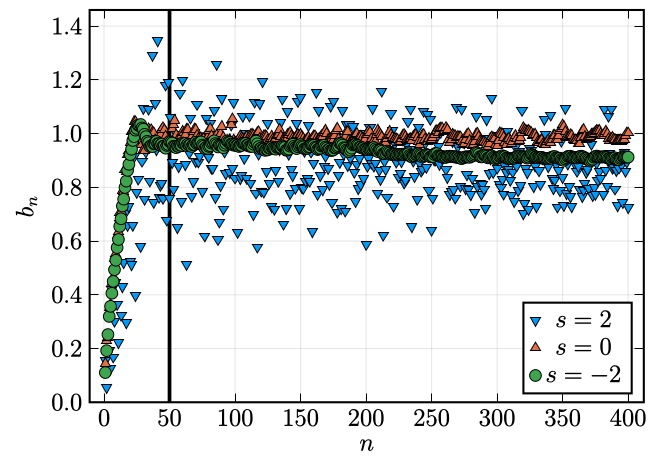}
		{\phantomsubcaption\label{fig:quantum_east_bns_nL}}
		\caption{
			Lanczos coefficients $\{b_n\}$ at $\beta = 0$ in the quantum East model (Eq.~\eqref{app.eq:quantum_east}) for $L=13$.
			The initial operator is $|\mathcal{O}_0) = \sigma^x_6$.
			We again mark the value of $n=50$ where the ramp in the Lanczos coefficients ends by a solid vertical line.
		}
		\label{fig:quantum_east_bns_operators}
	\end{figure}

	Naturally, due to the kinetic constraint, the quantum East model as in Eq.~\eqref{app.eq:quantum_east} can be divided into symmetry sectors:
	as spin-up sites can not facilitate dynamics on the previous lattice sites, a string of $l$ spin-down lattice sites, starting on the first site, will be kinetically disconnected from the dynamical part of the system.
	Therefore, their states must also be disconnected in Fock space, giving rise to a block structure of the Hamiltonian with the quantum number $l$.
	
	As all of these symmetry blocks are self-similar---in the sense that they just represent the same system but with a smaller system size---we naturally only want to study the largest symmetry block.
	To maximize the number of dynamical sites in our computational basis we modify the boundary conditions of the Hamiltonian~\eqref{app.eq:quantum_east} by inserting a spin-up lattice site on the $0^{\text{th}}$ lattice site (not simulated) and also allow for dephasing due to facilitation of dynamics on a lattice site at $L+1$ (also not simulated).
	The effective Hamiltonian with the aforementioned boundary conditions can be realized as
	\begin{equation}
		\Hh = 
		- \frac{1}{2}(e^{-s} \sigma_1^x - \mathbb{1}) - \frac{1}{2}\sum\limits_{i=1}^{L-1} n_i (e^{-s} \sigma^x_{i+1} - \mathbb{1}) - \frac{1}{2} n_L (e^{-s} - 1).
		\label{eq:quantum_east_effective}
	\end{equation}
	It was shown in \cite{Pancotti2020Jun} that by modifying the regular quantum East Hamiltonian in Eq.~\eqref{app.eq:quantum_east} like shown above, the system's first order transition can be captured by a sharp delocalization ($s<0$)/localization ($s>0$) transition in the ground-state at $s=0$.
	Furthermore, it was shown how in the regime $s > 0$, localized eigenstates can be constructed at arbitrary energy density.
	Due to this clear picture how we can understand the dynamics in the regime $s>0$, we consider a quantum East Hamiltonian with the boundaries Eq.~\eqref{eq:quantum_east_effective} instead of Eq.~\eqref{app.eq:quantum_east} throughout the main manuscript.
	
	In Fig.~\ref{fig:quantum_east_bns} of the main manuscript, we showed the plot for the Lanczos coefficient for the operator $n_6$ for system size $L=13$. Here we show for different system sizes the same plot for ergodic and ergodicity-broken regimes in Fig.~\ref{fig:quantum_east-bn-Ls}.
	This shows that our cut-off $n=50$ for the initial ramp holds across all considered system sizes.
	Next we show the Lanczos coefficients for the operator $\sigma^x_6$ at system size $L=13$ in Fig.~\ref{fig:quantum_east_bns_operators}. 
	This shows that also for different operators the chosen cut-off $n=50$ is valid.
	Naturally, we have also checked for all other combinations of operators and system sizes to confirm the validity of the scrambling cut-off.

\end{document}